\documentclass[structabstract]{aa}  
\usepackage{txfonts,graphicx,natbib,url}

\bibpunct{(}{)}{;}{a}{}{,} 

\begin{document}
   \title{The VMC survey\thanks{Based on observations made with VISTA at ESO Paranal Observatory under program ID 179.B-2003.}}

   \subtitle{VII. Reddening map of the 30 Doradus field and the structure of the cold interstellar medium} 

   \author{B. L. Tatton	\inst{1}
          \and
          J. Th. van Loon\inst{1}
          \and
          M.-R. Cioni\inst{2}$^{,}$\inst{3}\thanks{Research Fellow of the Alexander von Humboldt Foundation.}
          \and
          G. Clementini\inst{4}
          \and
          J. P. Emerson\inst{5}
          \and
          L. Girardi\inst{6}         
          \and
          R. de Grijs\inst{7} 
          \and
          M. A. T. Groenewegen\inst{8}
          \and
          M. Gullieuszik\inst{8}
          \and
          V. D. Ivanov\inst{9}
          \and
          M. I. Moretti\inst{4}
          \and
          V. Ripepi\inst{10}
          \and
          S. Rubele\inst{6}
          } 
   \institute{Astrophysics Group, Lennard-Jones Laboratories, Keele University, ST5 5BG, United Kingdom
         \and
             University of Hertfordshire, Physics Astronomy and Mathematics, Hatfield AL10 9AB, United Kingdom
             \and
             University Observatory Munich, Scheinerstra{\ss}e 1, D-81679 Munchen, Germany 
             \and
             INAF, Osservatorio Astronomico di Bologna, Via Ranzani 1, 40127 Bologna, Italy
             \and
             Astronomy Unit, School of Physics \& Astronomy, Queen Mary University of London, Mile End Road, London E1 4NS, United Kingdom
             \and
             INAF, Osservatorio Astronomico di Padova, Vicolo dell'Osservatorio 5, 35122 Padova, Italy
             \and
             Kavli Institute for Astronomy and Astrophysics, Peking University, Yi He Yuan Lu 5, Hai Dian District, Beijing 100871, China
             \and
             Royal Observatory of Belgium, Ringlaan 3, 1180 Ukkel, Belgium
             \and
             European Southern Observatory, Santiago, Av. Alonso de C\'ordoba 3107, Casilla 19, Santiago, Chile
             \and
             INAF, Osservatorio Astronomico di Capodimonte, Via Moiariello 16, 80131, Naples, Italy
             }
   \date{Received 31 January 2013 / Accepted - - 2013}
 
  \abstract
   {The details of how galaxies have evolved over cosmological times are imprinted in their star formation history, chemical enrichment and morpho-kinematic structure. Parameters behind these such as effective temperature and metallicity can be measured by combining photometric techniques with modelling. However, there are uncertainties with this indirect approach from the ambiguity of colour and magnitude and the effects of interstellar reddening.}
   {In this paper we present a detailed reddening map of the central 30 Doradus region of the Large Magellanic Cloud; for both community use and as a test of the methods used for future use on a wider area. The reddening, a measurement of dust extinction, acts as a tracer of the interstellar medium (ISM).}
   {Near infrared (NIR) photometry of the red clump stars is used to measure reddening as their fixed luminosity and intermediate age make extinction the dominant cause of colour and magnitude variance. The star formation history derived previously from these data is used to produce an intrinsic colour to act as a zero point in converting colour to reddening values $E(J-K_\mathrm{s})$ which are subsequently converted to visual extinction $A_V$.}
   {Presented is a dust map for the 30 Doradus field in both $A_V$ and $E(J-K_\mathrm{s})$. This map samples a region of $1\degr\times1\rlap{.}\degr5$, containing $\sim1.5\times10^5$ red clump stars which probe reddening up to $A_V\simeq6$ mag. We compare our map with maps from the literature, including optical extinction maps and radio, mid- and far-infrared maps of atomic hydrogen and dust emission. Through estimation of column density we locate molecular clouds.} 
   {This new reddening map shows correlation with equivalent maps in the literature, validating the method of red clump star selection. We make our reddening map available for community use. In terms of ISM the red clump stars appear to be more affected by the cooler dust measured by $70\mu$m emission because there is stronger correlation between increasing emission and extinction due to red clump stars not being located near hot stars that would heat the dust. The transition from atomic hydrogen to molecular hydrogen occurs between densities of $N_H\,\simeq\,4\times10^{21}$ cm$^{-2}$ and $N_H\,\simeq\,6\times10^{21}$ cm$^{-2}$.} 

\titlerunning{The VMC survey -- VII: Reddening map} \authorrunning{B. L. Tatton et al.} \maketitle

%

\section{Introduction}
\label{30d:int}
The details of how galaxies have evolved over cosmological times are imprinted in their star formation history (SFH), chemical enrichment and morpho-kinematic structure. However, such studies are impeded by attenuation of starlight by dust in the interstellar medium (ISM) and its variance with wavelength. By mapping the extinction we are able to reverse its effects and hence improve the reliability of the derived SFH and three-dimensional morphology. At the same time, the distribution of dust provides valuable insight into the structure of the ISM and its relation to recent, current and potential future star formation from dense molecular clouds.

The Magellanic System (MS) comprises of the Large Magellanic Cloud (LMC) and Small Magellanic Cloud (SMC), the Magellanic Bridge connecting them and the trailing Magellanic Stream and its leading arms. The LMC and SMC are gas-rich dwarf galaxies of irregular morphology; past interactions between them, and between them and the Milky Way galaxy, have led to the creation of the Bridge and Stream. The LMC, which is the focus of this work, is $\sim50$ kpc away (e.g. \citealt{walker11,ripepi12}). It has a rotating disc of gas and young and intermediate-age stellar populations, seen under an inclination angle of $i=26\pm2^{\circ}$ \citep{rubele12}\footnote{Recent determinations of $i$ range from $23^{\circ}$ to $37^{\circ}$ \citep{subramanian10}.}. This structure extends $\sim10^\circ$ on the sky, and has a depth along the line-of-sight of $\sim4$ kpc \citep{subramanian09}. The metallicity is $\frac{1}{3}Z_\odot-\frac{1}{2}Z_{\odot}$ \citep{kunth00,cioni11}, decreasing with distance to the centre \citep{cioni09}. The LMC also has an off-centred stellar bar which may \citep{nikolaev04} or may not \citep{subramaniam09} be elevated above (i.e.\ in front of) the plane of the disc.

The $Y$, $J$, $K_\mathrm{s}$ near infrared (NIR) VISTA Survey of the Magellanic Clouds (VMC; \citealt{cioni11}, hereafter Paper I) is currently performing observations of the Magellanic System (MS). The work presented here concerns the central LMC $6\_6$ tile\footnote{Centred at RA$=05^\mathrm{h}\,37^\mathrm{m}\,40\rlap{.}^{\mathrm{s}}008$\,(J2000), Dec$=-69\degr\,22^{\prime}\,18\rlap{.}^{\prime\prime}120$\,(J2000).}. This region contains the famous \object{Tarantula Nebula} (30 Doradus) mini starburst -- the most active star formation region in the Local Group -- centred around the compact massive cluster \object{RMC 136a} (hereafter R136 -- see \citealt{evans11}). The tile further contains several supernova remnants the most famous of which is SN1987A, a number of star clusters, H\,{\sc ii} regions and the southern molecular ridge \citep{wong11}. The south--west part of this tile covers part of the LMC bar, characterised by a high stellar density.

The red clump (RC) stars form a metal-rich counterpart to the horizontal branch. Stars in this phase of stellar evolution undergo core helium and shell hydrogen burning. RC stars are of intermediate age (1--10 Gyr) and mass (1--3 M$_\odot$, \citealt{girardi01}); following the first ascent of the red giant branch (RGB) the RC phase typically lasts $\sim0.1$ Gyr (up to $0.2$ Gyr). RC stars are useful for mapping interstellar reddening and 3D structure because of their large numbers and relatively fixed luminosity. The former gives us a large sample size that is well distributed (due to their ages they have mixed relatively well dynamically) and the latter means that changes in magnitude and colour due to reddening and distance (i.e.\ environment and structure) quickly dominate over those as a result of population differences. We here use the VMC NIR photometry of RC stars to construct interstellar reddening maps. Extinction coefficients such as those determined by \citet{girardi08} can be used to translate the NIR extinction to other wavelengths. The RC has been used in producing extinction maps for the LMC at optical wavelengths using OGLE data. The LMC bar region was covered by \citet{subramaniam05} and most recently, a much larger region of the LMC and SMC was covered by \citet{haschke11}. In the NIR, the RC has mainly been used in the Milky Way (e.g. \citealt{alves00}), with some distance determination done for the LMC (e.g. \citealt{sarajedini02}, \citealt{grocholski07}). Reddening maps of the Milky Way bulge have been made using the NIR VISTA Variables in the Via Lactea survey data in \citet{gonzalez12}.

Extinction due to dust acts as a measurement of dust column density which can be compared with the LMC  maps of H\,{\sc i} gas emission \citep{kim98} and far infrared (FIR) dust emission \citep{meixner06}. Dust is seen both in predominantly atomic ISM and in molecular clouds, and it can thus be used to probe the atomic--molecular transition. The dust-to-gas ratio in a galaxy is mainly correlated with metallicity \citep{franco86}; some additional variance is linked to the mass outflow rate \citep{lisenfeld98}. The Very Large Telescope FLAMES Tarantula Survey (VFTS) has used diffuse interstellar bands (DIBs) as a measurement of the environments in the Tarantula Nebula \citep{vanloon13}.

Section \ref{30d:dat} describes the VMC survey in general. How we extract the RC stars and derive reddening values for them is described in Section \ref{30d:meth}. Section \ref{30d:res} shows the resulting maps. In section \ref{30d:dis} we compare these with some of the previously mentioned reddening and ISM maps, determining the atomic--molecular transition. Finally, Section \ref{30d:conc} summarises and concludes this work.


\section{VMC observations \& data} \label{30d:dat}
VISTA is located at the Paranal Observatory in northern Chile, it has a $4$m mirror and Infrared (IR) camera composed of an array of $16$ Raytheon VIRGO $2048\times2048$ $20\,\mu$m pixel detectors in a $4 \times 4$ pattern. Each detector is spaced apart by $90\%$ detector width along the x-axis and $42\%$ along the y-axis. By combining 6 stepped exposures (paw-prints) into one tile, an area of $1.5\,$deg$\,^{2}$ is covered at least twice\footnote{Regions have at least $2$ pointings (after combining jitters) for each paw print. The underexposed regions ($\times1$) at two edges are not included in our regions but should recovered when adjacent tiles are observed. Other small regions have up to $6$ overlaps which are accounted for in the confidence maps.}, with a pixel size of $0\rlap{.}^{\prime\prime}34$.

The VMC survey is a uniform and homogeneous survey of the MS in the NIR using the Visible and Infrared Survey Telescope for Astronomy (VISTA; \citealp{emerson06,dalton06}) and has two main scientific goals; to determine the spatially-resolved SFH and the 3D structure of the Magellanic System. The VMC survey is performed in the $Y$, $J$ and $K_\mathrm{s}$ filters (centred at $1.02\,\mu$m, $1.25\,\mu$m and $2.15\,\mu$m respectively), in the VISTA (Vega mag) photometric system. Once complete, the survey will cover $\sim180\,\mathrm{deg}^{2}$ of the MS. The average $5\sigma$ magnitude limits\footnote{Estimated using ESO VIRCAM Exposure Time Calculator  v3.3.3 with settings of: blackbody with T=$5000$K, Airmass=$1.6$, Seeing= $1\arcsec(K_\mathrm{s})$, $1.1\arcsec(J)$ and $1.2\arcsec(Y)$, Point Source, DIT=1 sec, Total Exposure Time = $2400$s $(Y)$, $2400$s $(J)$ and $9000$s $(K_\mathrm{s})$.} are $Y=21.9$ mag, $J=22.0$ mag, $K_\mathrm{s}=21.5$ mag; about $5$ mag deeper than those of the 2 Micron All Sky Survey (2MASS)\footnote{\url{http://www.ipac.caltech.edu/2mass/overview/about2mass.html}}.

As part of the VMC goals each tile was observed at different epochs. Each epoch has $5$ jitters\footnote{Exposures taken at slightly different positions for two main reasons; firstly, to remove stars in producing a sky for sky subtractions and secondly, to minimise contributions from cosmic rays and bad pixels.} resulting in the median time for exposure per paw-print, per epoch being $800$s in $Y$ and $J$ and $750$s in $K_\mathrm{s}$. In total there are $3$ epochs for $Y$ and $J$ and $12$ for $K_\mathrm{s}$. The combined average total exposure, per paw-print, is $2400$s in $Y$ and $J$ and $9000$s in $K_\mathrm{s}$. A more thorough description of the VMC survey can be found in Paper I.

The data used are part of the v20120126 VMC release retrieved from the VISTA science archive (VSA). The VSA and the VISTA Data Flow System pipeline are described in detail by \citet{cross12} and \citet{irwin04}, respectively. 

\begin{figure}
\resizebox{\hsize}{!}{\includegraphics[angle=0,width=\textwidth,clip=true]{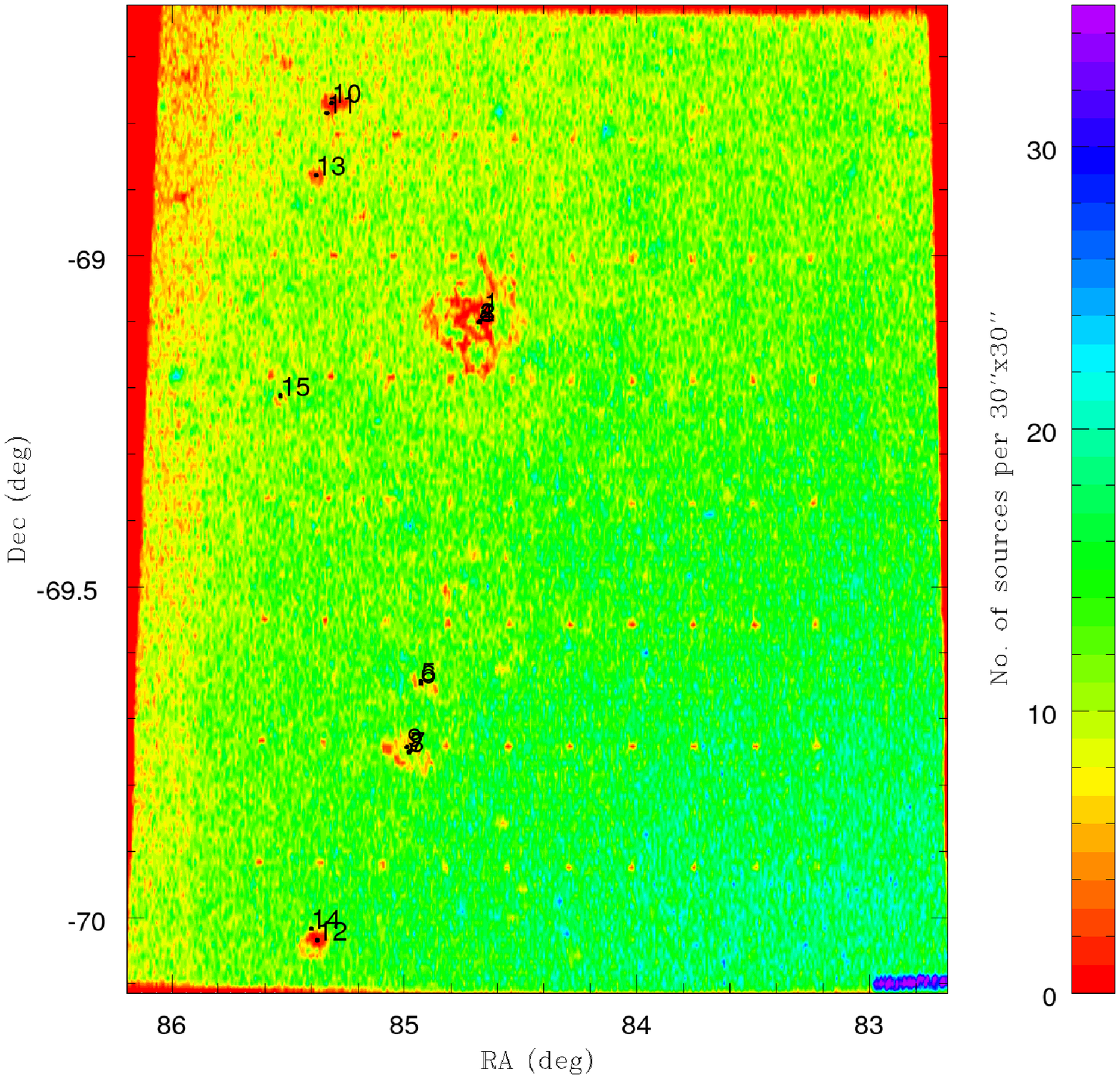}}\\
\resizebox{\hsize}{!}{\includegraphics[angle=0,width=\textwidth,clip=true]{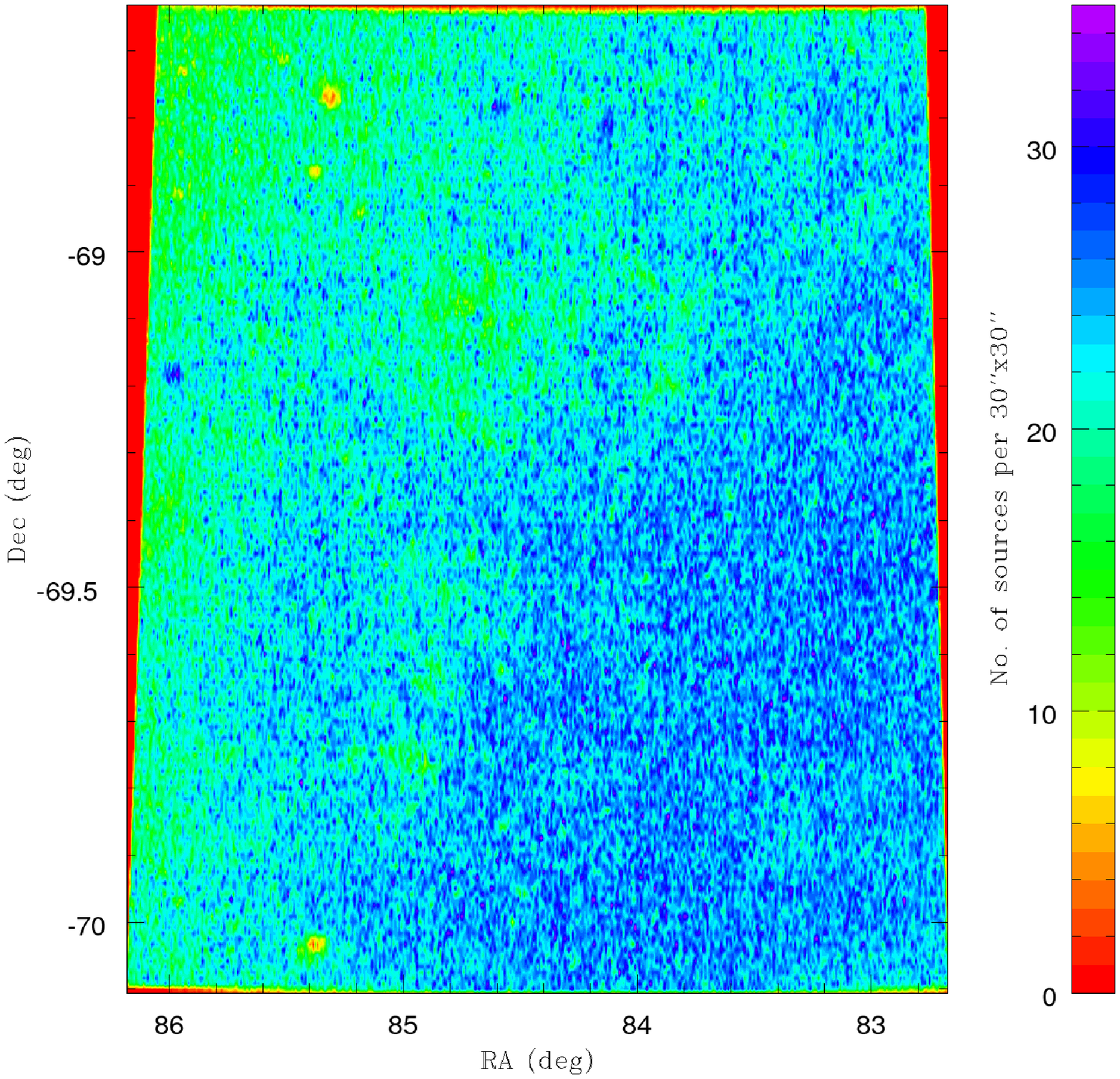}}
\caption{Density map containing all $K_\mathrm{s}$ band stars for tile $6\_6$ using aperture photometry (top), PSF photometry (bottom) from the VMC v20120126 release. The \textnormal{low density} gaps are described in Table \ref{tab:gap} and are much smaller in the PSF photometry. The $10\times7$ pattern of small gaps in the aperture photometry does not reflect a lack of data, but rather where the confidence is high due to contribution from multiple detectors. A software bug introduced as the VSA rescaled the confidence levels and pushed the values beyond the overflow limit so they show as low confidence rather than high, and hence not plotted. This bug has been fixed in future release\textnormal{s}.}
\label{fig:allstarmap}
\end{figure}

Point spread function (PSF) photometry (for details, see section 2.1 of \citealt{rubele12}  hereafter, Paper IV) is used instead of aperture photometry in this region because PSF photometry recovers more sources within crowded regions of which there are several in the $6\_6$ tile being studied in this paper. This is demonstrated in Figure \ref{fig:allstarmap} where density maps of aperture and PSF photometry are shown for this field and Table \ref{tab:gap} lists some of the known features associated with these incomplete regions, with star clusters and H\,{\sc ii} regions being the most prevalent.

\begin{table}
\caption{Objects associated with regions containing apparent gaps in Figure \ref{fig:allstarmap}.}
\label{tab:gap}
\[
\begin{tabular}{ccccc}
\hline
\hline
\noalign{\smallskip}
Object & no. & RA & Dec & Object type \\
 & & \degr & \degr & \\
\hline
Tarantula Nebula	&	1	&	84.658	&	 $-$69.085	&	H\,{\sc ii} (ionized) region	\\
NGC 2070	&	2	&	84.675	&	 $-$69.100	&	Cluster of stars	\\
R136	&	3	&	84.675	&	 $-$69.101	&	Cluster of stars	\\
BAT99 111	&	4	&	84.679	&	 $-$69.101	&	Wolf--Rayet star	\\
N 160A-IR	&	5	&	84.929	&	 $-$69.643	&	Young stellar object	\\
NGC 2080	&	6	&	84.929	&	 $-$69.647	&	Cluster of stars	\\
LH 105	&	7	&	84.975	&	 $-$69.748	&	Association of stars	\\
NGC 2079	&	8	&	84.979	&	 $-$69.750	&	Cluster of stars	\\
LHA 120-N 159	&	9	&	84.988	&	 $-$69.743	&	H\,{\sc ii} (ionized) region	\\
\textnormal{HD 38617}	&	10	&	85.312	&	$-$68.770	&	K3III star	\\ 
\textnormal{H88 310}	&	11	&	85.333	&	$-$68.785	&	Cluster of Stars	\\
\textnormal{HD 38706}	&	12	&	85.374	&	$-$70.034	&	M4III star	\\ 
\textnormal{HD 38654}	&	13	&	85.379	&	$-$68.879	&	M0/M1 star	\\
\textnormal{LHA 120-N 177}	&	14	&	85.400	&	 $-$70.017	&	H\,{\sc ii} (ionized) region	\\
\textnormal{NGC 2100}	&	15	&	85.533	&	 $-$69.212	&	Cluster of stars	\\
\noalign{\smallskip}
\hline
\end{tabular}
\]
\tablefoot{All data from SIMBAD (\url{http://simbad.u-strasbg.fr/simbad/}). Objects given in J2000 coordinates.\\
}
\end{table}

\section{Analysis} \label{30d:meth}

\subsection{Photometry}
\begin{figure}
\resizebox{\hsize}{!}{\includegraphics[angle=0]{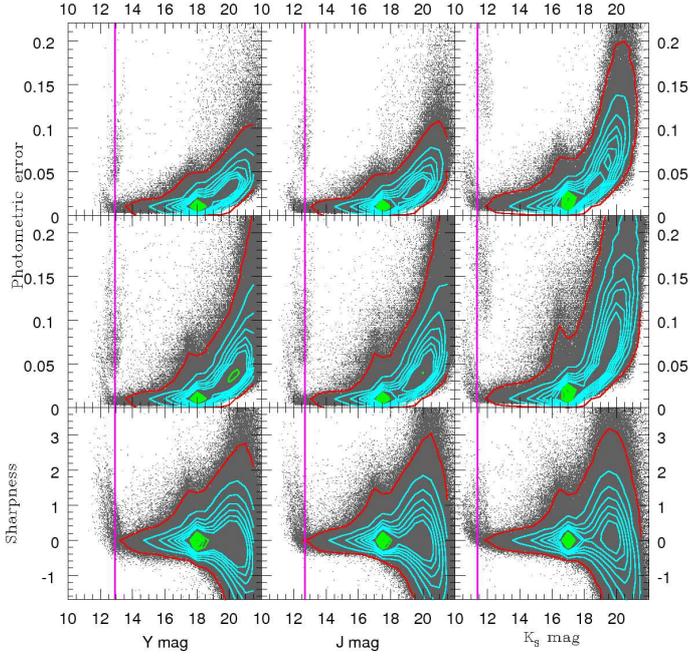}}
\caption{Sharpness (bottom) and error (middle and top; middle for all values of sharpness, top for sharpness between $-1$ \& $+1$) vs. magnitude for $K_\mathrm{s}$ (right), $J$ (middle) and $Y$ (left). Contours are overplotted in red, \textnormal{cyan} and green, saturation limit in \textnormal{magenta}.}
\label{fig:phot}
\end{figure}

In PSF photometry we can use photometric error and sharpness\footnote{An image-preculiarity statistic produced from the IRAF DAOPHOT allstar task used for producing PSF photometry (outlined in Paper IV).} for quality control. We plot photometric error and sharpness against magnitude for each band in Figure \ref{fig:phot}. Contour lines depict density of stars, relative to the densest region, for regions of $0.5$ mag and $0.01$ mag error or $0.2$ sharpness. Levels of $1\%$, $5\%$--$30\%$ (step $5\%$), $50\%$--$100\%$ (step $10\%$) are drawn in Figure \ref{fig:phot}, in red, \textnormal{cyan} and green, respectively. The saturation limit is shown in \textnormal{magenta}. Sources are extracted by the aperture-photometry pipeline up to $2$ magnitudes above this limit; see (\citealt{gullieuszik12}, hereafter Paper III, for details).

The patterns are very similar in each band; when sources are fainter the photometric error becomes larger and a similar change occurs in sharpness. A small number of sources exhibit a spike in error and sharpness at about $16.5$--$17.5$ mag, which precedes the more exponential rise. Their location with respect to the peak suggests this is caused by source blending or crowding effects. There is also a similar effect for bright objects with a large increase about the saturation limit and a separate, smaller increase seen beyond it. Both cases represent very low source density. We filter our selection to only include sources with sharpness between $-1$ and $+1$ in every band because very sharp points are likely to be bad pixels and un-sharp points are likely to be extended sources (such as cosmic rays). The top panel of Figure \ref{fig:phot} shows the effect this filtering has on photometric error. \textnormal{Using this filtering reduces the total number of sources to $9.35\times10^5$ (from $1.59\times10^6$).}

We are using the $J$ and $K_\mathrm{s}$ bands. Usage of the $Y$ band is covered in Appendix \ref{30d:app}.

\subsection{Colour--Magnitude Diagram} \label{30d:cmd}
\begin{figure}
\resizebox{\hsize}{!}{\includegraphics[angle=0,clip=true]{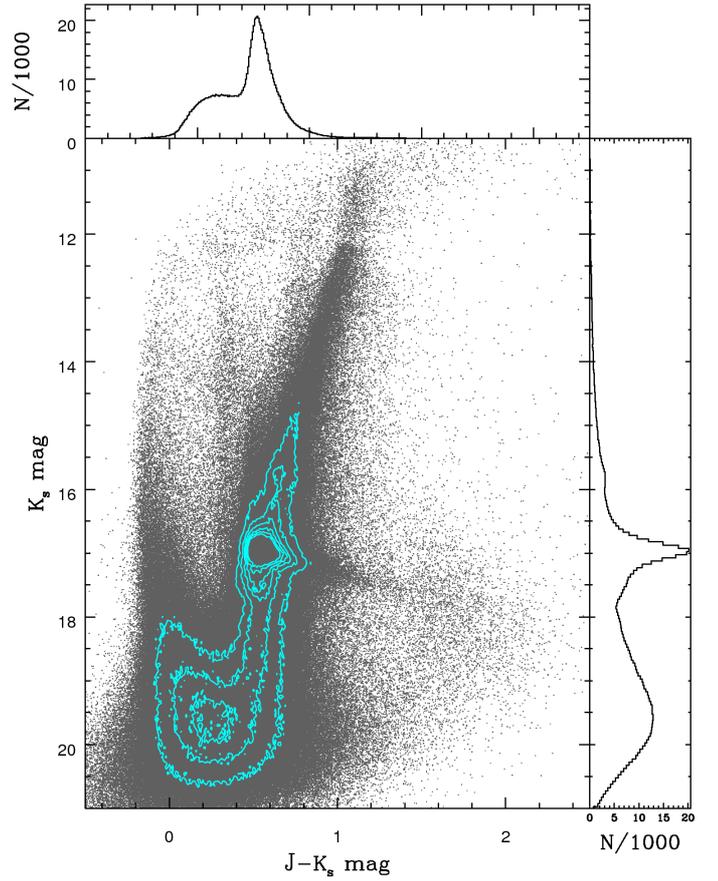}}
\caption{$K_\mathrm{s}$ vs. $(J-K_\mathrm{s})$ CMD of LMC field $6\_6$ accompanied by \textnormal{cyan} contours representing density, and histograms of $(J-K_\mathrm{s})$ (bin size $0.01$ mag) and $K_\mathrm{s}$ (bin size $0.05$ mag). It can be seen that this field is largely made up of main sequence and RC stars.}
\label{fig:cmd}
\end{figure}

The colour--magnitude diagram (CMD) is an important tool for evaluating the sources contained in the data for this region. Figure \ref{fig:cmd} shows the CMD for this tile, containing approximately $10^6$ stellar sources. This is an old population; the main sequence, RGB and RC populations are all resolved, and the asymptotic giant branch (AGB) is partly seen (the VISTA saturation limit prevents the full AGB population being seen; see Paper III for details) and the tip of the RGB (TRGB) is clearly seen at $K_\mathrm{s}=12$ mag, consistent with literature findings (e.g. \citealt{cioni00}). There is a noticeable \textnormal{amount of excess sources beyond $(J-K_\mathrm{s})=1.5$ mag and $K_\mathrm{s}=16.5$--$19$ mag}. These are typical magnitudes and colours of background galaxies \citep{kerber09}.

As the CMD is extremely dense in certain regions (the RC in particular), a contour map of stellar density is overplotted in \textnormal{cyan} on Figure \ref{fig:cmd}. The lines are built with a bin size resolution of $0.01$ mag in $K_\mathrm{s}$ and $0.05$ mag in $(J-K_\mathrm{s})$. The contour lines are drawn with the cell containing the most stars ($1172$) as the peak level and the \textnormal{cyan} contours are where the number of sources range from $5\%$--$30\%$ of the peak, with a step of $5\%$. 

The panels to the top and right of the CMD are histograms of colour (top) and magnitude (right) with the same bin sizes. The major peaks ($K_\mathrm{s}=17$ mag and $(J-K_\mathrm{s})=0.6$ mag) are within the RC region while the minor peaks ($K_\mathrm{s}=19$ mag and $(J-K_\mathrm{s})=0.2$ mag) are due to the main sequence.

\subsection{Defining the red clump} \label{30d:rcloc} 
\begin{figure}
\resizebox{\hsize}{!}{\includegraphics[angle=0]{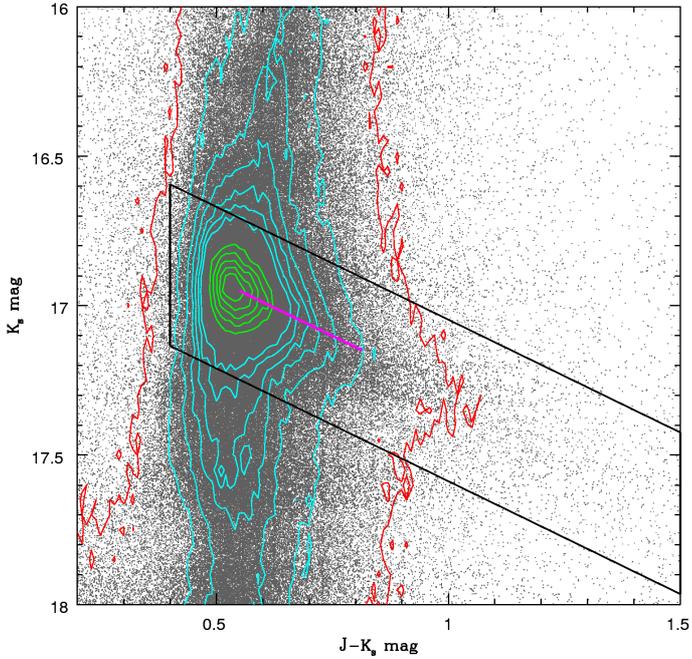}}
\caption{$K_\mathrm{s}$ vs. $(J-K_\mathrm{s})$ CMD with stellar density contour lines added (red: $1\%$ of peak, \textnormal{cyan}: $5\%$--$30\%$ of peak with step of $5\%$, green: $50\%$--$100\%$ of peak with step of $10\%$). Estimated reddening vector is in \textnormal{magenta} and RC selection box is marked in black.}
\label{fig:rcloc}
\end{figure}

Figure \ref{fig:rcloc} zooms in on an area of the CMD in which high stellar surface density is related to the RC. As the densest region is more compact we add green contours whose lines cover $50\%$--$100\%$ densities with steps of $10\%$; we also add a red contour at $1\%$ density. The green contours coincide with the RC and the $30\%$ and $25\%$ \textnormal{cyan} contours surround this feature. The $20\%$ contour extends to $K_\mathrm{s}=17.6$ mag. This is $K_\mathrm{s}\simeq0.4$ mag below the $30\%$ contour. This feature is made up of RGB stars and the secondary red clump (SRC) feature. SRC stars are helium burning (with enough mass to ignite helium in non-degenerate conditions), found up to $0.4$ mag away from the main clump and are the youngest RC stars with an age of 1 Gyr \citep{girardi99}. The smaller density of the SRC and the fact it may be less uniform across the whole LMC (due to the young age) mean we do not make efforts to include it in our RC star selection.

Several approaches can be taken for defining RC selection. A fixed range in colour and magnitude is one possibility. However, contamination issues (caused by RGB stars) arise due to covering a large magnitude range. From the contour lines we are able to see that for the bluer colours ($(J-K_\mathrm{s})<0.5$ mag) the density levels are essentially constant with respect to colour but for redder colours ($(J-K_\mathrm{s})>0.6$ mag) the contours stretch in colour and magnitude. This gradient is caused by the attenuating effects of dust and is described by the reddening vector. The magnitude range for the selection of RC stars depends on colour, tracing the reddening vector; in this way we can keep it narrow enough such as to minimise contamination by non-RC sources. To determine the $K_\mathrm{s}$ cuts we first determine the reddening vector as the gradient between the RC peak and the reddest point on the $5\%$ contour level (the $1\%$ contour level is too noisy and the $10\%$ contour level does not go to a point to the extent of the $5\%$ contour level). The line between these two points is shown in \textnormal{magenta} in Figure \ref{fig:rcloc} and the gradient is $0.754$. The magnitude range is determined from the highest and lowest $K_\mathrm{s}$ magnitudes of the $30\%$ contour. This region covers $\Delta\,K_\mathrm{s}=0.54$ mag for any given colour. \textnormal{The reason for choosing the 30\% contour is using lower \% contours makes the magnitude range too broad, increasing RGB star contamination while using higher \% contours makes the range too narrow, losing distance modulus variance.}

An alternative method to determine the gradient of the reddening vector is to use the VISTA filter application of the \citet{girardi08} extinction coefficients from Paper IV \footnote{This is described in further detail in Section \ref{30d:av}.}. For a typical RC star in the LMC ($T_\mathrm{eff}=4250\,$K, log$\,{g}=2$) the gradient is $0.741$, agreeing well with the previous measurement.

We exclude stars bluer than $(J-K_\mathrm{s})=0.4$ mag and redder than $(J-K_\mathrm{s})=1.5$ mag to minimise contamination from young main sequence stars and background galaxies, respectively. From Figure \ref{fig:rcloc}, the contribution of stars redder than $(J-K_\mathrm{s})>1.0$ mag is minor but is required to probe the reddest populations.

The final selection box plotted in black in Figure \ref{fig:rcloc} is the area of the CMD where the following equations are satisfied:
\begin{equation} 0.4 \leq (J-K_\mathrm{s}) \leq 1.5 \,\mathrm{mag} \end{equation}
\begin{equation}  K_\mathrm{s} \geq 0.754\times(J-K_\mathrm{s}) + 16.293 \,\mathrm{mag} \end{equation} 
\begin{equation}  K_\mathrm{s} \leq 0.754\times(J-K_\mathrm{s}) + 16.833 \,\mathrm{mag} \end{equation} 

\begin{figure}
\resizebox{\hsize}{!}{\includegraphics[angle=0,clip=true]{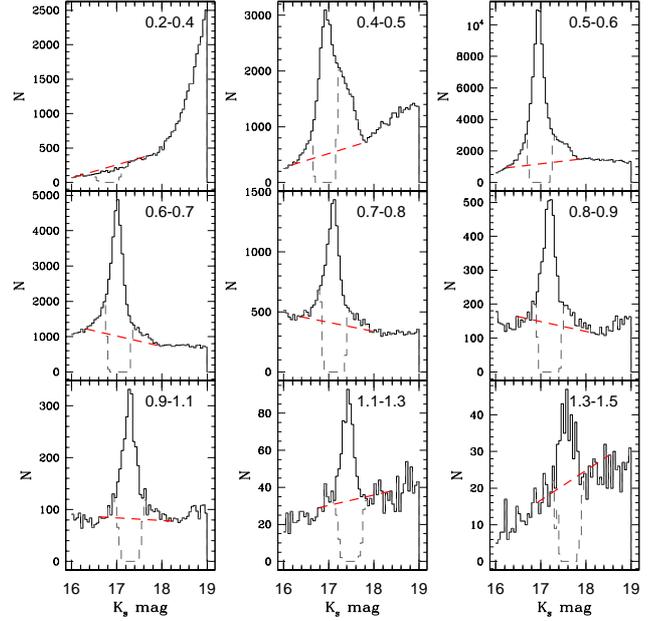}}
\caption{$K_\mathrm{s}$ band histograms (luminosity functions; bin size of $0.05$ mag) for $(J-K_\mathrm{s})$ colour given in panels. Black: all stars, dashed grey: excluding RC stars, dashed red: contamination interpolation.}
\label{fig:rcnon}
\end{figure}

\textnormal{Applying this selection yields a total of $150,328$ RC stars.} A problem with any selection for the RC is contamination from other components. These vary along our colour selection but include intermediate mass He-burning stars, young main sequence stars, background galaxies and, around the peak of the RC population, the first ascent of the RGB. In Figure \ref{fig:rcnon} we show histograms (luminosity functions) of $K_\mathrm{s}$ magnitude, for strips in $(J-K_\mathrm{s})$ colour; we show these histograms including all sources (in black), as well as after excluding stars satisfying the RC selection criteria (in dashed-grey).

Using the histogram we estimate the amount of non-RC population inside the RC selection by interpolating from points lying outside of the RC selection. In performing the interpolation, care was taken not to overestimate contaminants by using the distribution excluding RC selection (shown in dashed grey in Figure \ref{fig:rcnon}) and interpolating between $0.55$ mag from the brightest zero and $0.75$ mag away from the dimmest zero. This interpolation is shown in dashed-red in Figure \ref{fig:rcnon}.

These points excluded the fringes of the non-selected RC population as these have a large spread in magnitude, a larger degree of RGB contamination (for some colour ranges) and are minor compared to the selected RC population, which justifies their exclusion from the selection.

We can see in all panels, except the bluest, a bump where the RC occurs \textnormal{and this is covered by the magnitude range chosen}. The bluest colour bin ($(J-K_\mathrm{s})=0.2$--$0.4$ mag, \textnormal{not part of the RC selection}) lack this feature and the reddest colour bin ($(J-K_\mathrm{s})=1.3$--$1.5$ mag) start to have this feature become less distinguishable from contaminants.

In the panels covering $(J-K_\mathrm{s})=0.4$--$0.6$ mag there is an unusual knee-like bump at dimmer magnitudes compared with the more normal distribution of the other panels. This is a mixture of the RGB continuum (the RGB is found at $(J-K_\mathrm{s})\simeq0.5$ mag for these colours) and SRC.

\begin{figure*}[!t]
\includegraphics[width=0.33\textwidth,clip=true]{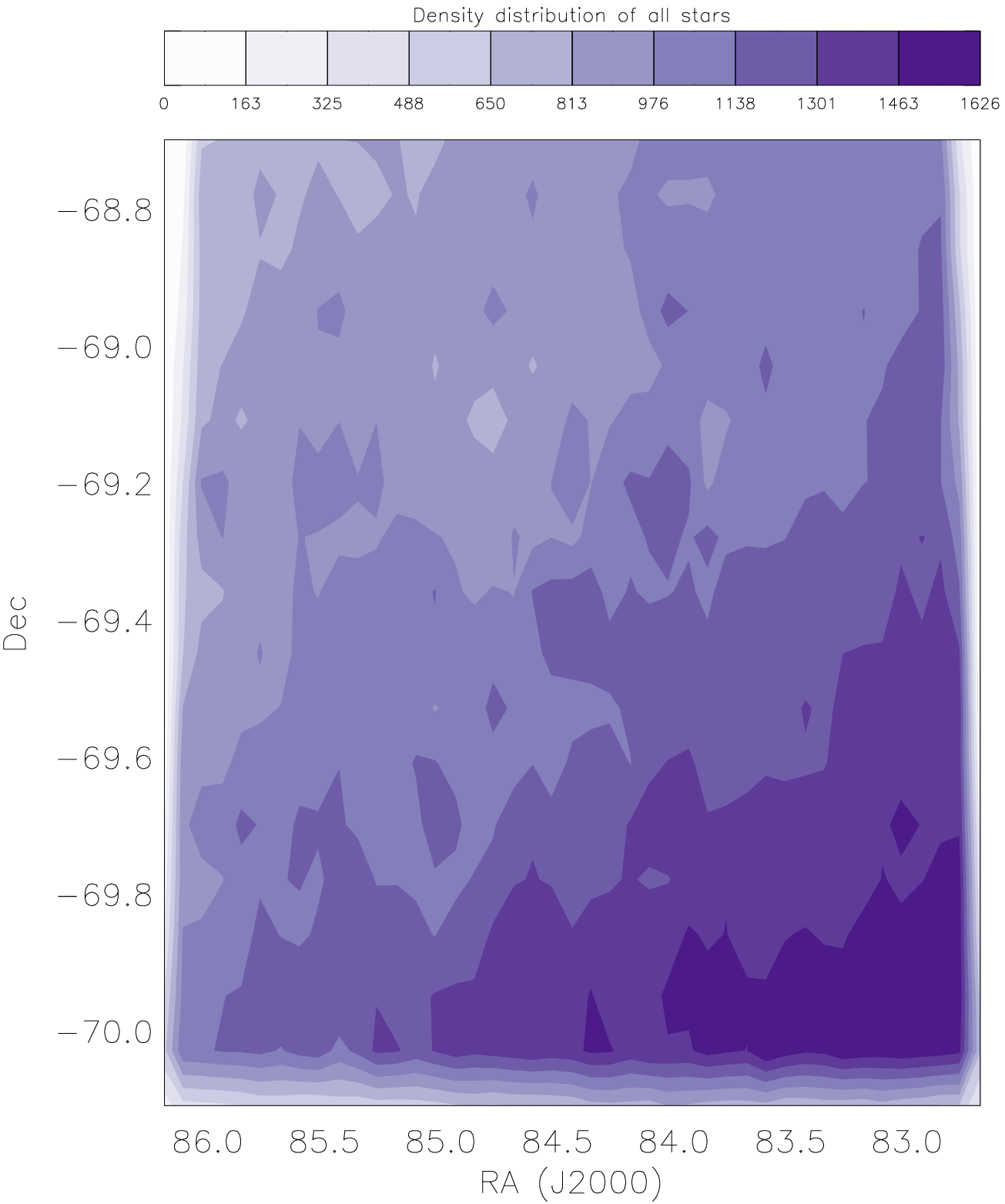}
\includegraphics[width=0.33\textwidth,clip=true]{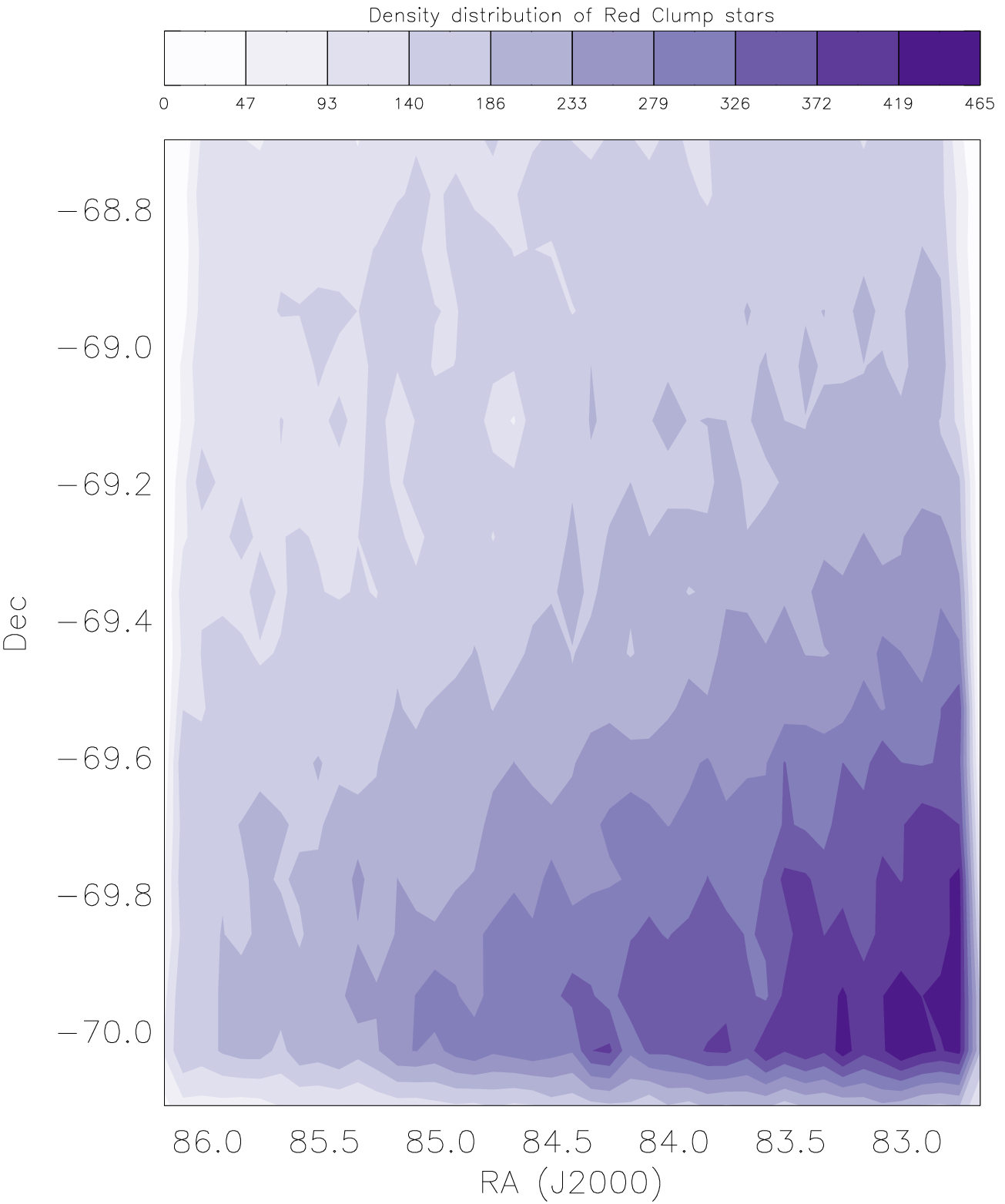}
\includegraphics[width=0.33\textwidth,clip=true]{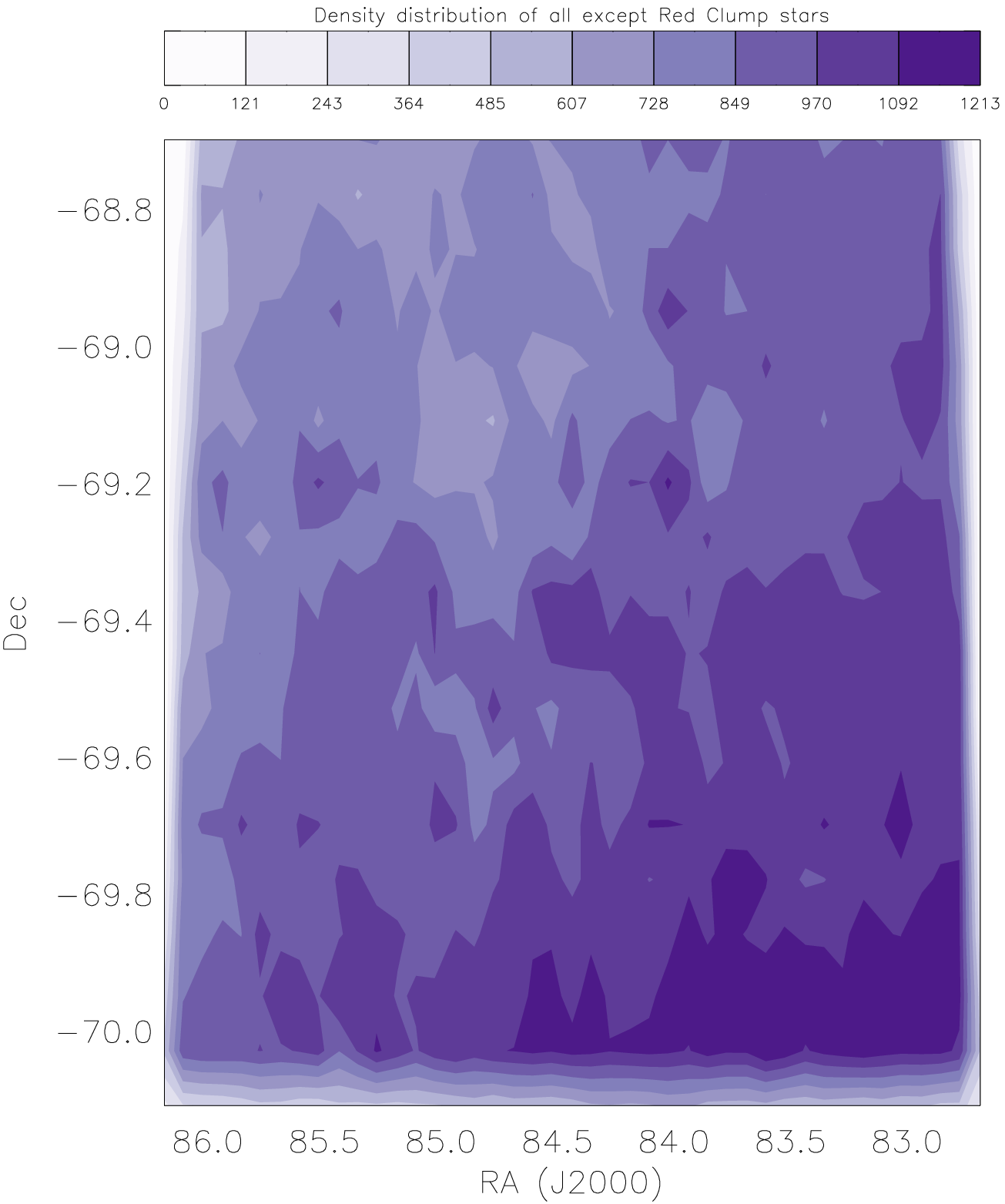}\\
\includegraphics[width=0.33\textwidth,clip=true]{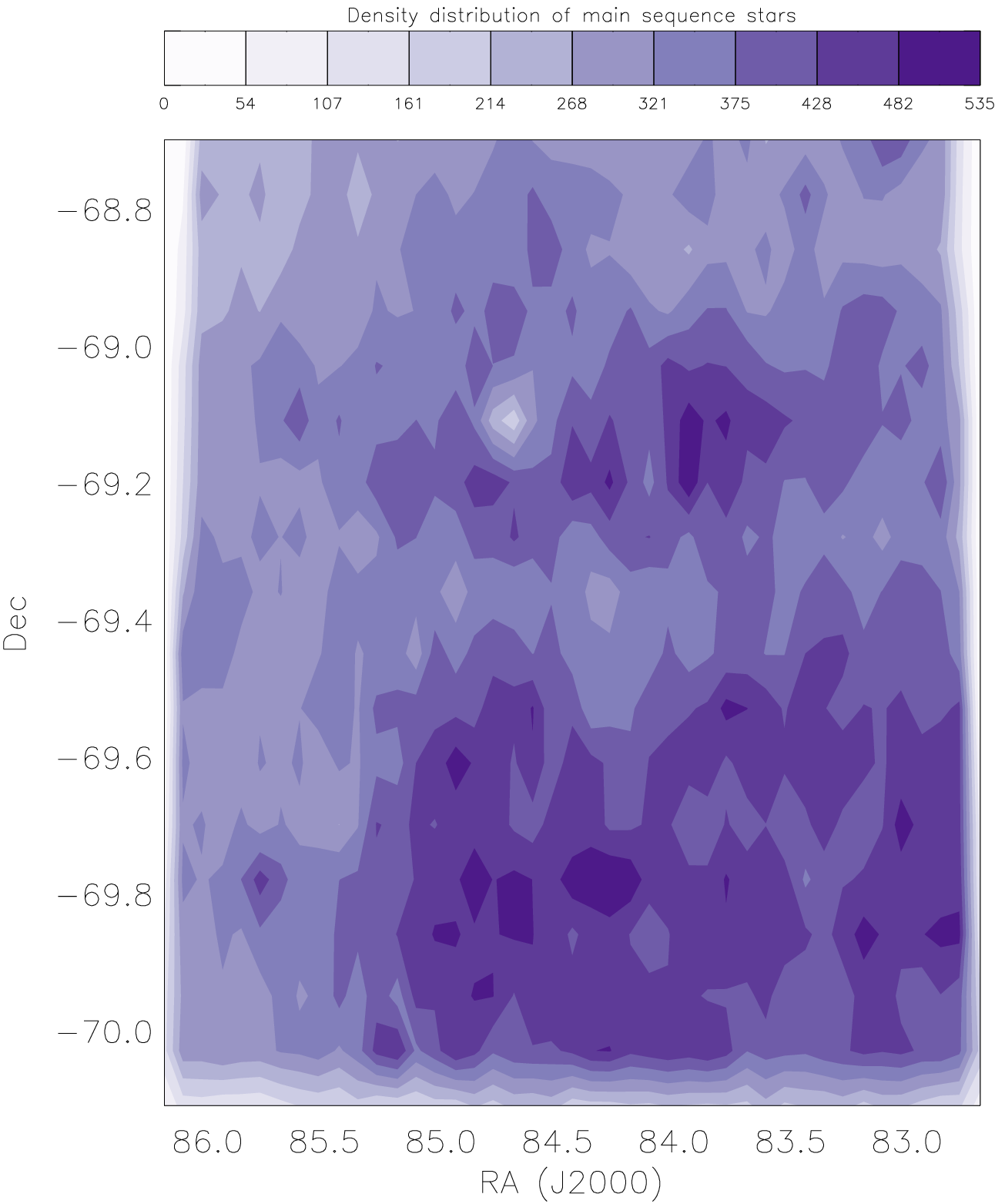}
\includegraphics[width=0.33\textwidth,clip=true]{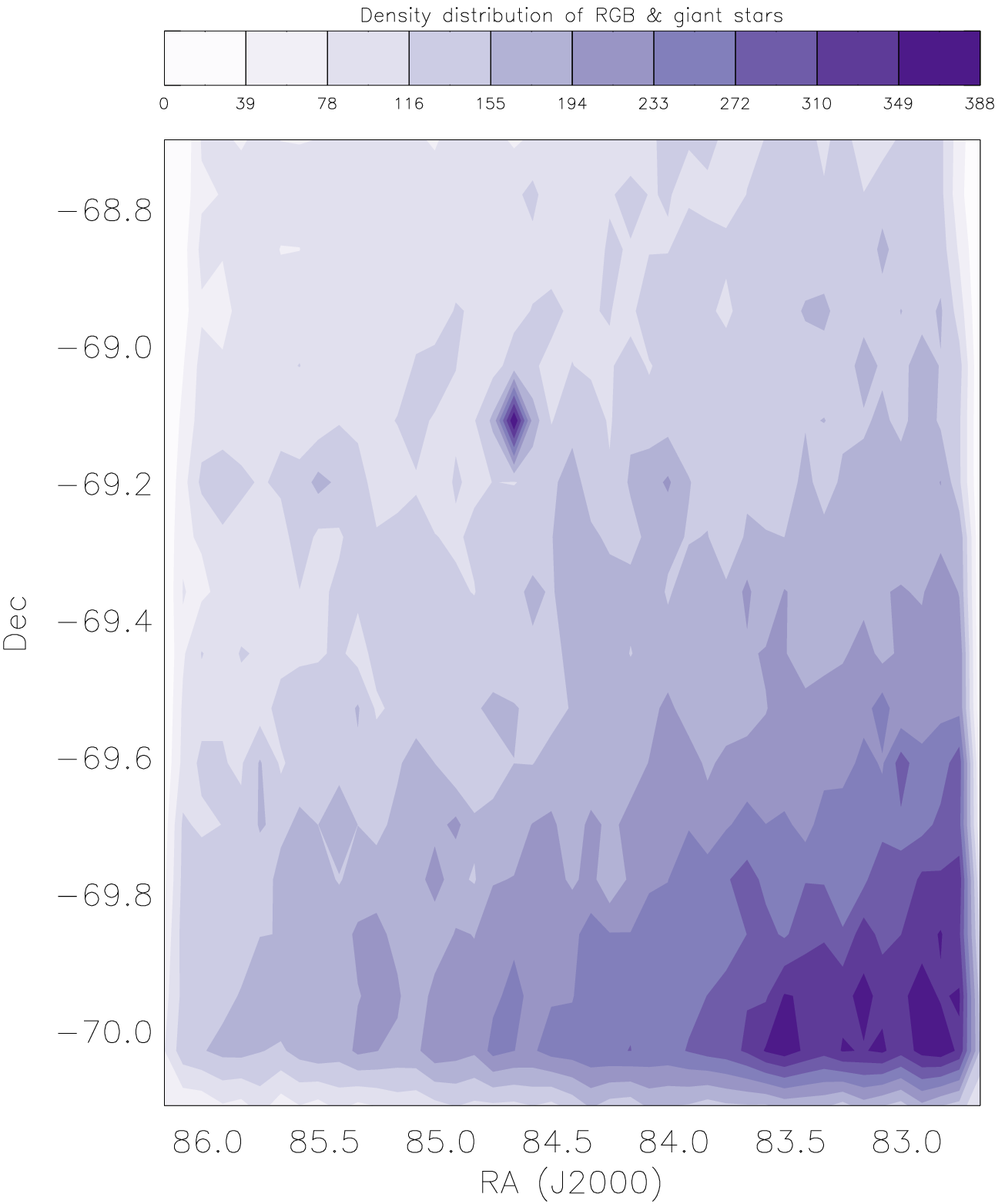}
\includegraphics[width=0.33\textwidth,clip=true]{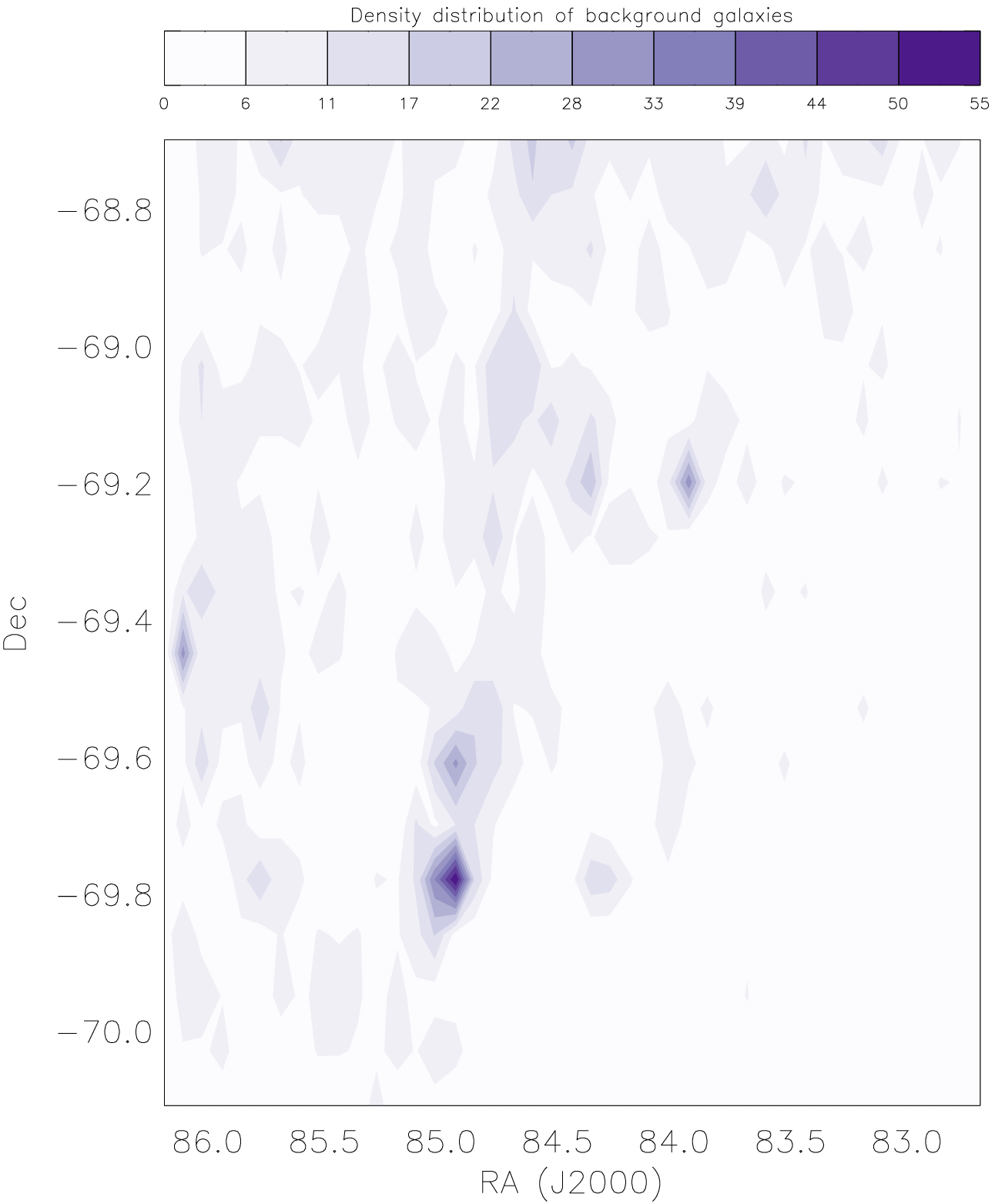}\\
\caption{Density of all stars (top--left), RC stars (top--middle), all except RC stars (top--right), main sequence stars (bottom--left) and RGB \& giant stars (bottom--middle) and background galaxies (bottom--right) within the field for regions of $5^{\prime}\times5^{\prime}$.}
\label{fig:dens}
\end{figure*}

\subsection{Stellar and red clump population density and distribution} \label{ssec:dens}
To gauge how well the RC (and hence, the estimated reddening) represents the stellar population as a whole, we compare the distribution on the sky of the RC against that of all stars and other populations using the following CMD selections:
\begin{itemize}
\item All stars: CMD range of $12{}\leq{}K_\mathrm{s}{}\leq{}20$ mag and $(J-K_\mathrm{s}){}\leq{}1.5$ mag.
\item RC stars: defined on the basis of the selection box in Section \ref{30d:rcloc} (using equations 1--3).
\item All except RC: All stars range excluding the RC selection box.
\item Lower main sequence: sources dimmer than $16.5{}\leq{}K_\mathrm{s}{}\leq{}20$ mag, bluer than $(J-K_\mathrm{s}){}\leq{}1.5$ mag and excluding the RC stars selection.
\item RGB and other giants: $12{}\leq{}K_\mathrm{s}{}\leq{}16.5$ mag, $(J-K_\mathrm{s}){}\leq{}1.5$ mag and excluding the RC selection.
\item Background galaxies: $(J-K_\mathrm{s}){}\geq{}1.5$ mag for $12{}\leq{}K_\mathrm{s}{}\leq{}20$ mag.
\end{itemize}
The magnitude range is chosen due to completeness of the CMD and photometric errors beyond these boundaries (see Figure \ref{fig:phot}; regrettably, this does exclude what can be seen of the AGB population) and the colour limit eliminates effects of background galaxies on the selection.

For each of those selections, we then produce contour maps of stellar density for regions sized $5^{\prime}\times5^{\prime}$. The output is displayed for $10$ evenly spaced contour levels between zero and the peak density. Figure \ref{fig:dens} shows these for the distribution of all stars (top--left), RC stars (top--middle), all stars except RC stars (top--right), main sequence stars (bottom--left), RGB and other giant stars (bottom--middle) and background galaxies (bottom--right).

The general trends seen are, an increase in sources towards the south--west region, (this region is where the LMC bar crosses the tile) and a decrease surrounding the 30 Doradus region (RA=$84\rlap{.}^{\circ}65$, Dec=$-69\rlap{.}^{\circ}08$). However, neither of these trends are identical for all populations. In particular the main sequence population (Figure \ref{fig:dens}, bottom--left) has less of an increase towards the LMC bar. This may be due to the main sequence stars being more abundant throughout the tile because the population distribution is flatter; as very few areas have contour levels below $50\%$ of the peak. The main sequence peaks in southern and western regions, the bar region is more abundant than the north--east.

The intermediate aged RGB and RC populations (Figure \ref{fig:dens}, top-- and bottom--middle) increase towards the bar, which suggests the bar itself is of intermediate age and not particularly metal poor. This matches the findings of \citet{cioni09}; the RGB star content of the bar is a few Gyr old and has a metallicity of [Fe/H]$=-0.66\pm0.02$ dex. The drop in the north--east region (clearer in these intermediate aged populations) correlates with areas of redder colour, such as regions $0$-$2$ and $1$-$2$ in Figure \ref{fig:gramcomp}.

The background galaxy distribution (Figure \ref{fig:dens}, bottom--right) is not flat or uniform as would be expected. If we ignore the lowest $2$ contour levels we find a few areas contain nearly all of the remaining $8$ contour levels. We investigated the possibility of these areas being galaxy clusters but did not find any to be known at the co-ordinates. However, it is also possible these are highly reddened RC stars, excluded by the colour cut-off (defined due to high amounts of noise). The presence of H\,{\sc ii} regions (LHA 120-N 159, LHA 120-N 157) in these areas suggests that this is the case. The RGB and other giant star distribution similarly has an unexpected find in the form of the peak region around R136 (in addition to the more typical peak around the LMC bar). This feature may be giant or pre-main sequence stars included in the CMD cut off as we did not define a lower colour limit. We return to this point in Section \ref{30d:r136}. \textnormal{From looking at these different populations, the density distributions should be taken into consideration when comparing RC to non RC populations, especially for the younger main sequence stars.}

\subsection{Reddening values} \label{30d:red}
The reddening in colour excess form, $E(J-K_\mathrm{s})$, is obtained using:
\begin{equation} E(J-K_\mathrm{s})=(J-K_\mathrm{s})_\mathrm{obs}-(J-K_\mathrm{s})_0 \end{equation}
Where $(J-K_\mathrm{s})_\mathrm{obs}$ is the observed colour of a RC star and $(J-K_\mathrm{s})_0$ is the colour of a non-reddened RC star known as the intrinsic colour.

\subsubsection{Determination of intrinsic colour $(J-K_\mathrm{s})_0$}
To determine the intrinsic colour, we make use of version 2.3 of the Padova isochrones\footnote{\url{http://stev.oapd.inaf.it/cgi-bin/cmd_2.3}}, using the evolutionary tracks of \citet{marigo08} and \citet{girardi10}, converted to the VISTA Vegamag system as described in Paper IV.

As intrinsic colour varies with metallicity and age, we constrain the possible age and metallicity ranges using the SFH produced in Paper IV on two corners of the tile (these areas of low extinction correspond to regions $0$-$0$ and $3$-$2$ in Table \ref{tab:clumppos}). Their SFH is fairly consistent for ages older than $1$ Gyr. 

We use their data for the best fitting age--metallicity relation (AMR) as input for the Padova isochrones. 
The metallicity is converted onto the $Z$ scale using the relation; log$\,Z=\mathrm{[Fe/H]}-1.7$ (Figure 5 of \citealt{caputo01} show this for LMC, SMC and Galactic Cepheids).
Two sets of isochrones are produced; the first, younger set covers log$\,(t/{\mathrm{yr}})=9.1$--$9.4$ with $Z=0.0125$ and the second, older set covers log$\,(t/{\mathrm{yr}})=9.4$--$9.7$ with $Z=0.0033$ (log$\,(t/{\mathrm{yr}})=9.4$ appears in both due to the SFH histogram bins overlapping). The age steps are $\Delta$log$\,(t/{\mathrm{yr}})=0.1$ for both. Foreground Galactic extinction was not accounted for as we want to find total reddening (when using the \citet{schlegel98} value; $A_V=0.249$ mag, the $(J-K_\mathrm{s})$ colour is reddened by $0.04$ mag).

As the isochrone tables are in absolute magnitude we apply an LMC distance modulus of $18.46\pm0.03$ mag \citep{ripepi12}\footnote{This is consistent with $18.48\pm0.05$ mag \citep{walker11}.} to the $K_\mathrm{s}$ magnitude.

\begin{figure}
\resizebox{\hsize}{!}{\includegraphics[angle=0,width=\textwidth]{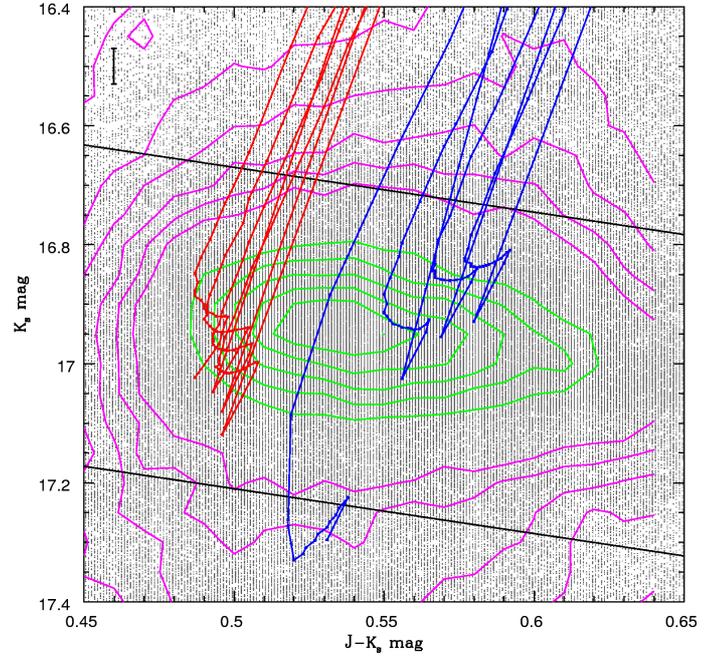}}
\caption{$K_\mathrm{s}$ vs. $(J-K_\mathrm{s})$ CMD with isochrone line for the helium burning sequence for two metallicities and the appropriate age ranges: Younger, more metal rich (log$\,(t/{\mathrm{yr}})=9.0$--$9.4$, $Z=0.0125$) are in blue and the older, more metal poor (log$\,(t/{\mathrm{yr}})=9.4$--$9.7$, $Z=0.0033$) are in red. The black error bar in the top left represents error in LMC distance. \textnormal{Magenta} and green contours are described in Figure \ref{fig:rcloc}. \textnormal{The RC peaks are consistent with the red isochrones when accounting for foreground Galactic extinction (estimated to be $J-K_\mathrm{s}=0.04$ mag from \citealt{schlegel98}).}} 
\label{fig:isoli}
\end{figure}

Figure \ref{fig:isoli} shows the CMD region of the RC stars and the isochrones for only the helium burning phase (after the RGB phase). Shown are the contours and RC selection box like in Figure \ref{fig:rcloc}, distance error for isochrones is shown in \textnormal{magenta}, and the isochrones are plotted in blue and red for the younger and older sets, respectively. 
The earlier, more age dependent phases (younger than $\sim2$ Gyr; log$\,(t/{\mathrm{yr}})<9.4$) do not make sense for a majority of the data as it would suggest unusually low or negative reddening. Also, the log$\,(t/{\mathrm{yr}})=9.1$ track lies outside our RC selection. The older tracks are contained within a narrower range of $\sim0.025$ mag in colour and all but the end stages of helium burning are within the RC selection. The average along the older tracks, $(J-K_\mathrm{s})_0=0.495$ mag, is adopted as the intrinsic colour. 

Comparing with Table \ref{tab:clumppos}, we see for regions $0$-$0$, $3$-$1$ and $3$-$2$ the median is around $E(J-K_\mathrm{s})\simeq0.035$ mag. This is consistent with the foreground Galactic extinction estimate of $(J-K_\mathrm{s})=0.04$ mag from \citet{schlegel98}. For the whole tile the mean is $E(J-K_\mathrm{s})=0.091$ mag, more than double the foreground Galactic estimate because \citeauthor{schlegel98} do not (and is not intended to) account for the additional reddening within the Magellanic Clouds.

\subsection{Reddening law \& conversion to $A_V$} \label{30d:av}
For consistency with the VMC survey and the Padova isochrones, we use the same extinction coefficients from \citet{girardi08} with the VISTA filter update from Paper IV. The extinction coefficients are based on the stellar spectra from \citet{castelli04}, in addition to the \citet{cardelli89} extinction curve. The VISTA filter transmission curves are fully accounted for giving it an advantage over Cardelli's curve alone (see Appendix \ref{30d:law}). Table \ref{tab:aav} gives the values relative to $A_V$ for a typical RC star ($T_\mathrm{eff}=4250\,$K, log$\,g=2$). We find $E(J-K_\mathrm{s})=0.16237\times A_V$.

\begin{table}
\caption{Extinction ratio $A_{\lambda}/{A_V}$ for the \citet{cardelli89} extinction law for VISTA passbands for a typical RC star.}
\label{tab:aav}
\[
\begin{tabular}{c|ccccc}
\hline
\hline
\noalign{\smallskip}
VISTA filters & $Z$ & $Y$ & $J$ & $H$ & $K_\mathrm{s}$ \\
\hline
$\lambda_\mathrm{eff}$ ($\mu$m) & 0.8763 & 1.0240 & 1.2544 & 1.6373 & 2.1343 \\
w$_\mathrm{eff}$ ($\mu$m) & 0.0780 & 0.0785 & 0.1450 & 0.2325 & 0.2600 \\
${A_{\lambda}/{A_V}}$ & 0.5089 & 0.3912 & 0.2825 & 0.1840 & 0.1201 \\
\hline
 \noalign{\smallskip}
\end{tabular}
\]
\end{table}


\section{Results}
\label{30d:res}

\subsection{Reddening map} \label{30d:redmap}
\begin{figure*}
\includegraphics[width=\textwidth]{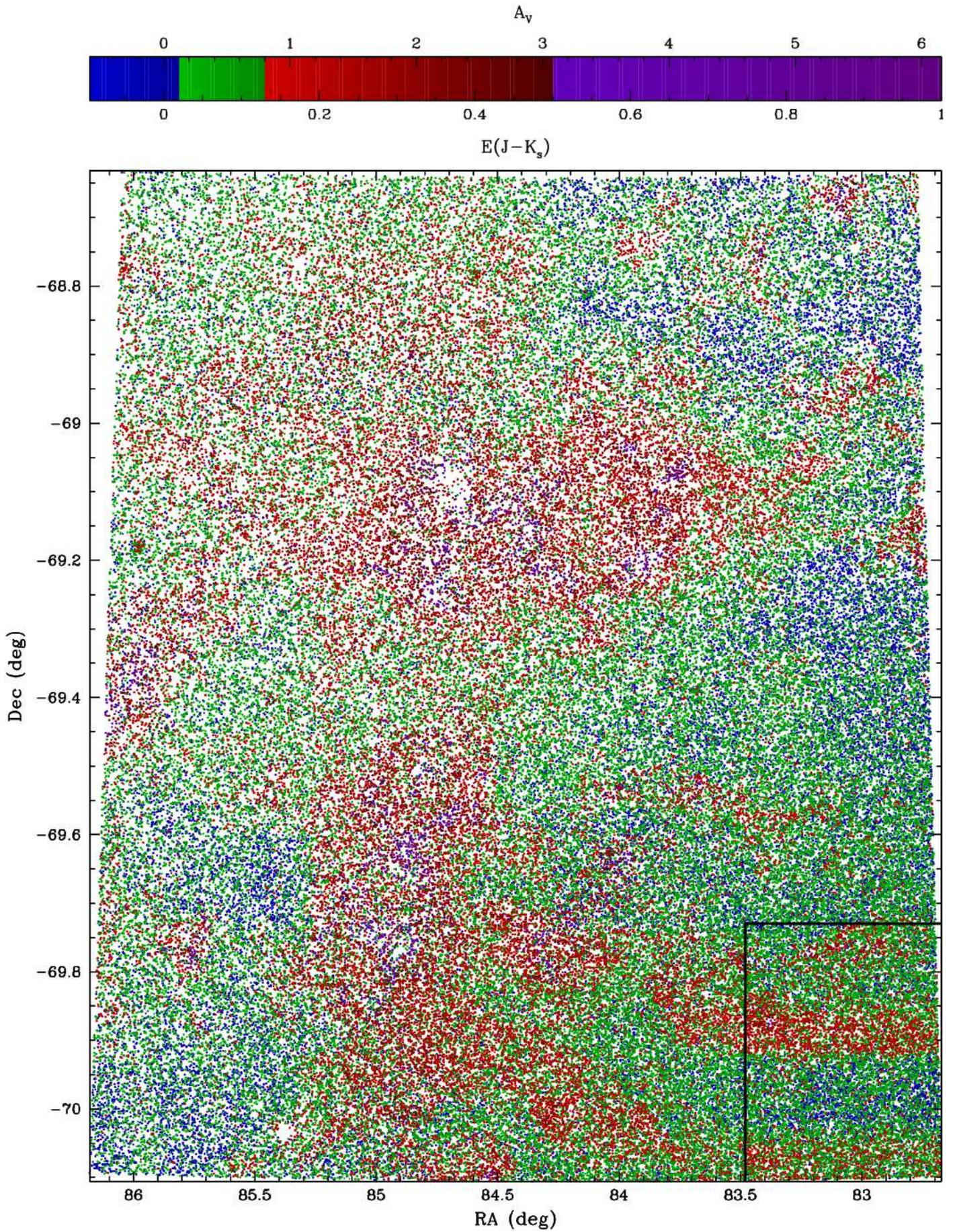}
\caption{Extinction map for whole tile. Colour scale is shown on top ($E(J-K_\mathrm{s})$ scale below and $A_V$ scale above) and reused in Figure \ref{fig:colgramexcite}. The black line marks the location including the effects of the two offset observations by detector 16 (see Section \ref{30d:issues} for more information).}
\label{fig:map}
\end{figure*}

The RC star extinction map for the tile is shown in Figure \ref{fig:map}; a histogram of the extinction scale (in $A_V$ and $E(J-K_\mathrm{s})$) is shown in Figure \ref{fig:colgramexcite}. There is a small gap in coverage around star forming region \object{NGC 2080} and little coverage within R136 (Section \ref{30d:r136} looks at R136 in more detail). As most locations contain high and low extinction stars (due to multi-layered structure) in close proximity, it is difficult to assess the impact from this map alone. In Section \ref{ssec:jksli} we look at eight maps covering more limited $E(J-K_\mathrm{s})$ extinction ranges.

\subsubsection{Issues with method and data} \label{30d:issues} 
Within the $K_\mathrm{s}$ vs. $(J-K_\mathrm{s})$ CMD there is some potential overlap between what we select as RC stars and the RGB. As these stars will also be affected by reddening it is important to know where they lie on the intrinsic CMD; if the RGB stars are at the same intrinsic colour as the RC stars then the extinction will be the same. However, if the RGB stars are intrinsically redder, the extinction is partially overestimated. Using the isochrones produced in Section \ref{30d:red} we find the RGB lies at $(J-K_\mathrm{s})=0.53$--$0.57$ mag (brighter magnitudes having redder colour), a difference of $\sim0.05\pm0.02$ mag from the mean RC star. However, the first ascent of the RGB covers a large magnitude range with stars moving at approximately $0.017$ mag/Myr which could mean a fairly low stellar density of these. Taking a typical RC lifetime of $0.1$ Gyr an RGB star moves a total of $1.7$ mag in this time. The RC selection box we use is about $0.54$ mag meaning it would take an RGB star $\sim32\%$ of the RC lifetime to cross the RC. 

In reality though, the first ascent of the RGB concentrates at the luminosity where the H-burning shells meets the composition discontinuity left by the first dredge-up. This is known as the RGB bump. For LMC metallicities this luminosity is at the same level of the RC, and the feature thus overlaps (and is hidden by) the RC. Also, the RGB evolution slows down at the RGB bump so that the movement outlined above is overestimated. The effect the RGB has on the reddening map is important for low reddening ($E(J-K)_\mathrm{s}\simeq0.1$ mag, $A_V\simeq0.6$ mag) where this might be overestimated in some instances but is less important for higher extinction.

The photometric errors and intrinsic spread in the RC lead to about $14\%$ of the RC selection having unphysical negative reddening.

On the instrument itself, detector $16$ has variable quantum efficiency in the top half, especially noticeable in low background conditions. As a result, flat fielding is worse for the regions covered by that detector, particularly for the outer half. The region that the two observations within detector 16 covers in our work is the dense LMC bar. The poor flat fielding is seen from the two strong horizontal lanes of apparently high extinction seen in the region, which correspond to the two offset observations from the outer half of detector 16. A method to gauge the effect is to see if any features in this region are also seen in non-VMC observations for this region. This is done in Section \ref{30d:dis:m}. The region affected is RA$<83\rlap{.}^{\circ}48$ and Dec$<-69\rlap{.}^{\circ}73$ and has been marked on Figure \ref{fig:map}.

\subsection{Distributions of reddening along the line-of-sight}
The aim of this subsection is to try and identify where dust lies in relation to RC stars. 

\subsubsection{Sub-maps: $J-K_\mathrm{s}$ slices} \label{ssec:jksli}
Figure \ref{fig:colsli} shows contour density maps for individual slices of extinction, to show a clearer picture of the regional distribution of extinction. The contours are drawn based on the density of RC stars per $5^{\prime}\times5^{\prime}$ split into 10 levels. The legend gives the number of stars at each contour level (on the bottom) and the probability of stars in that region having the same extinction (on the top). Table \ref{tab:colslin} describes the number of stars contained in each slice and the extinction range covered by each slice. Slice $2$ is the most numerous containing $\sim26\%$ of all the RC stars while slices $1$--$4$ cover $\sim88\%$ of all the RC stars.

The $1^{\mathrm{st}}$ and $2^{\mathrm{nd}}$ slices are virtually \textnormal{inverted} in the $4^{\mathrm{th}}$ and $5^{\mathrm{th}}$ slices. This suggests the latter slices are the reddened population from the former slices. However, this view is complicated by slices $6$, $7$ and $8$ also containing populations absent from slices $1$, $2$ and $3$.

The probabilities given in the legend of Figure \ref{fig:colsliprob} are an estimate of the probability of a star in a given sightline, having extinction within that extinction range, based on the total number of stars in all slices. We see that around R136 nearly all the sources are found in slices $5$--$8$, suggesting that the dust lies in front of the stars here. If the assumption is made that the RC decently represents the tile as a whole, the probability maps can be used for reddening estimates of other star populations. From Section \ref{ssec:dens} we see other intermediate aged populations match the RC star distribution well while earlier phases correlated less. In particular we found relatively few RC sources in the north--west and relatively many RC sources for small regions of the south--west. We see that both of these regions are almost exclusively comprised of stars in slices $1$--$4$. The south--west region contains the LMC bar which is likely to be in front of the LMC, or has little dust obscuring it.

Finally, the peaks in the background galaxy distribution in Figure \ref{fig:dens} correspond with regions in slices $7$ and $8$. This suggests those peaks are further reddened RC stars missed from our selection due to the background as a whole being more dominant when looking at the whole tile.
\begin{figure}
\resizebox{\hsize}{!}{\includegraphics[angle=0,clip=true]{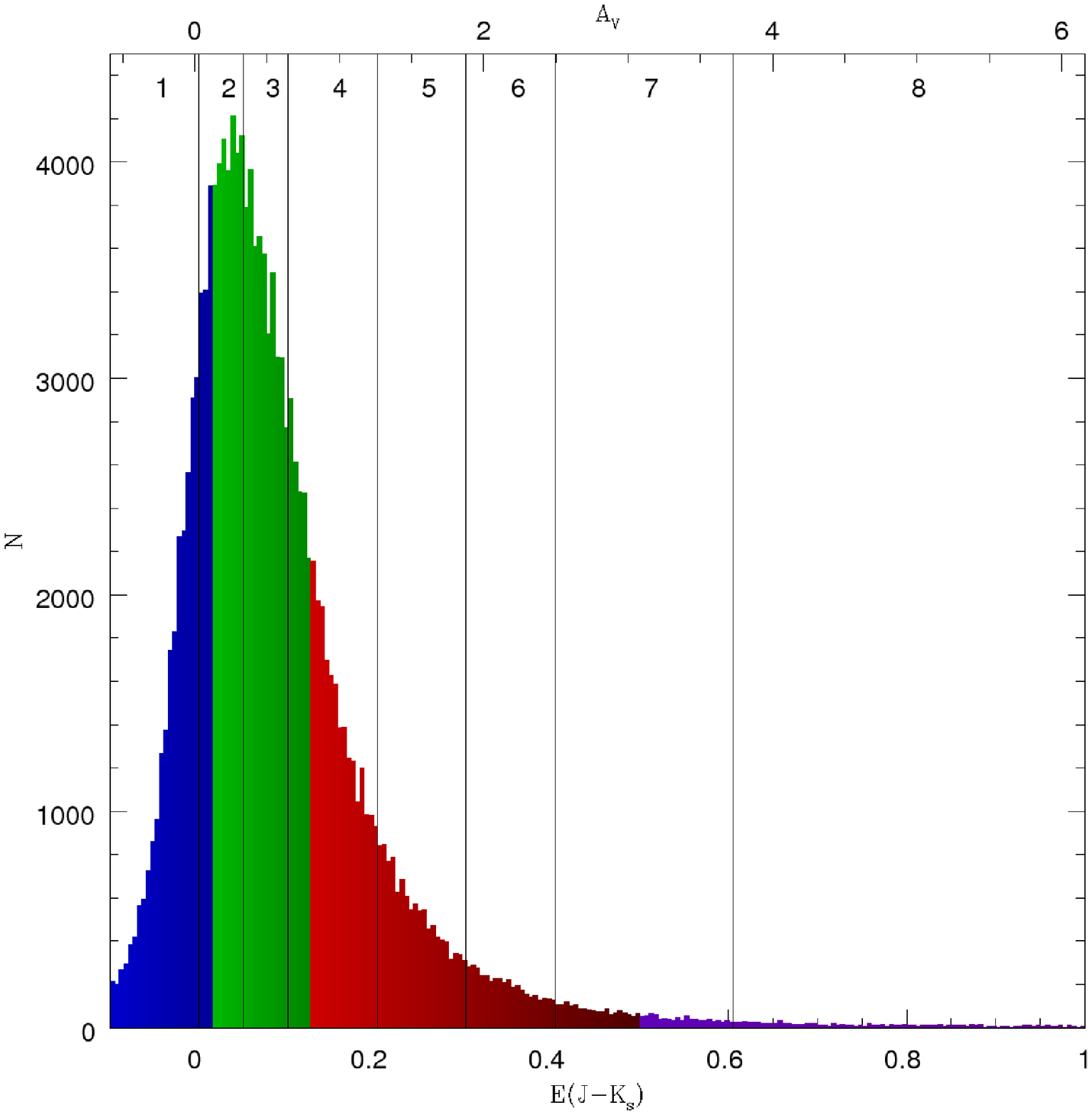}}
\caption{Histogram of $E(J-K_\mathrm{s})$ (bottom) and $A_V$ (top) distribution with bin size of $0.005$ mag. Green is the region covering $50\%$ of the sources centred on the median where darker blue is lower than this and brighter red higher and \textnormal{purple} extremely high (covering the sparsely populated upper $50\%$ of the reddening range). The numbers and vertical lines detail the colour slice regions used in Section \ref{ssec:jksli}.} 
\label{fig:colgramexcite}
\end{figure}
\begin{figure*}
\includegraphics[width=0.24\textwidth,clip=true]{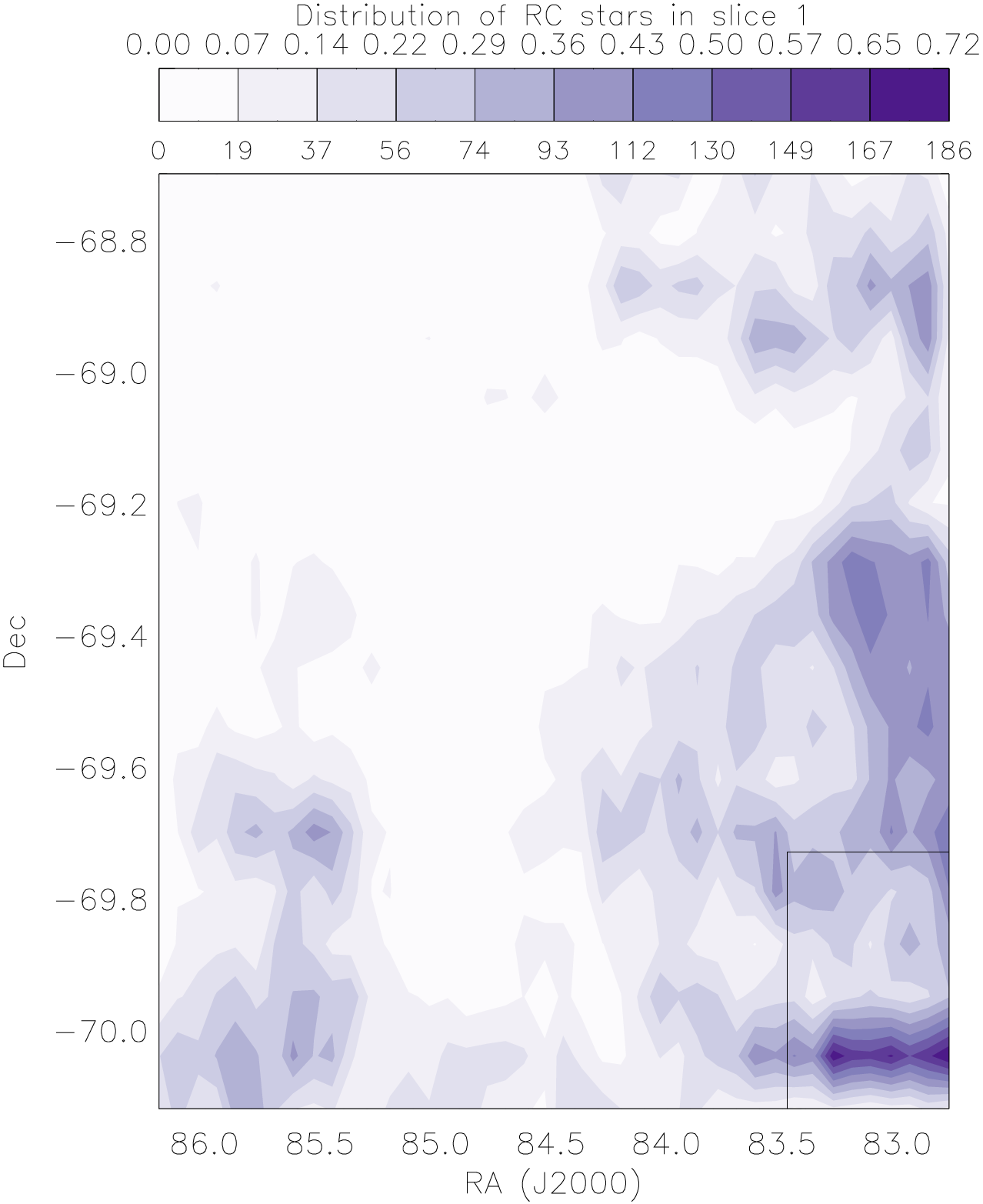}
\includegraphics[width=0.24\textwidth,clip=true]{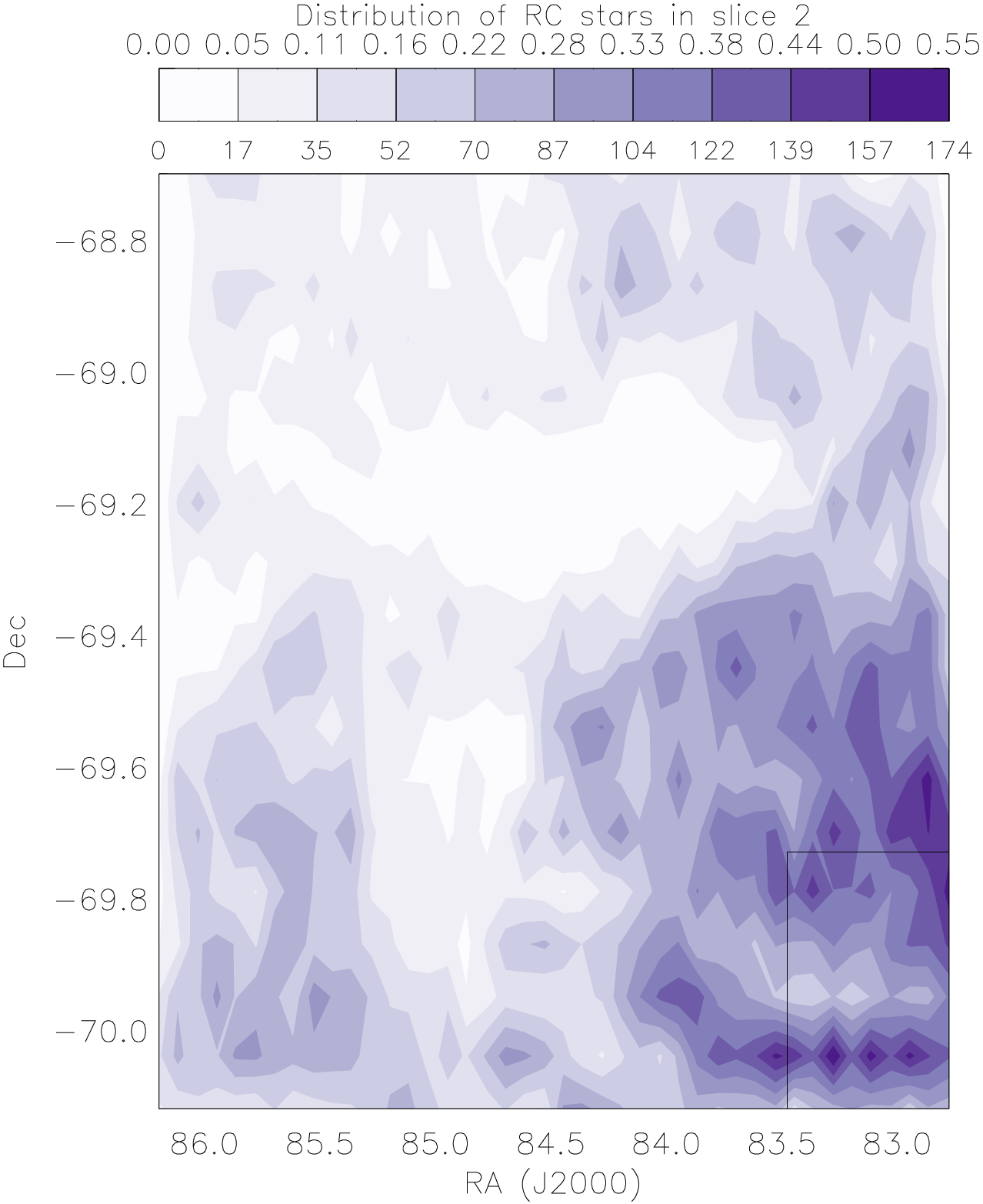}
\includegraphics[width=0.24\textwidth,clip=true]{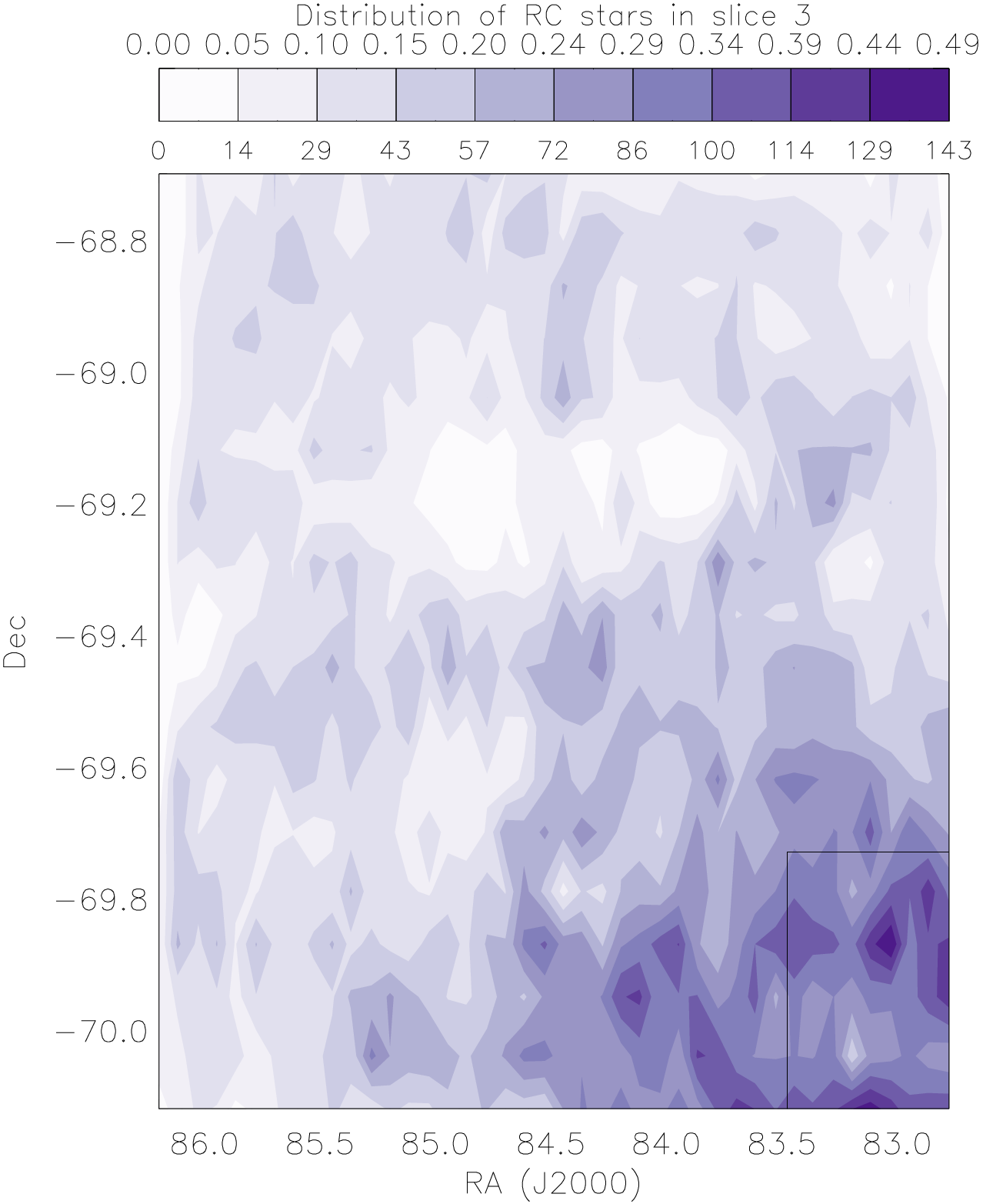}
\includegraphics[width=0.24\textwidth,clip=true]{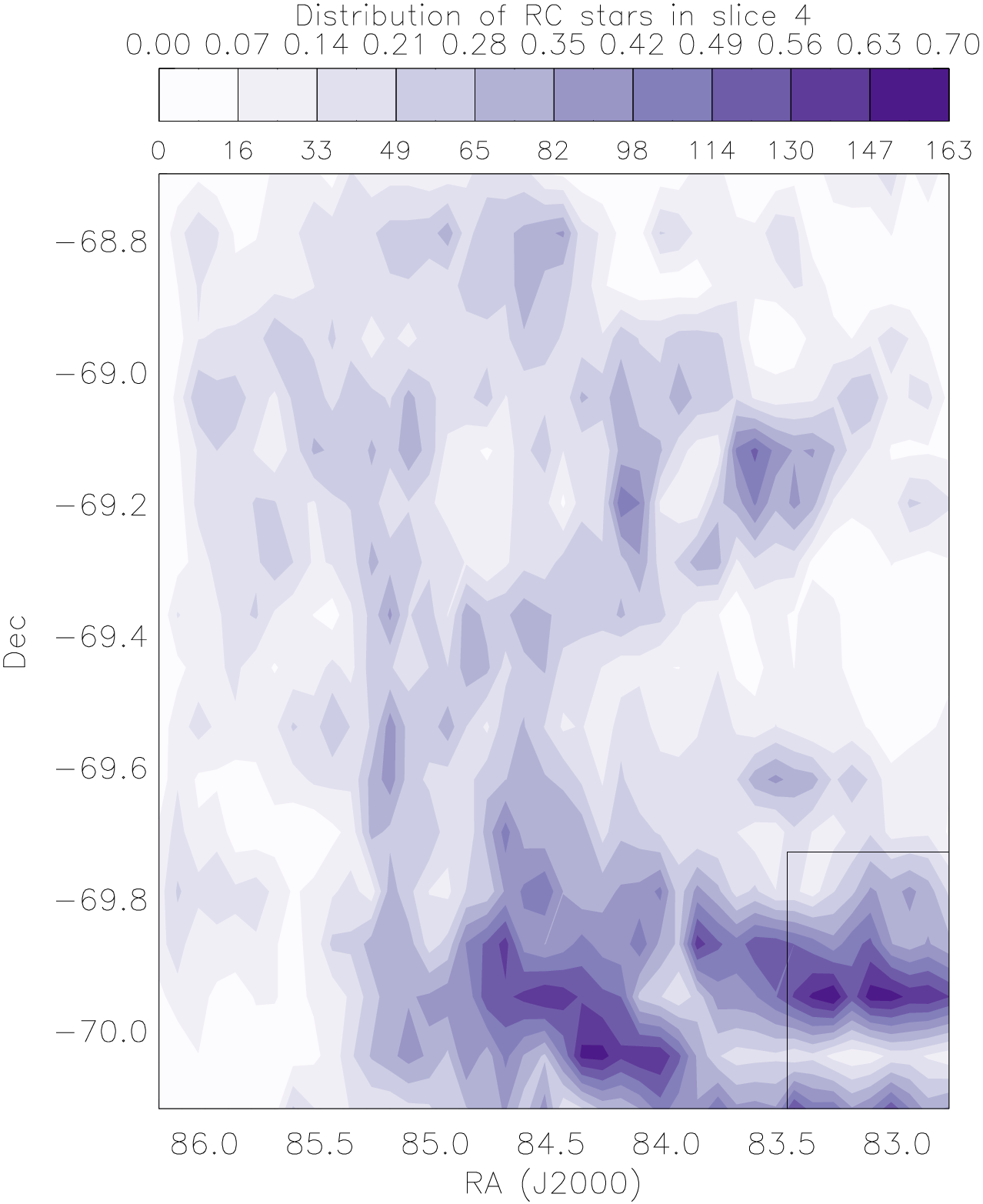}\\
\includegraphics[width=0.24\textwidth,clip=true]{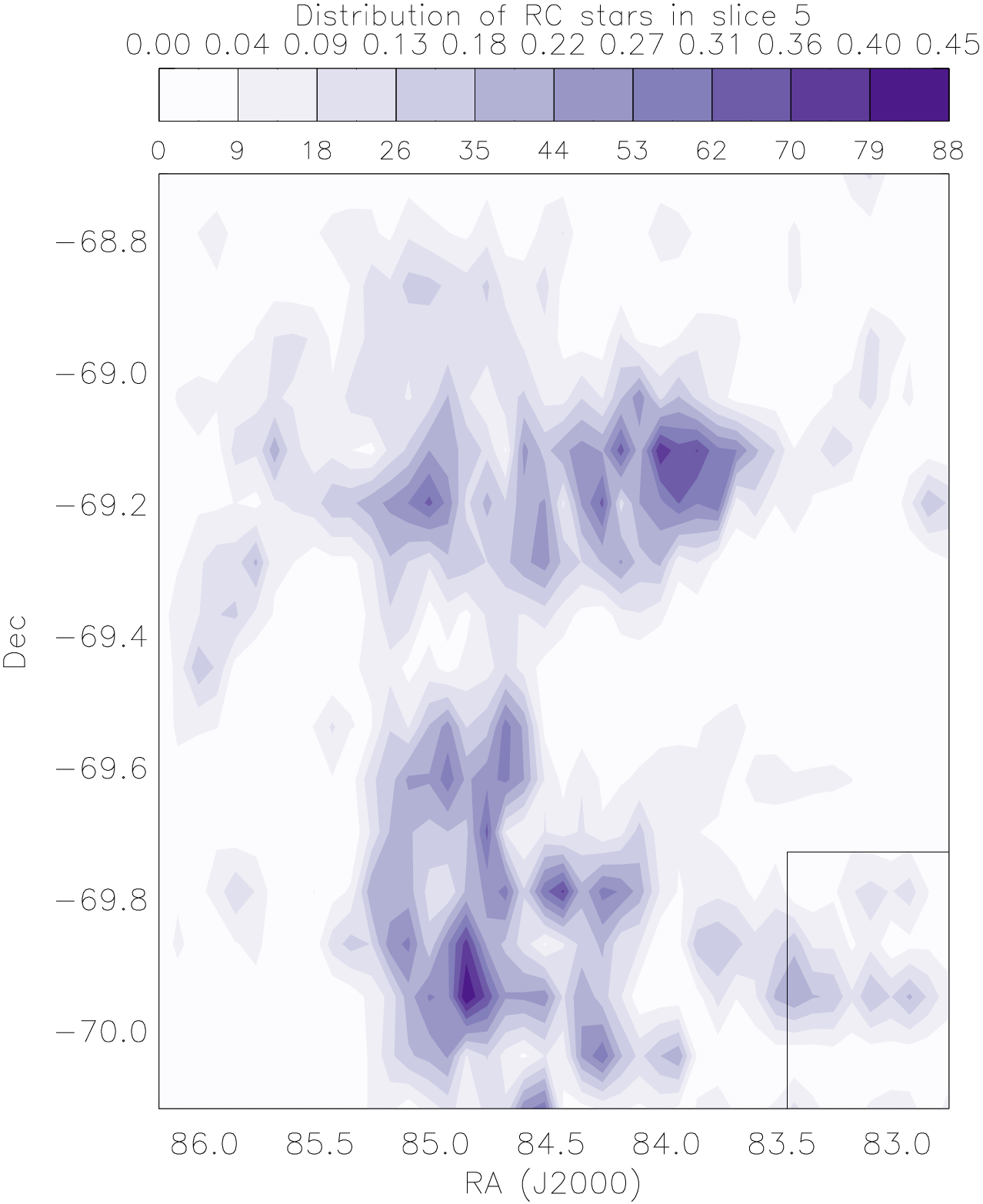}
\includegraphics[width=0.24\textwidth,clip=true]{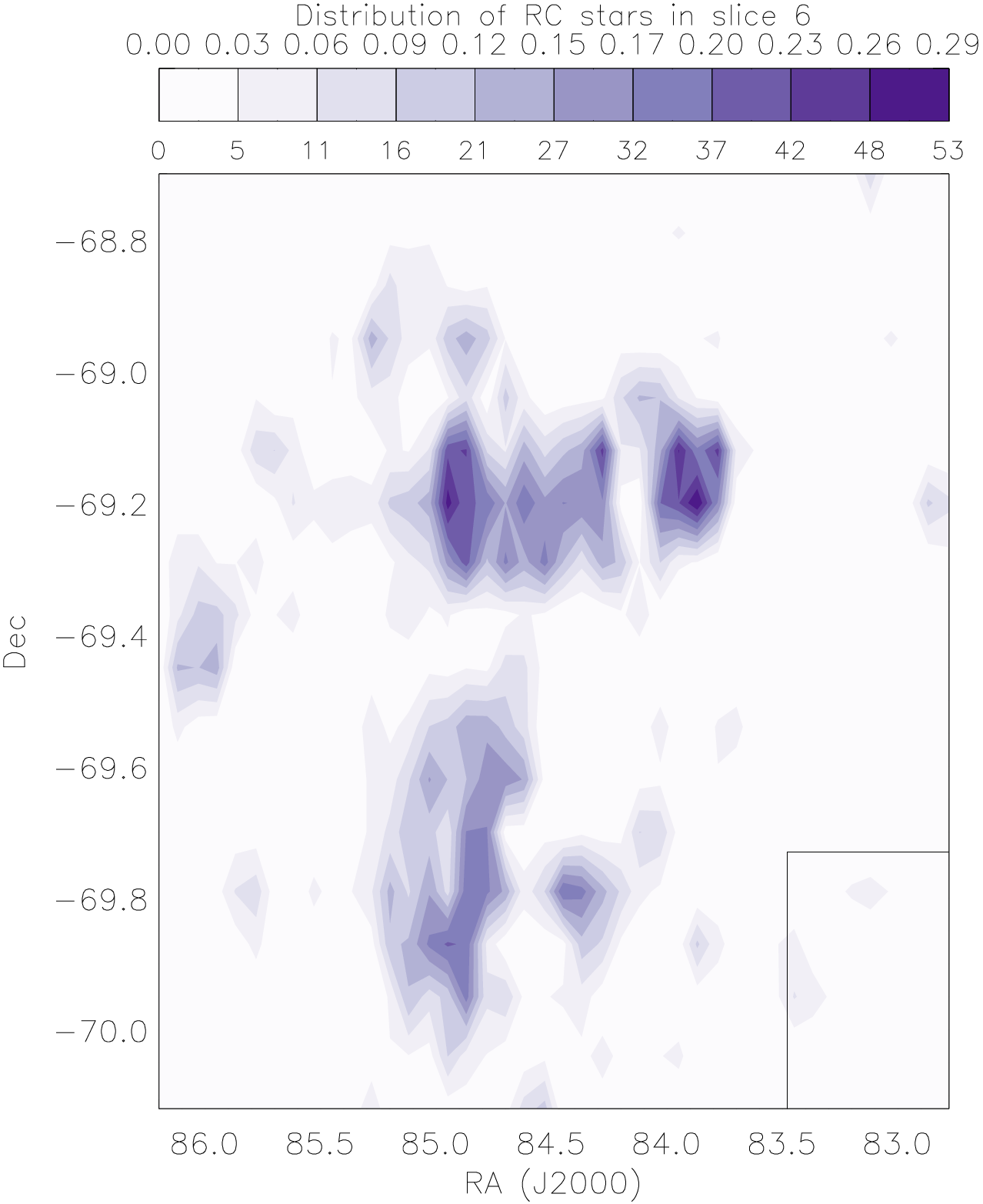}
\includegraphics[width=0.24\textwidth,clip=true]{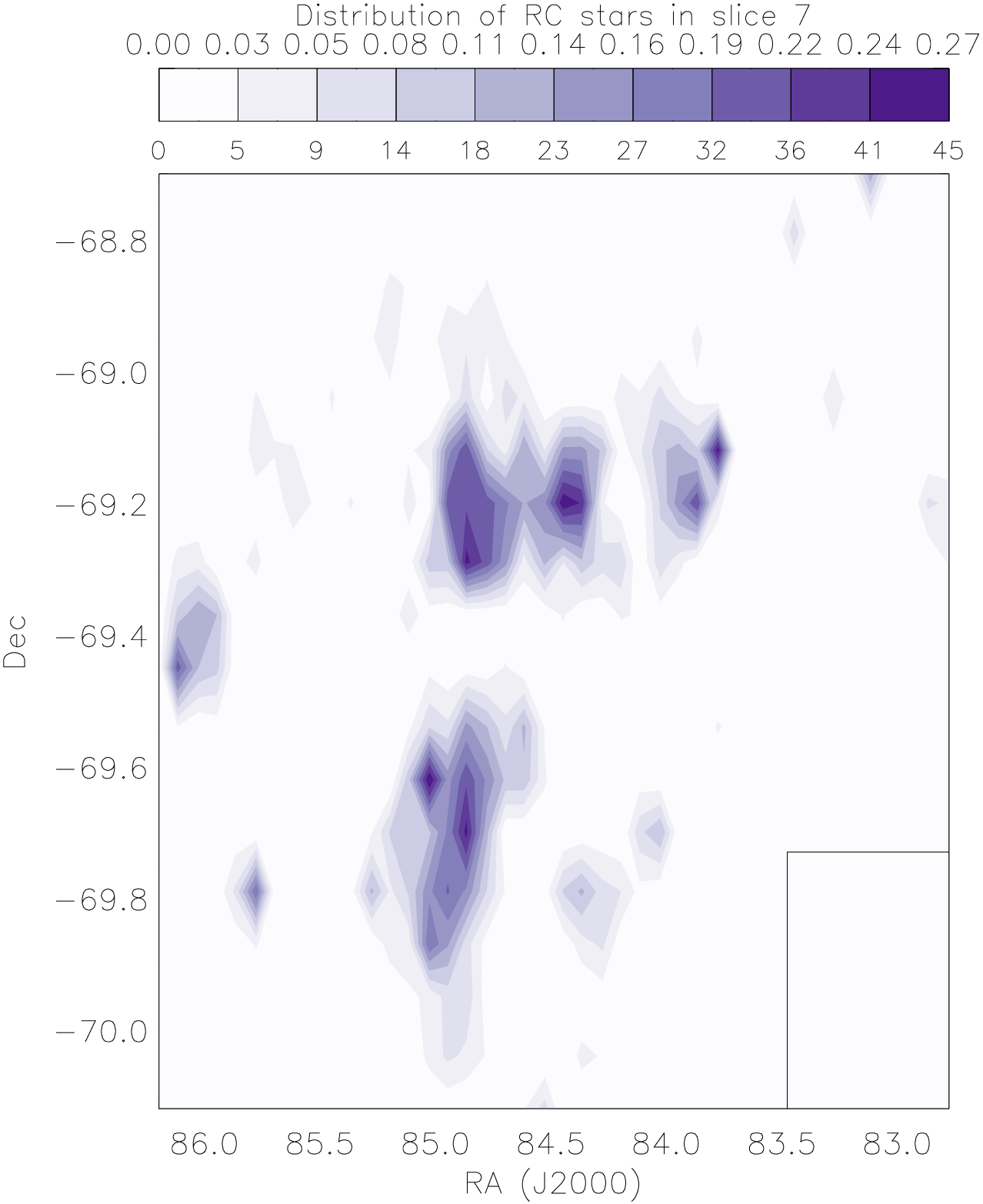}
\includegraphics[width=0.24\textwidth,clip=true]{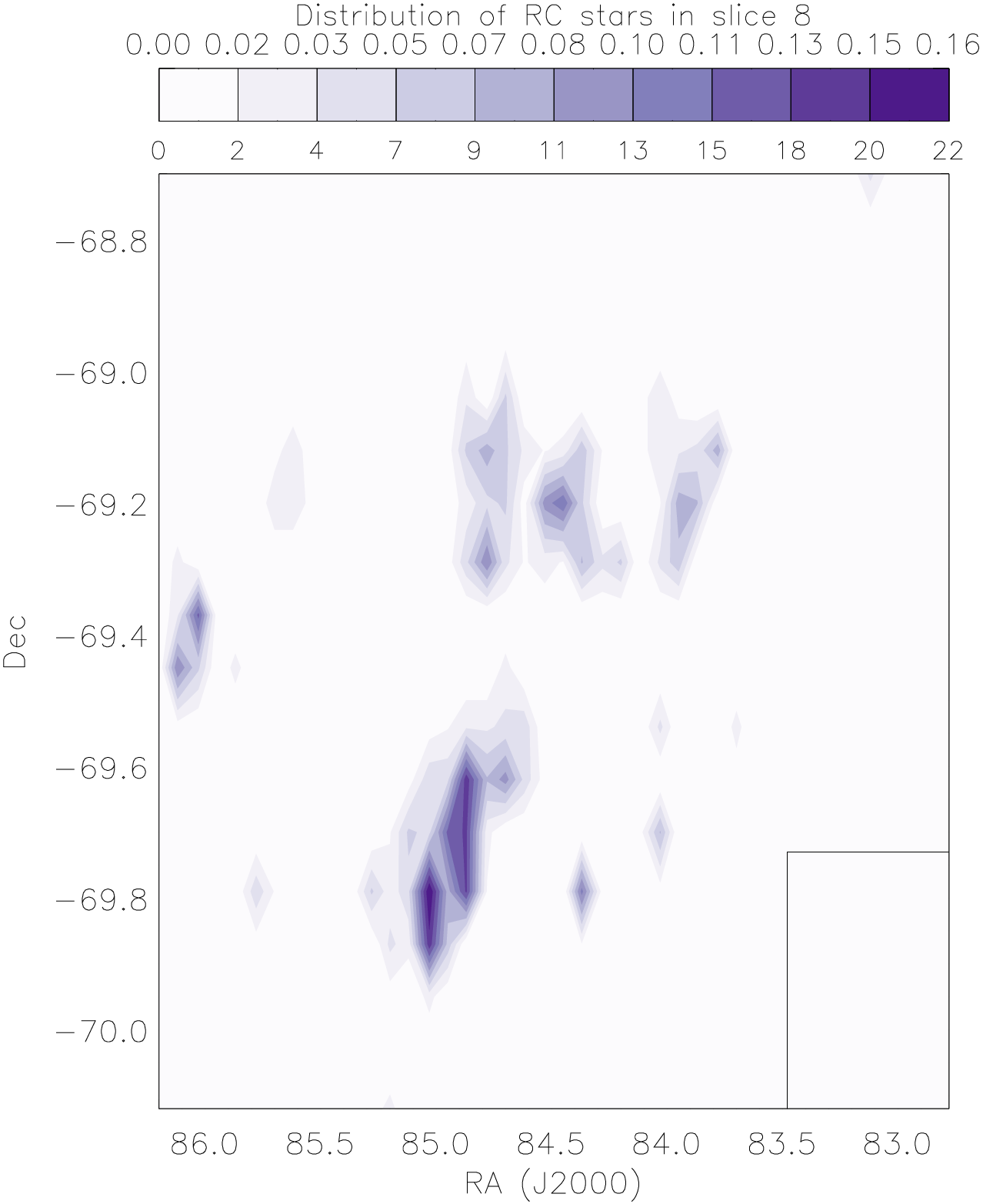}\\
\caption{Contour map of number of RC stars (bottom legend) and probability of RC star (top legend) within $5^{\prime}\times5^{\prime}$ bins for each colour slice. Region containing detector 16 is outlined in black. $E(J-K_\mathrm{s})$ and $A_V$ range of each colour slice are given in Table \ref{tab:colslin}.}
\label{fig:colsliprob}
\label{fig:colsli}
\end{figure*}

\begin{centering}
\begin{table}
\caption{Number of RC stars in each colour slice (extinction ranges given) and total number of RC stars.}
\label{tab:colslin}
\[
\begin{tabular}{cccccc}
\hline
\hline
\noalign{\smallskip}
Slice & N$_\mathrm{RC}$ & \multicolumn{2}{c}{$E(J-K_\mathrm{s})$} & \multicolumn{2}{c}{$A_V$} \\
 & & min & max & min & max\\
\ldots & \ldots & \multicolumn{2}{c}{mag} & \multicolumn{2}{c}{mag} \\
\hline
1	&	24738	&	\llap{$-$}0.095	&	0.005	&	\llap{$-$}0.584	&	0.031	\\
2 	& 	38868 	&	 0.005	&	 0.055  &	 0.031 	&	 0.339	\\
3	&	34381	&	0.055	&	0.105	&	0.339	&	0.645	\\
4	&	34014	&	0.105	&	0.205	&	0.645	&	1.260	\\
5	&	10851	&	0.205	&	0.305	&	1.260	&	1.875	\\
6	&	3968	&	0.305	&	0.405	&	1.875	&	2.489	\\
7	&	2439	&	0.405	&	0.605	&	2.489	&	3.719	\\
8	&	1069	&	0.605	&	1.005	&	3.719	&	6.177	\\
Total & 150328 & \hrulefill & \hrulefill & \hrulefill & \hrulefill \\
\noalign{\smallskip}
\hline
\end{tabular}
\]
\end{table} 
\end{centering}

\subsubsection{Reddening dispersion} \label{30d:disp}
\textnormal{We measure dispersion of reddening for RC stars to try and learn about the relative locations of stars and dust. A population with low dispersion will contain less dust while higher dispersion can be related to stars being embedded within the dust or it can show a significant fraction lies behind dust clouds. Investigating higher moments\footnote{Measures of shape of a distribution of points.} can help distinguish between these possibilities}.

\begin{figure*}
\includegraphics[width=\textwidth,clip=true]{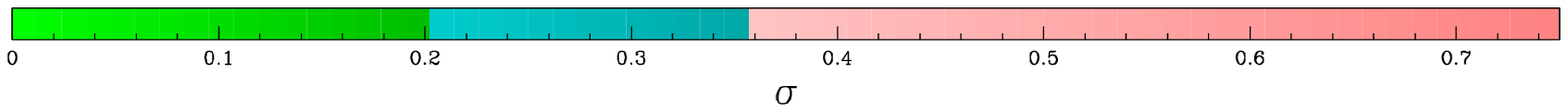}\\
\includegraphics[width=0.33\textwidth,clip=true]{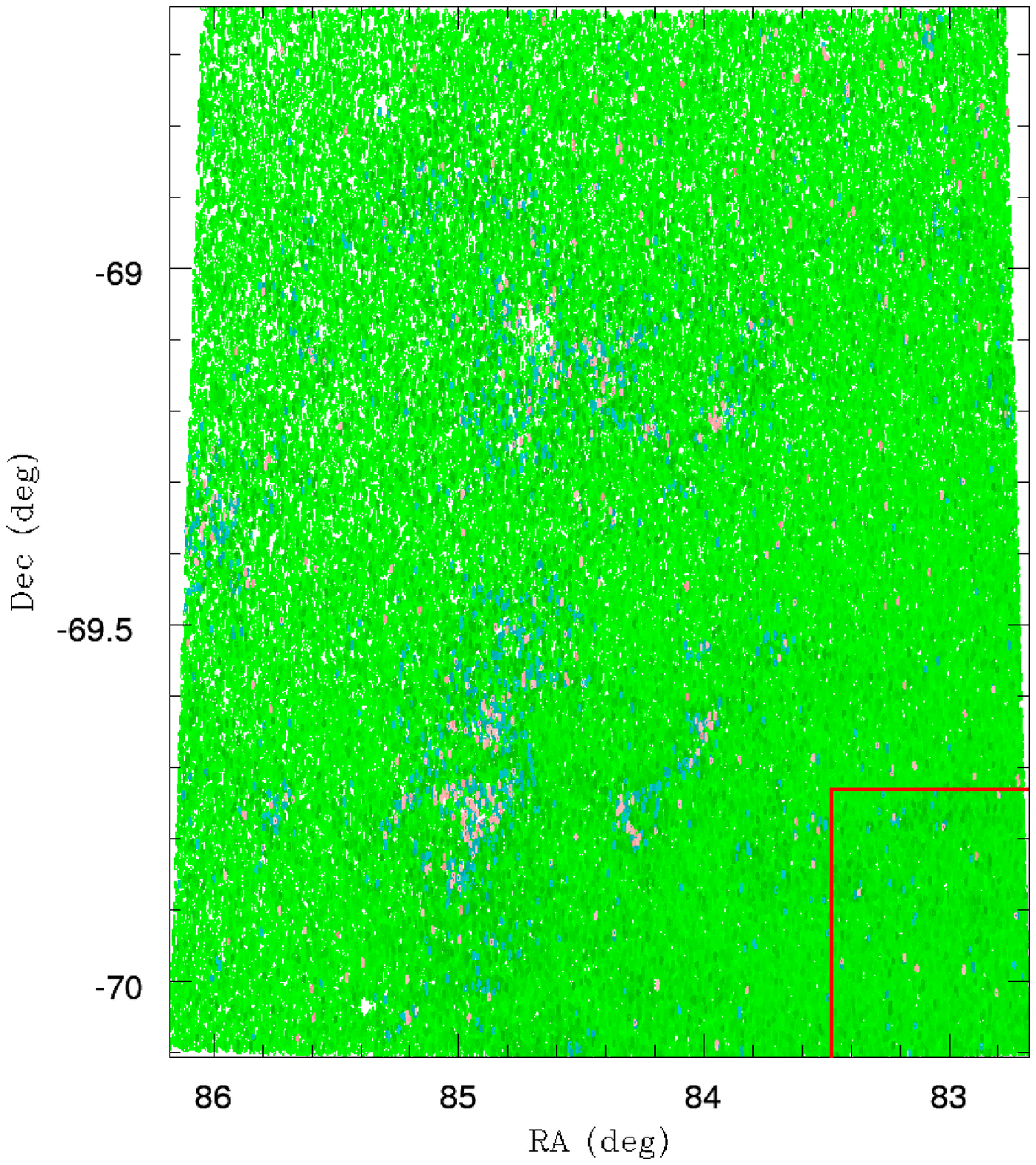}
\includegraphics[width=0.33\textwidth,clip=true]{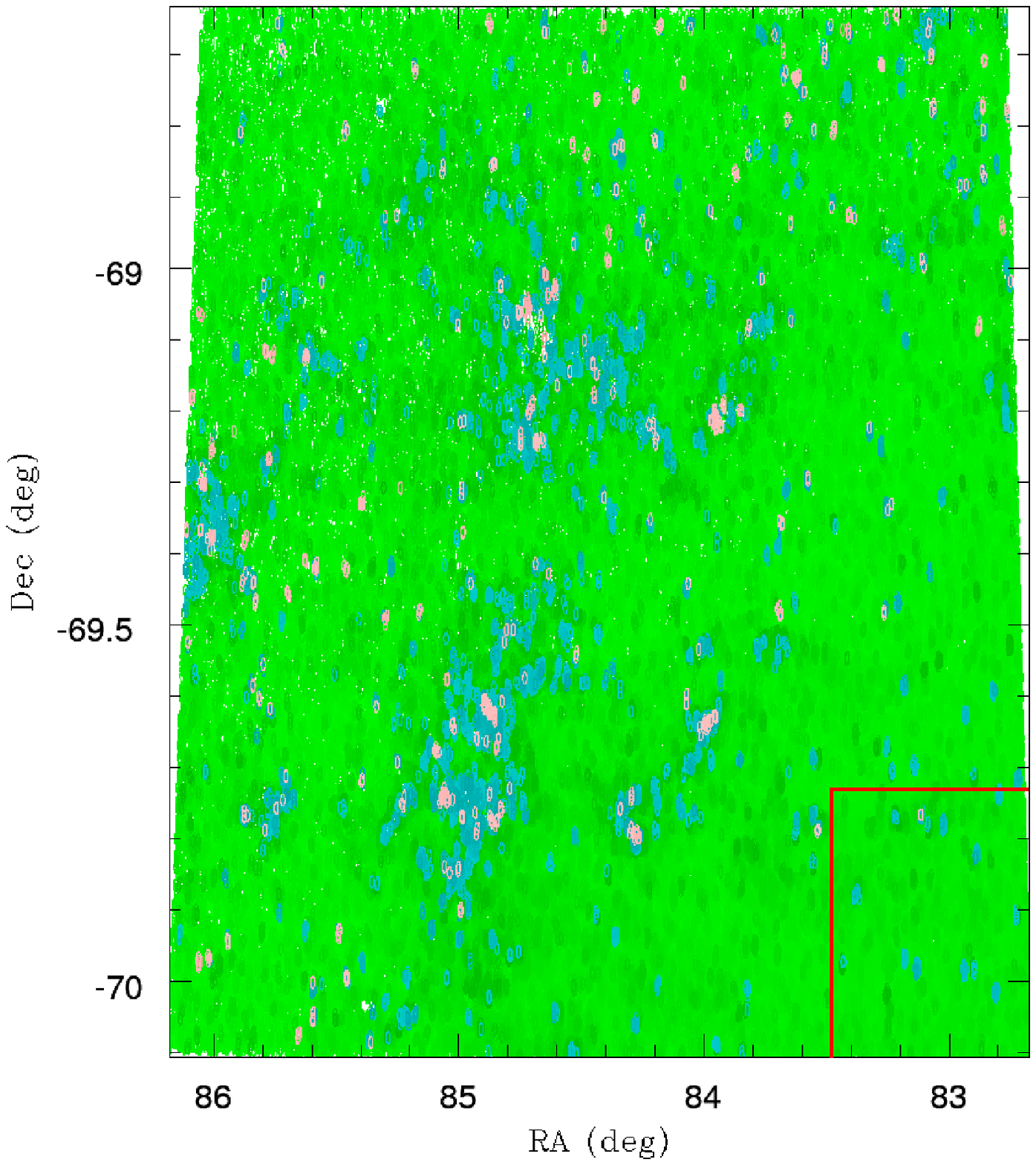}
\includegraphics[width=0.33\textwidth,clip=true]{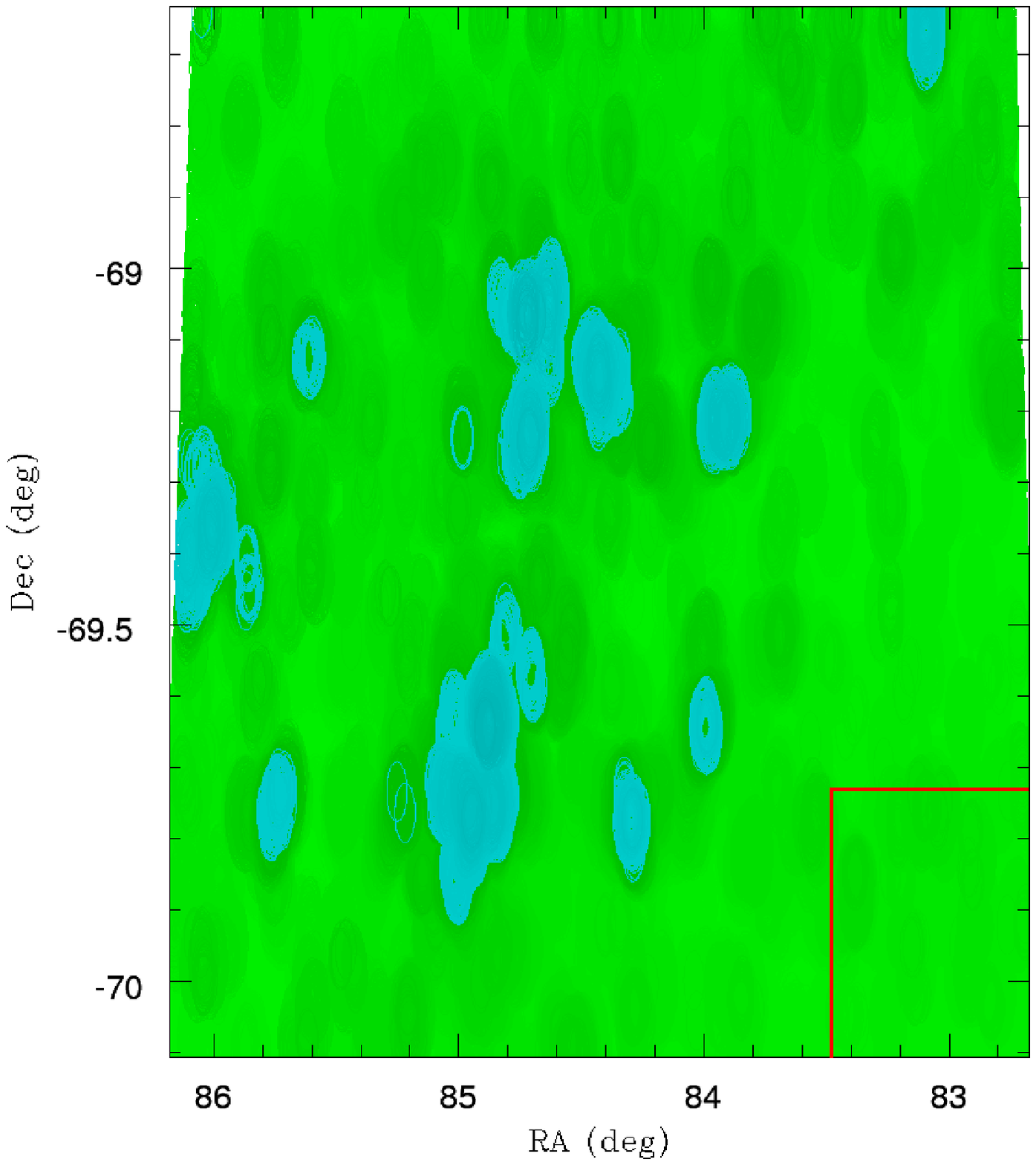}
\caption{\textnormal{Maps of $\sigma$ for RC stars within $0.5\arcmin$ (left), $1\arcmin$ (middle), $5\arcmin$ (right)} diameter of each RC star. Values shown in legend.}
\label{fig:catmap}
\end{figure*}

For a map based comparison of reddening variance we select RC stars within a $30\arcsec$, $1\arcmin$ and 5$\arcmin$ diameter of every RC star. We then find the standard deviation ($\sigma$) of these selections. Figure \ref{fig:catmap} shows \textnormal{the resulting maps. These have} some similarity to the reddening map in that $\sigma$ is higher in regions covered by higher extinction slices. Less similarity is seen in the $30\arcsec$ map though the regions of greatest $\sigma$ are also the regions of greatest reddening (but the regions surrounding these have much lower values than in the $1\arcmin$ and 5$\arcmin$ maps, this is due to $30\arcsec$ diameter resulting in small samples and $\sigma$ being zero). Likewise, the $5\arcmin$ diameter map is more similar to the reddening map \textnormal{where the higher dispersion is associated with the regions containing the populations of reddening slices $5$--$8$ (in Figure \ref{fig:colsli})}. This increase in $\sigma$ is due to high extinction regions covering a much larger range of extinction values than the low extinction regions.

The larger extinction range is caused either by the RC stars being in front and behind the dust or them being embedded within the dust. To investigate this we looked at a higher moment, the kurtosis, finding for the $5\arcmin$ diameter there is a high kurtosis for low extinction regions (as the data is largely around the peak) while high extinction regions had a lower kurtosis, but these values were still close to those of a normal distribution that you would expect from embedding within the dust.
 
\subsubsection{Dependence of reddening on distance}
Another measure of the relative location of dust and stars is the change in reddening with respect to luminosity. RC stars near the bright edge of the reddening selection box will in general be nearer to us than stars near the faint edge of the reddening selection box, reflecting differences in distance modulus. The more distant stars are more likely to lie behind more dust than stars near the front of the LMC, and we thus expect the more distant stars to be more heavily reddened.

When splitting our RC selection into $5$ equal magnitude ranges, which follow the reddening vector, we observe a decrease in mean and median reddening with respect to increasing magnitude (decreasing luminosity). This is caused by the intrinsic slopes of the RC, RGB and SRC rather than the reddening (which would be expected to cause the opposite effect). This effect is more clearly shown in Figure \ref{fig:rcnon} where the $0.4{}\leq{}(J-K_\mathrm{s}){}\leq{}0.5$ mag panel has a bump at dimmer magnitudes from the SRC while the $(J-K_\mathrm{s})=0.5$--$0.7$ mag panels have a slight bump at brighter magnitudes.

\begin{figure}
\resizebox{\hsize}{!}{\includegraphics[width=0.50\textwidth]{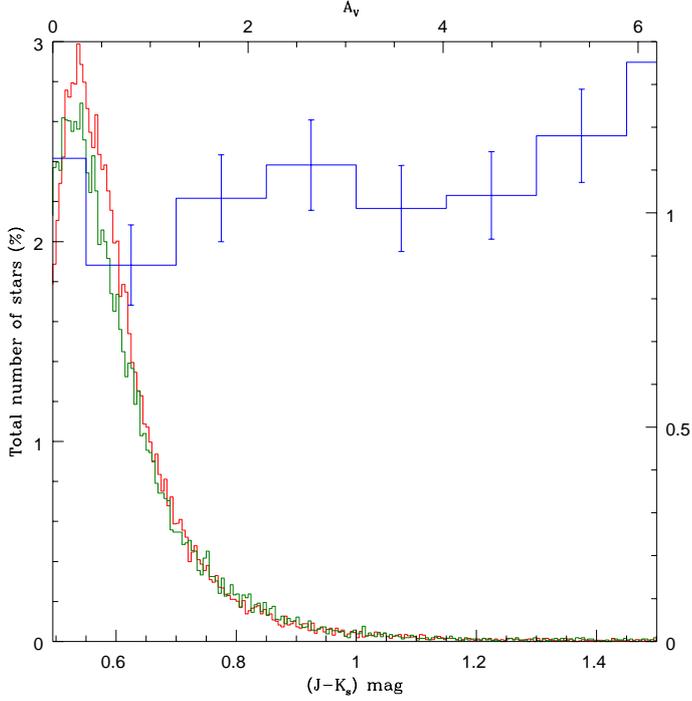}}
\caption{Histograms of more luminous (red) and less luminous (green) RC stars, normalised to total number of stars in selection. In blue is the ratio of less luminous to more luminous stellar numbers using a larger bin size (ratio given on right y-axis). \textnormal{Sources within detector 16 region have been excluded from selection.}}
\label{fig:hisdisp}
\end{figure}

To examine the effect of magnitude on the reddening distribution\footnote{Absolute magnitude does not affect reddening, but reddening may be dependant on apparent magnitude as these have some relation to location within line-of-sight.} we use histograms of colour on the $5$ magnitude ranges. We choose the $2^\mathrm{nd}$ and $4^\mathrm{th}$ parts (which have a difference of $K_\mathrm{s}=0.21$ mag, a distance of $5$ kpc) as a contrast for comparison because the outermost parts are more affected by the intrinsic slope and RGB stars (while the central part will not show as much contrast). These histograms are normalised to the total number of RC stars included in the parts (rather than to the peak as the height of the peak may vary) and also plotted are the ratio of the less luminous part divided by the more luminous part. Figure \ref{fig:hisdisp} shows these plots. We see that for higher reddening the further, less luminous component has up to $35\%$ more highly reddened stars than the closer, more luminous component and this effect becomes stronger with extinction. While there is some small number statistics it shows to a small extent that the most highly reddened stars will be more frequently found to be less luminous and most distant.

\subsection{Total hydrogen column density} \label{30d:dust}
It is possible (making assumptions about the composition and radius distribution of dust particles) to use extinction to estimate the total hydrogen gas column density ($N_\mathrm{H}$ in units of cm$^{-2}$) from its relation to $A_V$;
\begin{equation} \frac{N_\Sigma\,\mathrm{H}}{A_V}=x\times10^{21}\,\mathrm{cm}^{-2} \end{equation}
Where $N_\Sigma\,\mathrm{H}$ is the sum of molecular, atomic and ionised hydrogen content and $x$ varies throughout the literature; \citet{guver09} find $x=2.21\pm0.09$. However, this is for Galactic regions where the dust-to-gas ratio (and metallicity) is higher than it is in the LMC. To account for this ($Z_\mathrm{LMC}=0.5\times\,Z_\mathrm{MW}$), we halve $A_V$. This is more more easily expressed by (and has the same effect as) doubling the \citet{guver09} value for $x$.

We use the above equation to estimate total hydrogen column density \footnote{\textnormal{Referred hereafter as Inferred total hydrogen column density and Inferred total H.}} for regions of $1\arcmin\times1\arcmin$ using three different measures of density. The first measure (maximum) selects the star with the highest reddening in a region because this would be affected by (and hence measure) the most gas. A drawback of this method is that it might provide an overestimate for regions of small scale structure (as these small scale results are magnified to arcminute scales). As a countermeasure we also look at the mode and median (the former is defined from histograms of bin size $0.44\times10^{21}\,\mathrm{cm}^{-2}$ where the peak is taken as the mode) although these are skewed by stars in front of the ISM and gas column.

\begin{figure*}
\includegraphics[width=0.25\textwidth,clip=true]{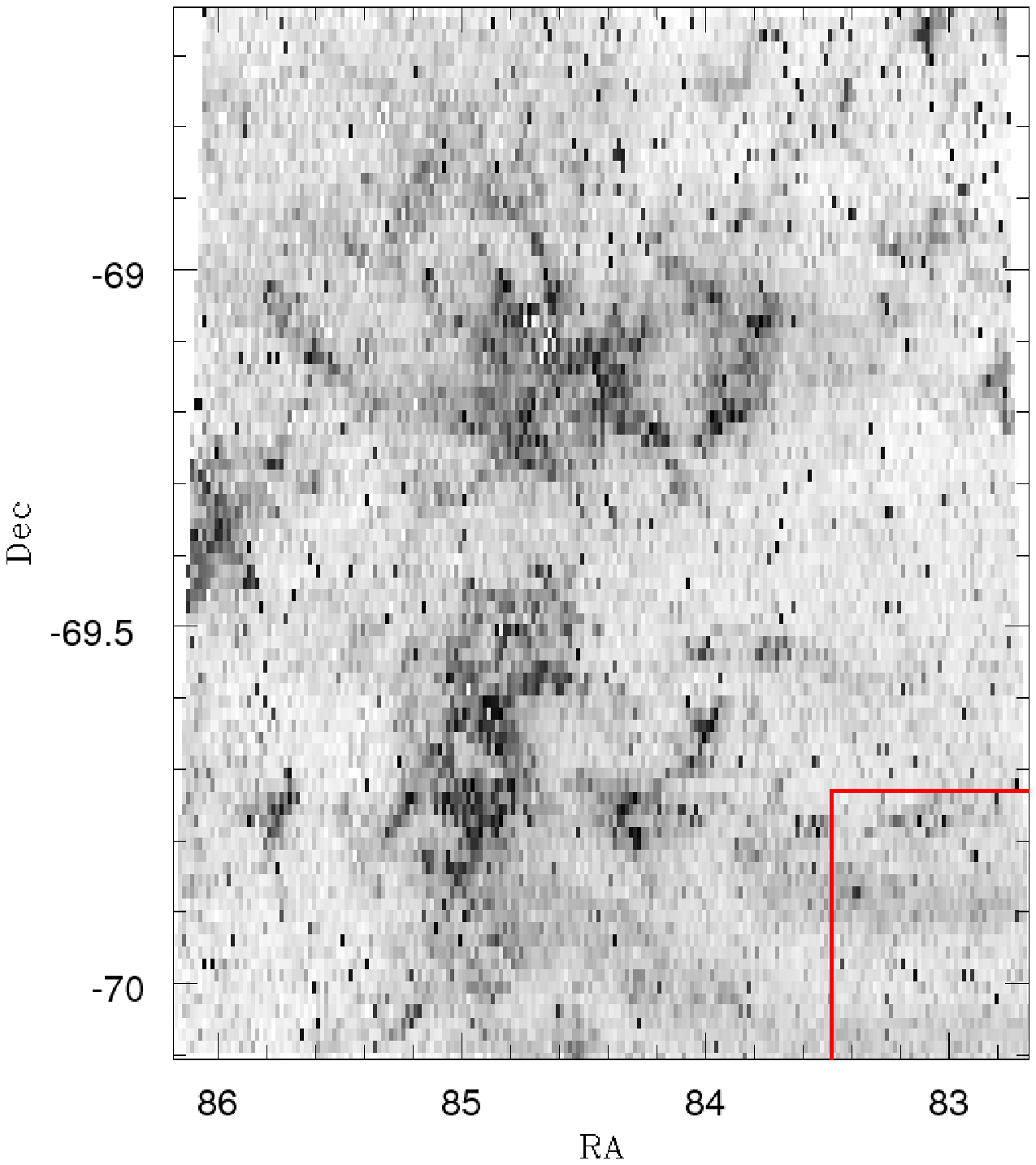}
\includegraphics[width=0.25\textwidth,clip=true]{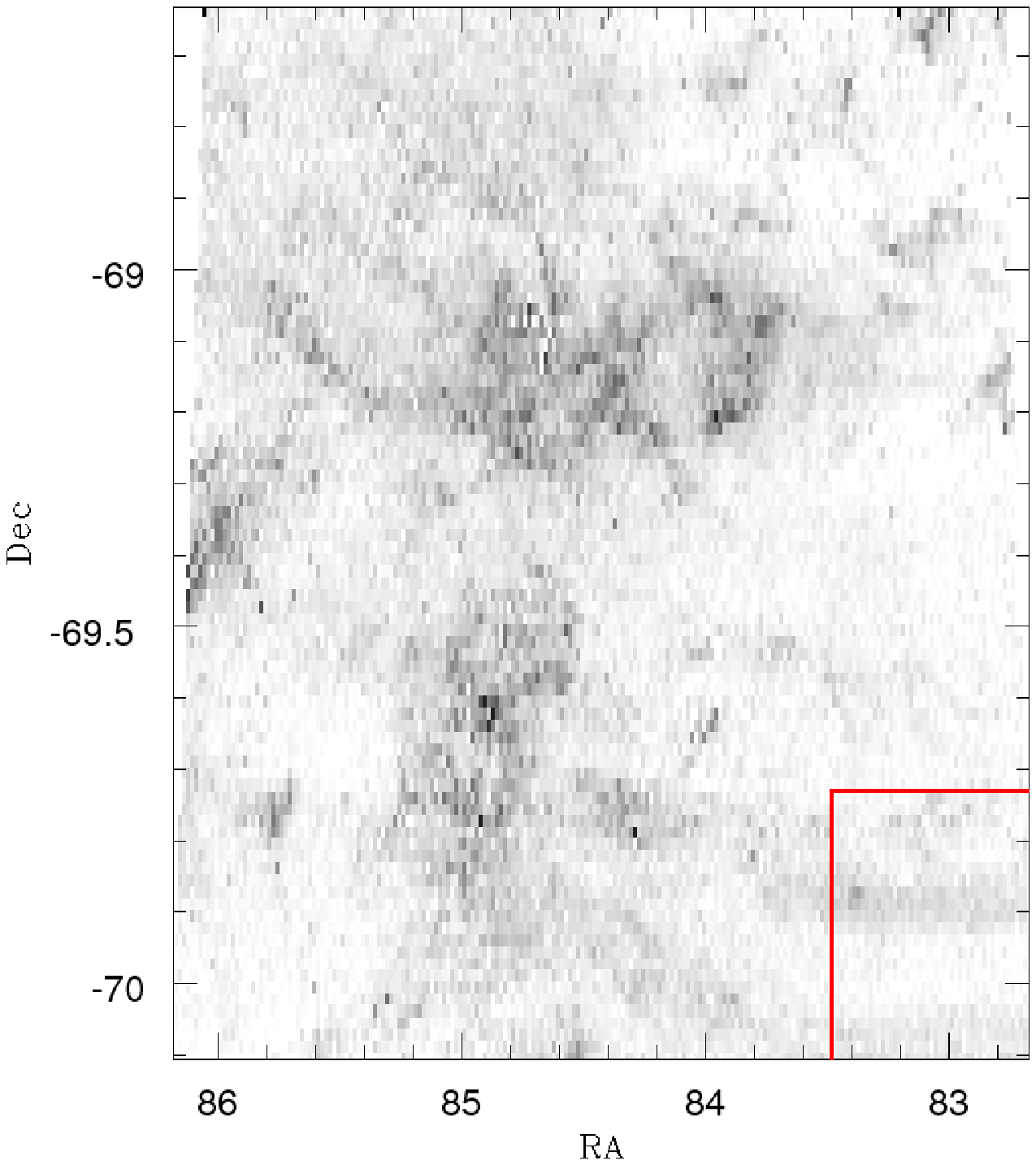}
\includegraphics[width=0.25\textwidth,clip=true]{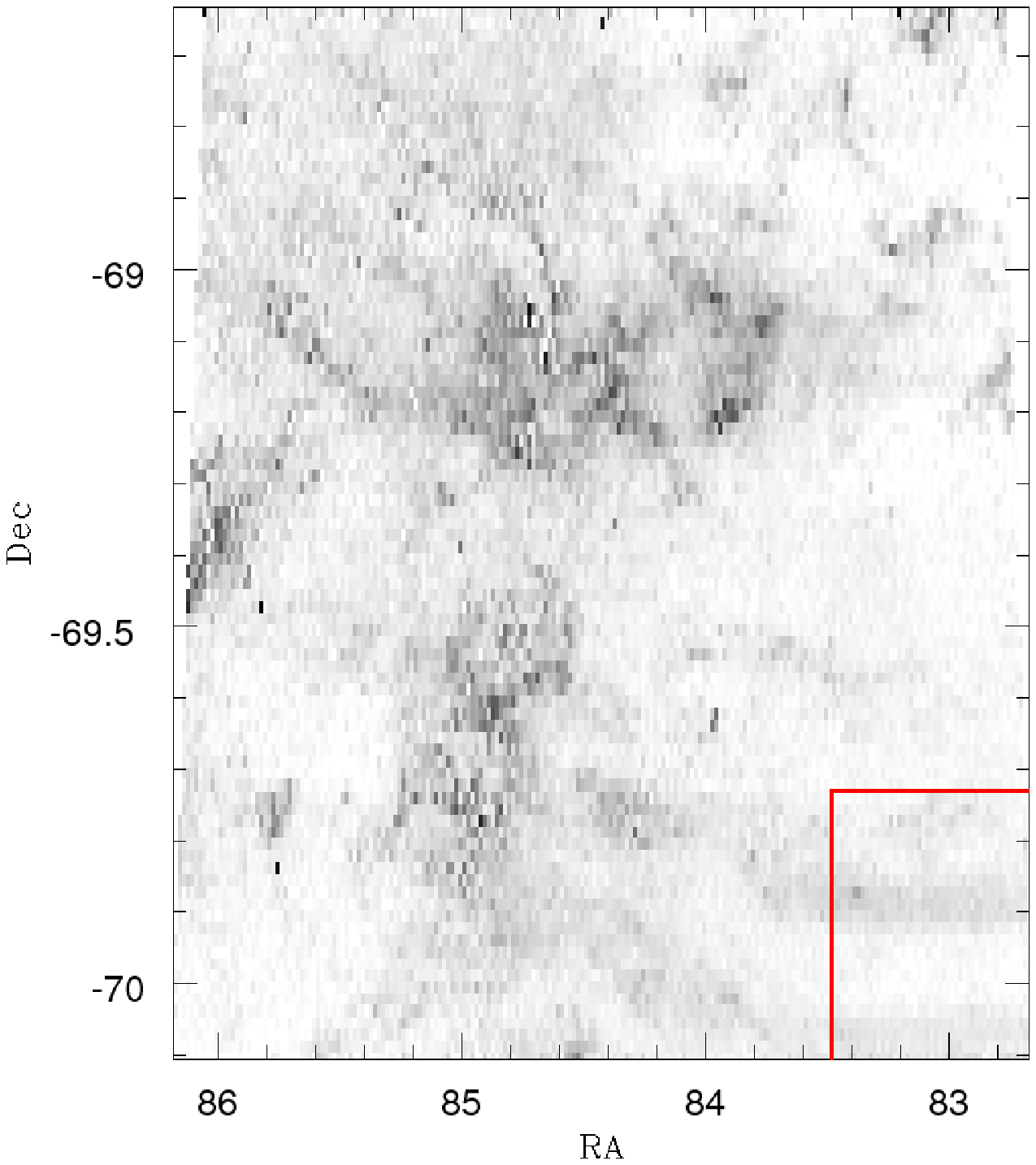}
\includegraphics[width=0.25\textwidth,clip=true]{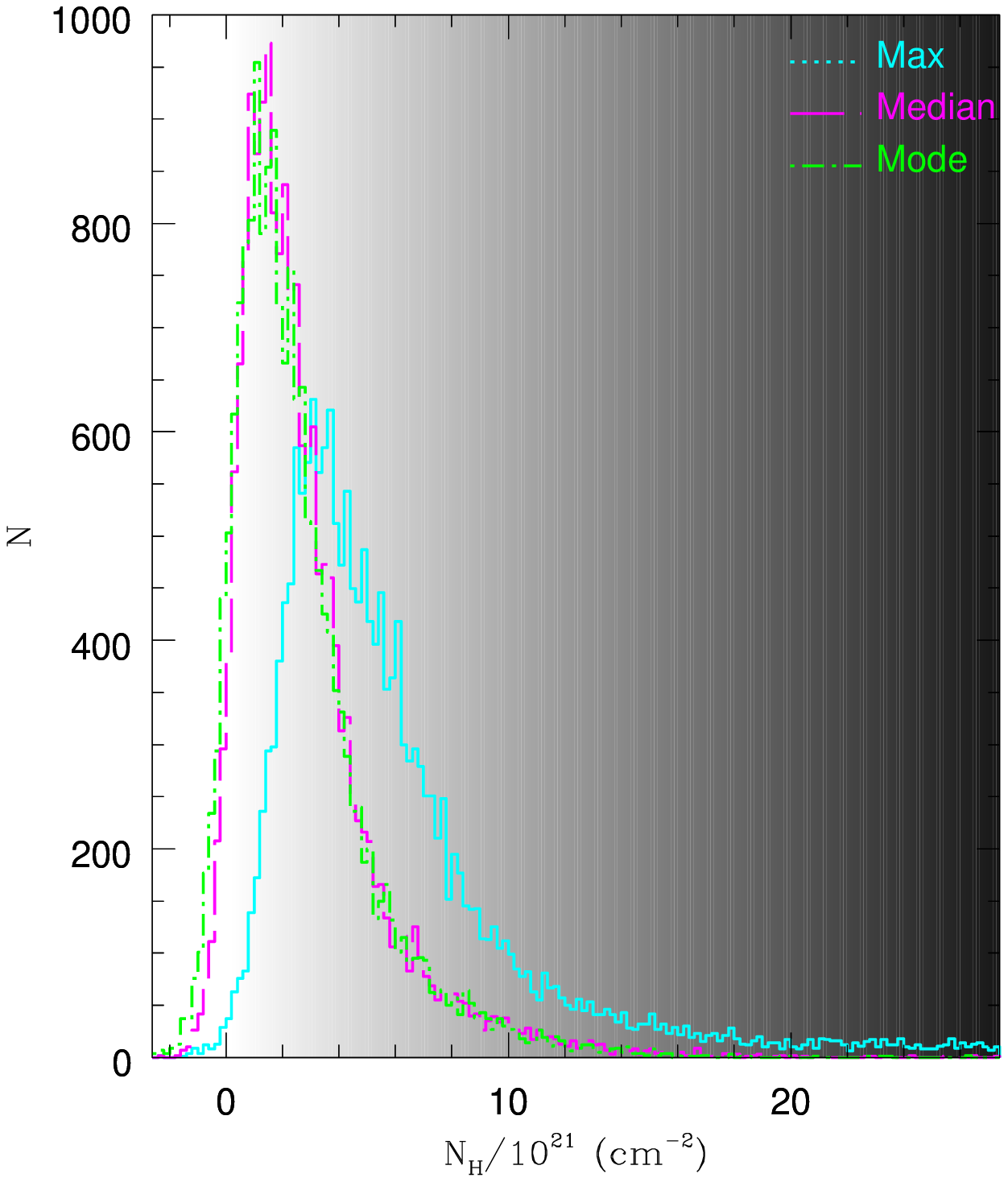}\\
\caption{Maps of inferred total H column density for $1\arcmin\times1\arcmin$ regions of the $6\_6$ tile. Maps are based on \textnormal{(from left--right)} maximum, mode and median extinction. The region containing detector 16 is outlined in red. \textnormal{Histograms of inferred  total H column density are shown in the right panel}.}
\label{fig:30d:dustcol}
\end{figure*}

Figure \ref{fig:30d:dustcol} shows maps for the three methods (\textnormal{from} left -- maximum, -- mode,  -- median), \textnormal{the right panel} shows the density represented by the pixel count and  \textnormal{histograms} of the source distributions \textnormal{are overplotted}. We see that the peak extinction is affected somewhat by random noise from the reddest RC star selection, possibly including some background galaxies. However, small-scale structure with a dense column density (such as molecular clouds) may be responsible. We also see that Figure \ref{fig:30d:dustcol} has maximum column densities as high as $25\times10^{21}\,\mathrm{cm}^{-2}$, such concentrations are no longer atomic hydrogen but almost exclusively molecular hydrogen, meaning that the maps can be compared with molecular clouds surveys. We revisit this point in Section \ref{30d:dis:mc}.

When comparing the maximum with the mode and median, we notice that a small region around RA$=84$\degr,\, Dec$=-69.6$\degr\, is not present in the latter two, implying that the stars are mostly in front of the feature here. There are known H\,{\sc ii} regions (\object{LHA 120-N 154} and \object{DEM L 248}) and a molecular cloud nearby (\object{[FKM2008] LMC N J0536-6941}; see Section \ref{30d:dis:mc}).

We also use these maps in Section \ref{30d:dis:r} where we compare them with H\,{\sc i} column densities measured from observation.

\section{Discussion} \label{30d:dis}

\subsection{Using the data} \label{30d:rel}
In order to make the data products useful to the community we are making reddening maps available in several formats. One of these is a complete catalogue containing each RC star as well as means, medians and maximum values (as well as dispersions) for regions of sizes $30\arcsec$, $1\arcmin$ and 5$\arcmin$. The probability map for reddening in those regions will also be made available as a cumulative probability data cube. 

This is available from CDS. A sample of the table is shown in Table \ref{tab:cdsdat}. The data cube is in the form of FITS files linearly projected in 16 slices.

\begin{table*}
\centering
\caption{Table of RC stars, giving position, reddening and statistics on RC stars surrounding it for 3 diameters; 30\arcsec, 1\arcmin and 5\arcmin. Two sample lines are displayed; the full table is available electronically from the CDS. Ellipsis indicate missing columns.}
\label{tab:cdsdat}
\[
\begin{tabular}{|ccc|ccccc|ccccc}
\hline
\hline
RA & Dec & $E(J-K_\mathrm{s})$ & \multicolumn{5}{c|}{30\arcsec} & \multicolumn{5}{c}{1\arcmin}\\
\multicolumn{2}{|c}{(J2000)} & mag & N & $\sigma$ & Mean & Median & Maximum & N & $\sigma$ & Mean & Median & \ldots \\
\hline
84.4710 & $-$70.1051 & 0.1575 & 1 & 0.0000 & 0.1575 & 0.1575 & 0.1575 & 6 & 0.0472 & 0.1108 & 0.0945 & \ldots \\
84.7594 & $-$70.1042 & $-$0.0395 & 2 & 0.1945 & 0.0980 & 0.2355 & 0.2355 & 6 & 0.1165 & 0.0666 & 0.0995 & \ldots \\
\hline
\end{tabular}
\]
\end{table*}

\subsection{Comparison with optical reddening}\label{30d:dis:h}
\cite{haschke11} produced optical reddening maps using OGLE III data for the LMC and SMC and have made these data available via a webform\footnote{\url{http://dc.zah.uni-heidelberg.de/mcx}}. Their selection criteria define the RC as being in the CMD region of $17.50\leq I \leq19.25$ mag and $0.65\leq(V-I)\leq1.35$ mag for sub fields containing at least $200$ RC stars, where the sub field sizes range from $4\rlap{.}^{\prime}5\times4\rlap{.}^{\prime}5$ to $36^{\prime}\times36^{\prime}$, based on number of RC stars. A histogram is then taken which has a Gaussian plus second order polynomial fitted to it, the peak of which is taken to be the mean and is tested against the reduced $\chi^2$; where fields with values greater than $3$ were flagged for inspection by eye. For the LMC $4$ fields were affected, which was caused by the histogram being double peaked due to the RGB.

Their overall LMC average was $E(V-I)=0.09\pm0.07$ mag; although, for the region we are comparing this average is $E(V-I)=0.13$ mag. They also inspected, by eye, regions where $E(V-I)$ exceeded $0.2$ mag (these are $2\%$ of the LMC fields; totalling $60$ regions). There were $23$ regions which had $E(V-I)\geq0.25$ mag; $16$ are in the region we are comparing. These regions were better represented by moving the RC box to $17.5\leq$$I$$\leq19.5$ mag and $0.9\leq (V-I) \leq1.8$ mag. The mean is calculated as before. For most of these regions the extinction remains unchanged but the $1\sigma$ width becomes much larger (more so in the red than the blue). The extinction values in these recalculated regions are not given in their output tables but are given in Table 2 of \citet{haschke11}.

Their selection box follows a linear selection (rather than a reddening vector) which is between $3$--$4$ times larger than ours and because of this there is greater RGB contamination (largely accounted for by the use of a second-order polynomial).

We compared their $E(V-I)$ values to our $E(J-K_\mathrm{s})$ using two methods. The first method (fixed range) uses the same RC selection as Section \ref{30d:rcloc} and the second method (sliding range) re-defines the RC selection as a fixed box as follows:\\
\begin{equation}  [(J-K_\mathrm{s})_\mathrm{P}-0.25]{}\leq{}(J-K_\mathrm{s}){}\leq [(J-K_\mathrm{s})_\mathrm{P}+0.45]{}\,\mathrm{mag} \end{equation}
\begin{equation}  [(K_\mathrm{s})_\mathrm{P}-0.5]{}\leq{}K{}\leq{}[(K_\mathrm{s})_\mathrm{P}+0.5]{}\,\mathrm{mag} \end{equation}
$(J-K_\mathrm{s})_\mathrm{P}$ and $(K_\mathrm{s})_\mathrm{P}$ are the densest $0.05\times0.05$ mag areas of the CMD for that region.

For both methods we calculate the mean, median and standard deviation ($\sigma$) for each region. The average value for each region is summarised in Table \ref{tab:hasbeen}. The $E(J-K_\mathrm{s})$ median is found to be consistent because both selections cover the colour range where the highest density of RC stars lie. The mean and $\sigma$ $E(J-K_\mathrm{s})$ have greater differences because the sliding range is narrower in colour than the fixed range.
However, the narrower colour in the sliding range is more consistent with the work we are comparing with.

We compare the optical extinction with the average of the mean and median results for the second method. The values are then all converted into $A_V$ using \citet{schlegel98} for the optical and Section \ref{30d:av} for our NIR $E(J-K_\mathrm{s})$. These are referred to as $A_V$\,(Haschke) and $A_V$\,(This Work; hereafter TW), respectively.

\begin{table}
\centering
\caption{Average $E(J-K_\mathrm{s})$ for comparison methods.}
\label{tab:hasbeen}
\[
\begin{tabular}{ccc}
\hline
\hline
 & \multicolumn{2}{c}{Range} \\
 & Fixed & Sliding \\
 & mag & mag \\
\hline
Mean & 0.081 & 0.074 \\
Median & 0.068 & 0.066 \\
Sigma & 0.087 & 0.082 \\
\hline
\end{tabular}
\]
\end{table}

\begin{figure}
\resizebox{\hsize}{!}{\includegraphics[width=0.49\textwidth,clip=true]{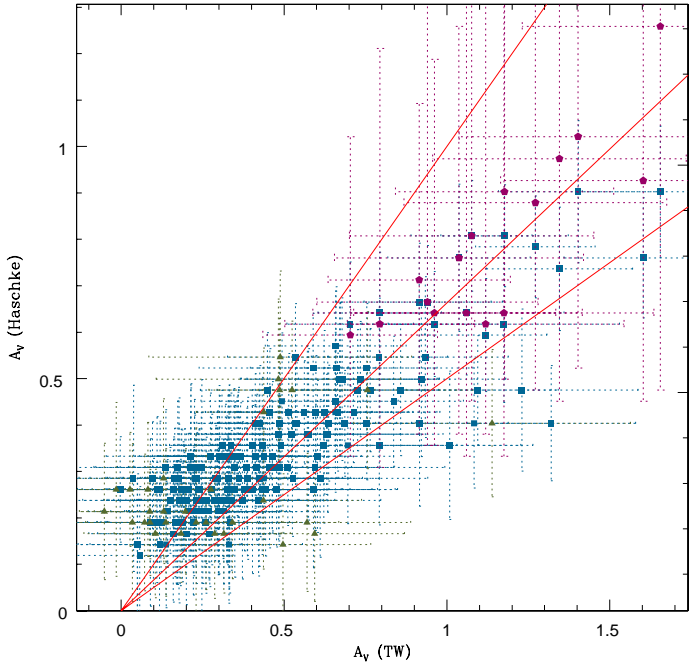}}
\caption{Comparison between $A_V$ (Haschke) and $A_V$ (TW), conversion from \citet{schlegel98} and Section \ref{30d:av}, respectively. Squares indicate regions with over 200 stars while triangles indicate regions with less than 200 stars. Purple hexagons are the values for the recalculated regions. \textnormal{Regions within detector 16 have been excluded from the plot.} Red lines represents gradients equal to $1$, $0.75$ and $0.5$.}
\label{fig:hascompeve}
\end{figure}

Figure \ref{fig:hascompeve} compares the extinction values with the $1\sigma$ widths plotted as error bars. Squares and triangles are for points where our selection has more than and less than $200$ stars, respectively. The purple hexagons show the $A_V$ values of the recalculated $E(V-I)$ values (and their error bars) from \citet{haschke11} against the $A_V$ of our work. Three red lines are drawn showing gradients of $1$, $0.75$ and, $0.5$. A positive trend is observed but considering both are measuring $A_V$ there is considerable deviation from a linear fit beyond $A_V=0.4$ mag (and up to $A_V=0.4$ mag there is considerable scatter but the middle ground is a linear fit). The best fitting red line for higher values of $A_V$ has a gradient of $0.75$, representing the \citeauthor{haschke11} $A_V$ being $75\%$ of our $A_V$, provided we use the recalculated reddening values. In the intermediate range of $0.5<A_V\mathrm{(TW)}<0.8$ mag we see the greatest deviation with Haschke et al. reporting low extinction. Looking at the sky location of these points we find the majority of these points are in the region affected by Detector $16$ (see Section \ref{30d:issues}) where our data is unreliable.

There are some factors to consider regarding differences:\\
$[1]$ Some difference might arise from the extinction within the NIR being much less (approximately a factor of 3 less), meaning the NIR is a less sensitive probe of lower extinction and hence an under estimate.\\
$[2]$ The Galactic foreground may have some influence. According to \citet{schlegel98} this contributes $E(J-K)=0.04$ mag and $E(V-I)=0.103$ mag. This could explain why the zero points do not join exactly at $E(V-I)=E(J-K_\mathrm{s})=0$.\\
$[3]$ Also, we must not ignore the difference in intrinsic colour. \citeauthor{haschke11} assume a metallicity of $Z=0.004$ and obtain $E(V-I)_0=0.92$ mag \citep{olsen02}. We, on the other hand, used a metallicity of $Z=0.0033$. To test the difference, we produced isochrones (as we did in Section \ref{30d:red}, but instead for the $V$ and $I$ bands), finding an intrinsic colour of $E(V-I)_0=0.90$ mag. Using this would redden the Haschke et al. values slightly with the largest effect seen for the lowest reddening.\\

\begin{figure*}
 {\label{fig:hasmapb1}\includegraphics[width=0.33\textwidth]{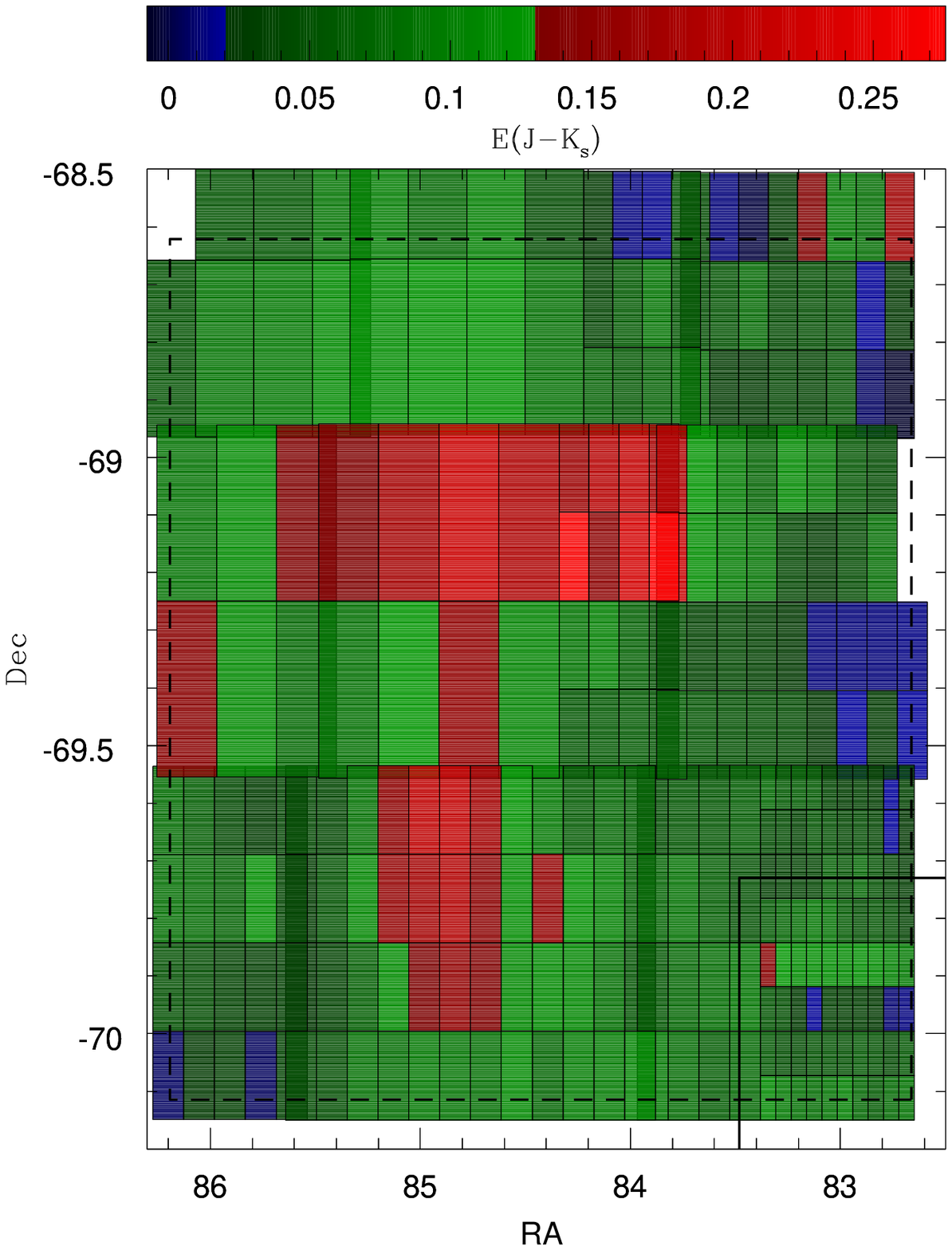}}
  {\label{fig:hasmapb3}\includegraphics[width=0.33\textwidth]{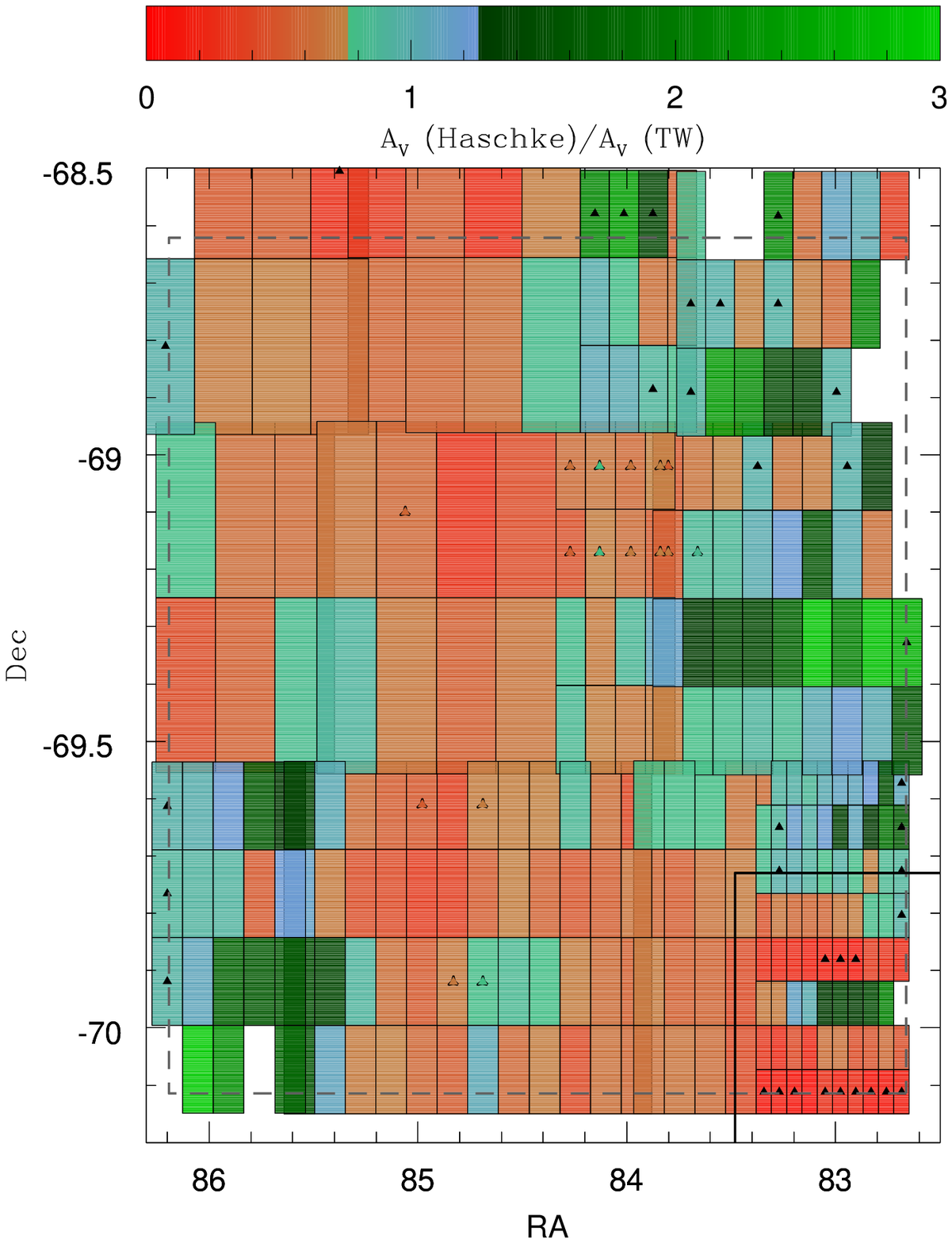}}
 {\label{fig:hasgram}\includegraphics[width=0.33\textwidth]{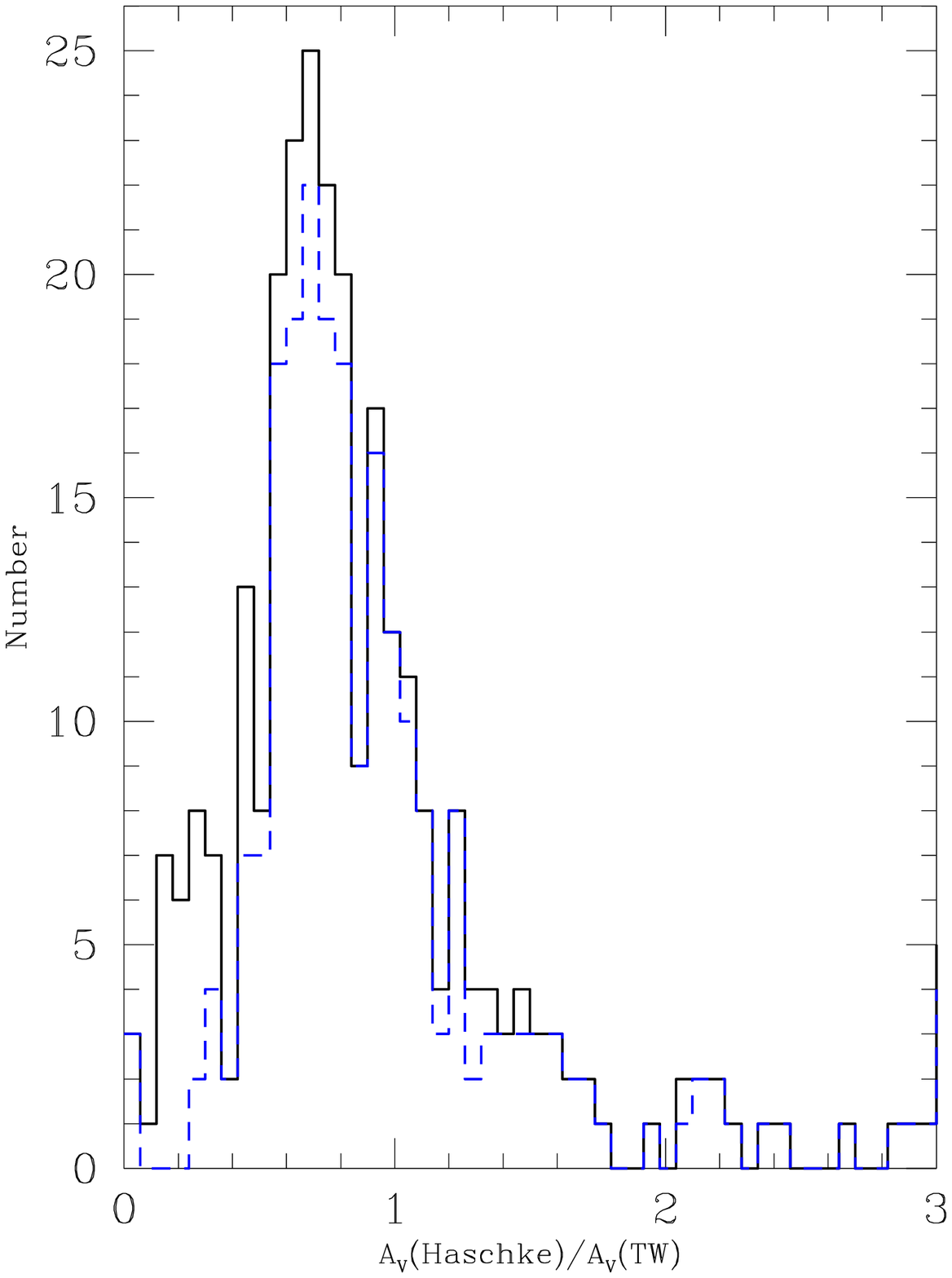}}\\
\caption{Map of sky showing extinction values (left) and $A_V$(Haschke)/$A_V$(TW) (middle). The dotted outline is the boundary of our tile. Black triangles on a cell indicate that particular cell was made using data that comprise of fewer than $200$ stars, detector 16 region outlined in black. Coloured triangles on a cell indicate recalculated extinction ratio (where colour if triangle is recalculated value). Empty cells are where $E(J-K_\mathrm{s})\simeq0$ mag (excluded due to producing extreme values in the ratio range). Right: histogram of $A_V$(Haschke)/$A_V$(TW), the dashed blue histogram excludes that detector 16 region.}
\label{fig:hascomp}
\end{figure*}

Next we examine the effect of sky position on relative extinctions. This is shown using two maps in Figure \ref{fig:hascomp}; the first map is an $E(J-K_\mathrm{s})$ map using the Haschke et al. regions and the second map is the extinction ratio of $A_V$(Haschke)/$A_V$(TW). For the second map we omit regions where $A_V$(TW)$\simeq0$ because these outliers yield a very high ratio in comparison with the rest of the data. These maps are shown in the left and middle panels of Figure \ref{fig:hascomp}. A histogram of the $A_V$ ratios is shown in the right panel of Figure \ref{fig:hascomp}, the peak is at $A_V$(Haschke)/$A_V$(TW)$=0.75$ with a fairly even distribution on both sides of the peak. In dashed blue is the histogram when excluding the detector $16$ region. The distribution is more positively skewed, caused by less lower extinction regions sampled.

When we compare the two maps we see a lower $A_V$(Haschke)/$A_V$(TW) ratio at higher extinction (and the detector 16 region) and the opposite for lower extinction. The cause is likely to arise from a combination of the optical being more sensitive to lower reddening and the intrinsic colour used being bluer than ours. Meanwhile, for higher reddening; the OGLE-III data are prevented from detecting the stars with the most reddening, as these lie outside of the \citeauthor{haschke11} selection box. 

\citeauthor{haschke11} already consider the latter by inspecting these regions by eye and re-drawing the star selection boxes to better cover these regions. The values in these regions have been over plotted as outlined, coloured triangles in the middle panel of Figure \ref{fig:hascomp}. Little change is seen because this does not take into account the $1\sigma$ widths associated with these values (which tend to be skewed toward higher extinction).

Note that the criteria for re-drawing the selection do not consider regions which have both low and high extinction populations. This could lead to some cases of the high extinction component being completely missed.

\citeauthor{haschke11} also used the OGLE III data to produce an RR Lyrae extinction map (shown in their Figures $9$ and $10$) which visually, appears to agree better with our result. However, tabular data is not provided because the stars are sparsely distributed (with $120$ having estimated extinction, using the mean of the surrounding cells due to no RR Lyrae stars contained in them). \textnormal{This suggests that in the optical wavelengths, extinction derived from RR Lyrae stars is more reliable than that of the Red Clump.}

The VMC survey has also made use of the RR Lyrae stars OGLE III database \citep{soszynski09} in studying reddening (Moretti et al. 2013, in prep), producing two maps. 
The first map shows reddening derived from the relation between intrinsic colour, $(V-I)_0$ and V-band amplitude \citep{piersimoni02}. As OGLE III provides I-band amplitudes these were scaled to V-band amplitudes using a fixed scaling factor (1.58, \citealt{dicriscienzo11}).
The second map uses the visual V Period--Luminosity relation (which has large dispersion due to varied reddening range) paying attention to stars with $V>20.5$ mag and stars with $20<V<20.5$ mag. Regions containing these stars trace structures that could be associated with reddening excess due to dust in the regions. These regions correspond with the high reddening regions of the first map suggesting this is the case.

Both of these maps agree with what is shown in Figure \ref{fig:map}. \textnormal{The median reddening value of the first map, $E(V-I)=0.11\pm0.05$ mag agrees excellently with the \cite{haschke11} RR Lyrae reddening median value of $E(V-I)=0.11\pm0.06$ mag. For the RC stars of \citet{haschke11} the difference in reddening values was calculated (for each RR Lyrae star within a RC field) finding a median difference of $0.00\pm0.05$ mag.}

\subsection{Comparison with H\,{\sc i}} \label{30d:dis:r}
The H\,{\sc i} observations come from the Australia Telescope Compact Array (ATCA) observations presented in \citet{kim98}. They are in the form of a ($1998\times2230\times120$) data cube containing 120 channels covering a heliocentric velocity range of $190$--$387$ km\,s$^{-1}$, each covering a velocity dispersion of $1.65$ km\,s$^{-1}$. The resolution per pixel is approximately $1\arcmin\times1\arcmin$. The area that covers our tile is $243\times270$ pixels. Due to the H\,{\sc i} emission being a combination of column density, H\,{\sc i} nuclear spin temperate and background emission this will not always be a perfect mirror of the column density.

In a few of the regions surrounding and within 30 Doradus the emission features become absorption features. This is due to nearby bright sources affecting the instrument normalisation leading to overcompensation, or self-absorption (Figure 1 of \citealt{marx-zimmer00}, identifies these regions as absorption sites). However, the total number of pixels affected by this ($12\times8$ and $3\times3$) are a small minority of the total ($243\times270$) and have been excluded from analysis. Figure \ref{fig:mipsmaps} shows the total H\,{\sc i} emission in the top--right panel.

We convert the emission from (Jy beam $^{-1}$) into column densities (cm$^{-2}$) using equations 1 and 5 from \citet{walter08}. This is then compared with the \textnormal{inferred} total column density derived in Section \ref{30d:dust}. For convenience in comparing Figure \ref{fig:30d:dustcol} with the top--right panel of Figure \ref{fig:mipsmaps}, column densities are given in Table \ref{tab:h1cols}.

\begin{table}
\centering
\caption{Table of total H\,{\sc i} emission and corresponding total H\,{\sc i} column density.}
\label{tab:h1cols}
\[
\begin{tabular}{cc}
\hline
\hline
Emission & Column density \\
(Jy beam $^{-1}$) & ($\mathrm{cm}^{-2}$) \\
\hline
4 & 2.03$\times10^{21}$ \\
8 & 4.06$\times10^{21}$ \\
11 & 5.58$\times10^{21}$ \\
15 & 7.61$\times10^{21}$ \\
19 & 9.64$\times10^{21}$ \\
\hline
\end{tabular}
\]
\end{table}

We compare the measured H\,{\sc i} and \textnormal{inferred} total H (derived from $A_V$) column densities using two methods. The first (individual) method compares the total column density of each RC star against the measured column density at the location of that RC star and the second (group) method compares mean and maximum total column densities for a region of RC stars equal to a one pixel of the H\,{\sc i} data. As the $A_V$ is a surrogate of all the H content rather than just the H\,{\sc i} we expect to find agreement for the two data between .

\begin{figure}
\centering
\resizebox{\hsize}{!}{\includegraphics[angle=0,width=0.48\textwidth]{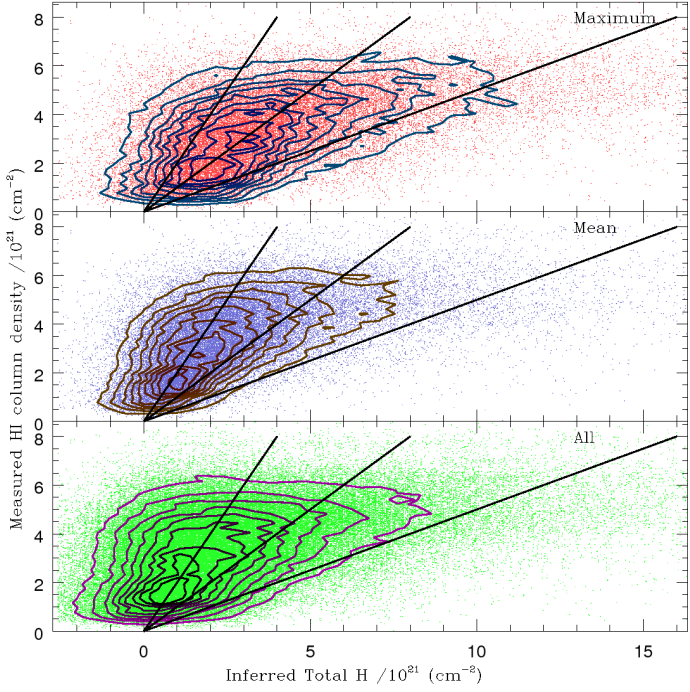}}
 \caption{Measured H\,{\sc i} column density vs. \textnormal{inferred} total column density (converted from extinction). Points depict individual values, contours show increasing source density ($10$ levels ranging from $10\%$--$100\%$ of highest density for bin size of $2\%\times2\%$ for each axis length), darker lines are higher density. Lines show gradients of \textnormal{2}, 1 (i.e. $N_\mathrm{H}=N_\mathrm{H}$) and \textnormal{0.5}. \textnormal{Analysis excludes regions within detector 16.}}
\label{fig:30d:comphydro}
\end{figure}

The measured H\,{\sc i} column density vs. \textnormal{Inferred} total H column density for both methods is shown in the panels of Figure \ref{fig:30d:comphydro}. Overplotted are gradients of 2:1, 1:1 and 1:2 and contours ($10$ density levels of $10\%$--$100\%$ of maximum, for bin sizes of $2\%$ in each axis data range). We see the best fit to the measured H\,{\sc i} vs. maximum \textnormal{inferred} total H is the 1:1 line up to about $N_H=6\times10^{21}$ cm$^{-2}$ where the measured H\,{\sc i} levels off. The best fit to the measured H\,{\sc i} vs. mean \textnormal{inferred} total H is the 2:1 line, supporting the hypothesis that for a given sightline, on average half the stars lie in front of the H\,{\sc i} column. The levelling off is once again observed.

There are a number of possible causes for this effect. To highlight a few:\\
$[1]$ The $1\arcmin$ resolution of the emission does not resolve the small structure causing the extinction; \\
$[2]$ The instrumentation only detects up to $0.8$ Jy\,beam$^{-1}$ for a given velocity channel. As the spectrum peaks in a certain range this might mean the instrumentation becomes saturated, which is reflected in the observed plateau. \\
$[3]$ Hydrogen becomes molecular in the dustiest clouds and thus ceases to contribute to the column density. Additionally, the levelling off effect begins at $N_H=4\times10^{21}$ cm$^{-2}$ marking this as the start point of atomic--molecular transition.

The latter is the most likely and can be checked by focusing on the upper end of maximum extinction and seeing where these lie in relation to known molecular clouds. This is performed in Section \ref{30d:dis:mc}.

\begin{figure}
\centering
\resizebox{\hsize}{!}{\includegraphics[angle=0,width=0.50\textwidth]{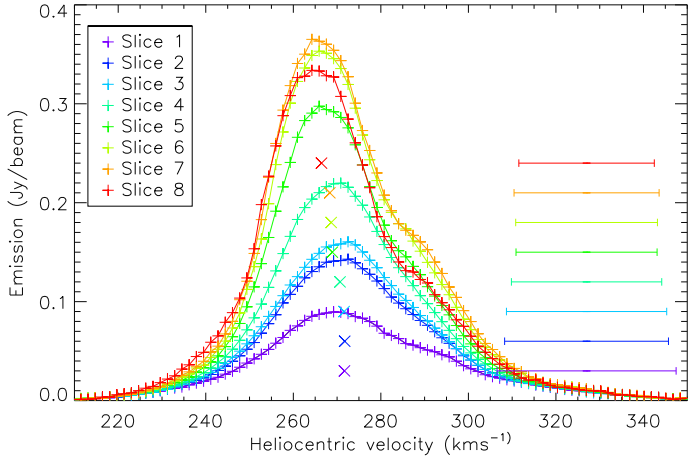}}
 \caption{Mean H\,{\sc i} emission for each velocity slice and extinction slice for mean extinction of each pixel. Extinction ranges of each slice given in Table \ref{tab:colslin}. Peak of emission tends slightly to early slice and distribution of emission narrows for highest extinction.}
\label{fig:shit}
\end{figure}

We also measured the velocity of the peak emission finding this lies between the $40^{\mathrm{th}}$ and $60^{\mathrm{th}}$ channel ($256$--$289$ km\,s$^{-1}$), a range which got slightly narrower for higher extinction. To demonstrate this, we split the H\,{\sc i} map into 8 extinction slices (using the ranges from Table \ref{tab:colslin}) by using the maximum extinction calculated for Figure \ref{fig:30d:comphydro}. For each slice we then calculate spectra by averaging the emission for each of the $120$ velocity channels contained in each extinction slice. The resulting 8 spectra are shown in Figure \ref{fig:shit}.We see the distribution narrowing, the average grows and the peak of the graph tends to appear in slightly lower channels with increasing extinction, this is demonstrated with the FWHM and central peak plotted in Figure \ref{fig:shit}. This reinforces what can be deduced from Figure \ref{fig:30d:comphydro}; for high extinction, the distribution narrows and a higher ratio of sources reach the plateau value.

\subsection{Comparison with molecular clouds} \label{30d:dis:mc}
In Figure \ref{fig:30d:comphydro} we saw that the observed H\,{\sc i} started to level off after $N_H\,\simeq\,4\times10^{21}$ cm$^{-2}$ (i.e. saturation is sufficient to begin formation of H\,{\sc ii}) and generally finished off around $N_H\,\simeq\,6\times10^{21}$ cm$^{-2}$ where in between is a mixture of H\,{\sc ii} and H\,{\sc i} due the transition being incomplete and observing some H\,{\sc i} in front of the site of transition. Figure 1 of \citet{schaye01} shows Hydrogen column densities as a function of density and the fractions of H\,{\sc i} and H\,{\sc ii} compared to the total hydrogen and the H\,{\sc ii} dominates this at $N_H\,\simeq\,5.6\times10^{21}$ cm$^{-2}$. Figure 2 of \citet{schaye01} shows the metallicity vs. total H\,{\sc i} content and there is little change between Galactic and LMC metallicity.

\begin{figure}
\centering
\resizebox{\hsize}{!}{\includegraphics[angle=0,width=0.5\textwidth,clip=true]{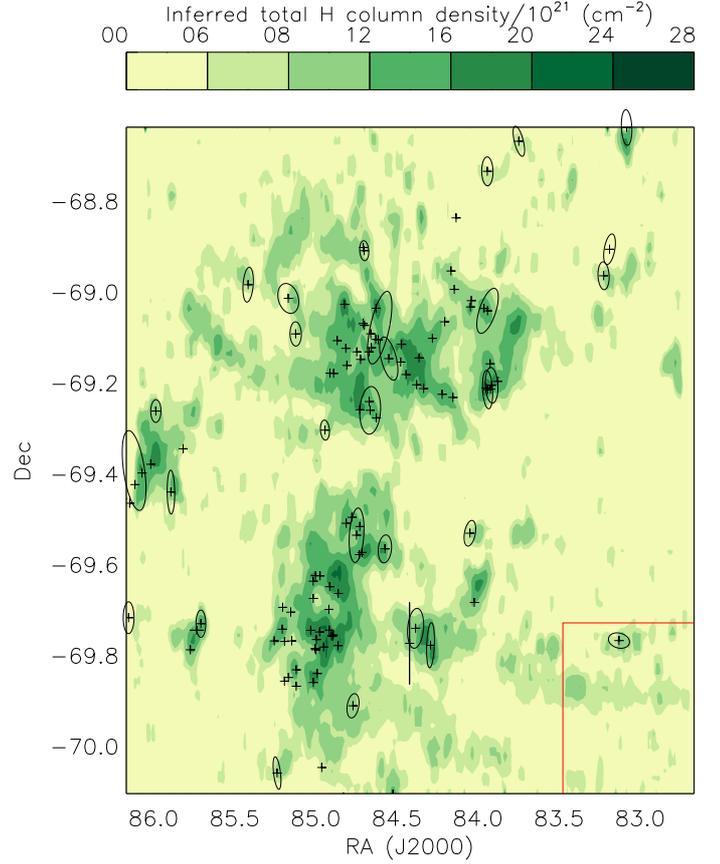}}
 \caption{Smoothed total H column density map identifying regions where $N_{H\,{_I}}>8\times10^{21}$\,cm$^{-2}$. Points in black are molecular clouds identified in the literature with ellipses drawn where \citet{fukui08} estimate properties.}
\label{fig:molcpot}
\end{figure}

The maximum total H emission map produced in Section \ref{30d:dust} is used to identify potential molecular cloud regions. This is done by boxcar smoothing the data (to remove noise) and then applying a cut of $N_H>8\times10^{21}$\,cm$^{-2}$. This cut is required because only at high densities does hydrogen become molecular. We also look at sites where $N_H>6\times10^{21}$\,cm$^{-2}$ because we have seen the atomic--molecular transition to be largely complete at this level. These are then compared with identified molecular cloud sites from the literature obtained via a SIMBAD\footnote{\url{http://simbad.u-strasbg.fr/simbad/}} query. A majority of these originate from the work of \citet{fukui08} (used CO as a surrogate for H\,{\sc ii}) who also provide size estimates. Figure \ref{fig:molcpot} shows the map with these identified sites overplotted and their radii drawn where possible.

On the whole we see good agreement with increasing column density showing scaled reddening to be an effective tool for finding molecular clouds. However, it is not an absolute tool as we have regions of high density where there are no nearby molecular clouds and a few regions in the north where \citet{fukui08} has identified clouds which we find to be below $N_H=8\times10^{21}$\,cm$^{-2}$. These regions are identified as LMC N J0532-6838, LMC N J0535-6844, LMC N J0532-6854 in table 1,3 of \citet{fukui08}. They have fairly low CO luminosities. Also, one is a small cloud; LMC N J0536-6850, which \citet{fukui08} do not derive properties for.

When comparing the top--right panel of Figure \ref{fig:mipsmaps} with Figure \ref{fig:molcpot} we see that \textnormal{in some regions,} where the H\,{\sc i} emission levels off ($N_H>11$ Jy beam $^{-1}$) does not correspond with molecular clouds sites. The north-east region (which is also a region of low RC star density; see top-middle panel of Figure \ref{fig:dens}) is \textnormal{one} notable example.

\subsection{Comparison with dust emission} \label{30d:dis:m}
Reddening is caused by dust in front of a star (dust extinction); a comparison between the extinction and emission may reveal the relative location of the stars and dust. A complicating, but in itself interesting factor is that reddening probes a very narrow column -- a stellar diameter wide or $10^{-7}$ pc for a typical RC star -- whereas the angular resolution of images is typically $0.1$--$10$ pc at the distance of the LMC; comparison between reddening and emission maps can therefore reveal information about the smallest scales at which the dust clouds present structure. 

To this aim we compare our extinction map with the Multiband Imaging Photometer for Spitzer (MIPS) maps from the Surveying the Agents of Galaxy Evolution (SAGE) survey \citep{meixner06}. SAGE is a uniform and unbiased survey of the LMC in the IRAC and MIPS bands of the Spitzer Space Telescope (covering a wavelength range of $3.6$--$160\,\mu$m).
The MIPS data we are using are the $24\,\mu$m and $70\,\mu$m maps which have pixel sizes of $2\rlap{.}^{\prime\prime}49$ and $4\rlap{.}^{\prime\prime}80$, respectively. The $160\,\mu$m map is not used because the dominant background source (the complex Milky Way cirrus) has not been removed.

The shorter wavelength $24\,\mu$m data trace warmer and smaller dust grains while the longer wavelength $70\,\mu$m data trace larger, cooler dust grains. The maps being used have had the residual background and bright point sources removed.

Like in Section \ref{30d:dis:r}, we are comparing our extinction map to these emission maps in two ways. The first way is by obtaining the corresponding MIPS emission value for each of our RC stars (individual method) and the second way is by obtaining extinction over a range of MIPS pixels with means, modes and maximum values produced (group method). The pixel ranges chosen result in similar resolutions for both $24\,\mu$m and $70\,\mu$m data (see sections \ref{30d:dis:m24} \& \ref{30d:dis:m70} for exact numbers).

For both methods we produce two diagrams with contour plots overlaid comparing emission (presented on a logarithmic axis due to the wide range of values) against extinction. The reason for overlaying contours is to identify where the majority of the data lies more easily and the significance of outliers is overestimated from looking at diagrams alone (because in high density regions sources heavily overlap, lowering their visual impact). The reason for two plots is that the second one applies boxcar smoothing with a filter size of $3$ to the emission and extinction. The smoothing reduces noise and outliers. We also fit linear plots (emission increase per $0.01$ mag of extinction increase) to the Maximum extinction.

For the group method we also produce maps of the comparison, for easier regional inspection of data and comparison with prior maps.

The average extinction is expected to generally increase with respect to emission. However, when it does not this either means that the star population lies in front of that traced ISM feature or; the extinction sightlines miss the ISM because its structure is very clumpy. The maximum extinction values should show a tighter, more continuous correlation with surface brightness.

\subsubsection{$24\,\mu$m emission} \label{30d:dis:m24}
The data we are using are the second MIPS data release from early $2008$. For these data the individual MIPS observations were mosaicked to match the size and positions of the $1^{\circ}\times1^{\circ}$ Spitzer InfraRed Array Camera (IRAC) tiles the SAGE programme also observed. MIPS tiles $22$, $23$, $32$ and, $33$ contain the area of our extinction map. These tiles were cropped to only cover the areas within our extinction map. The emission range is much greater in the eastern tiles than in the western tiles (due to the southern molecular ridge and 30 Doradus). However, the overall majority of the emission lies within a similar emission range for all tiles. For the second method each comparison pixel in our reddening map is equal to $13\times13$ pixels of the MIPS map making each of our comparison pixels sample a $32\rlap{.}\arcsec37\times32\rlap{.}\arcsec37$ region.

To account for the range of emission values for the map we define contour levels using a power law ($10^{x/5.6}$) rather than linear values. There are some gaps in the emission map arising from bright source subtraction leading to over subtraction in some regions (e.g. centre of 30 Doradus at RA$=84.6$\degr, Dec$=-69.1$\degr). The map is shown in the top--left panel of Figure \ref{fig:mipsmaps} for the emission range of $1$--$2471$ MJy\,sr$^{-1}$.

\subsubsection{$70\,\mu$m emission} \label{30d:dis:m70} 
For this wavelength we used the $2009$ release of the full $70\,\mu$m mosaic containing the combined Epoch 1 and Epoch 2 data. This being a full mosaic, unlike the $24\,\mu$m data, all of the observing area is contained in a single combined file which we again cropped. However, the effects of detector variance are more visible in this format and at this wavelength. These effects were our main reason for opting to apply boxcar smoothing to the group method described in Section \ref{30d:dis:m}.

With the second method each comparison pixel in our reddening map is equal to $7\times7$ pixels of the MIPS map (this was chosen due to the pixel count of the cropped MIPS map for this tile, also tested were $1\times1$, $2\times2$ and $14\times14$). This means the second method covers areas of $33\rlap{.}\arcsec6\times33\rlap{.}\arcsec6$.

Unlike the $24\,\mu$m data a linear emission scale of $15$--$315$ MJy\,sr$^{-1}$ with steps of $20$ MJy\,sr$^{-1}$ suffices for showing the data. This range allows the map shown in the top--middle panel of Figure \ref{fig:mipsmaps} to cover approximately the same regions as the $24\,\mu$m map.

\subsubsection{The comparisons} \label{30d:dis:mcomp}
Note that the FIR emission is a combination of dust emissivity, $N_\mathrm{dust}$ and $T_\mathrm{dust}$ (via the Planck function) and $T_\mathrm{dust}$ will be higher and emission brighter in the immediate vicinity of H\,{\sc ii} regions.

\begin{figure*}
\centering
 {\label{fig:mips24map1}\includegraphics[width=0.32\textwidth,clip=true]{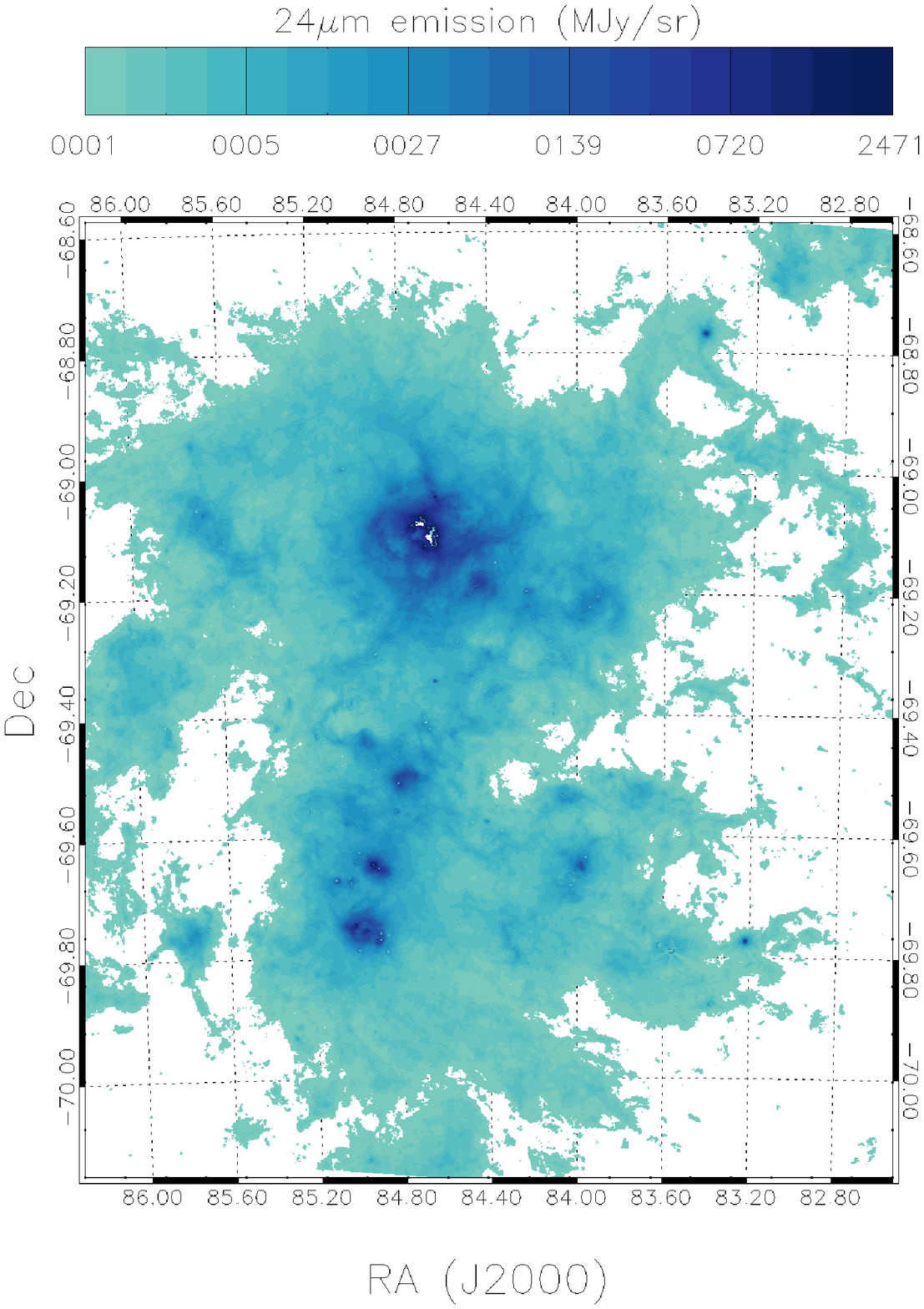}}
 {\label{fig:mips70map1}\includegraphics[width=0.32\textwidth,clip=true]{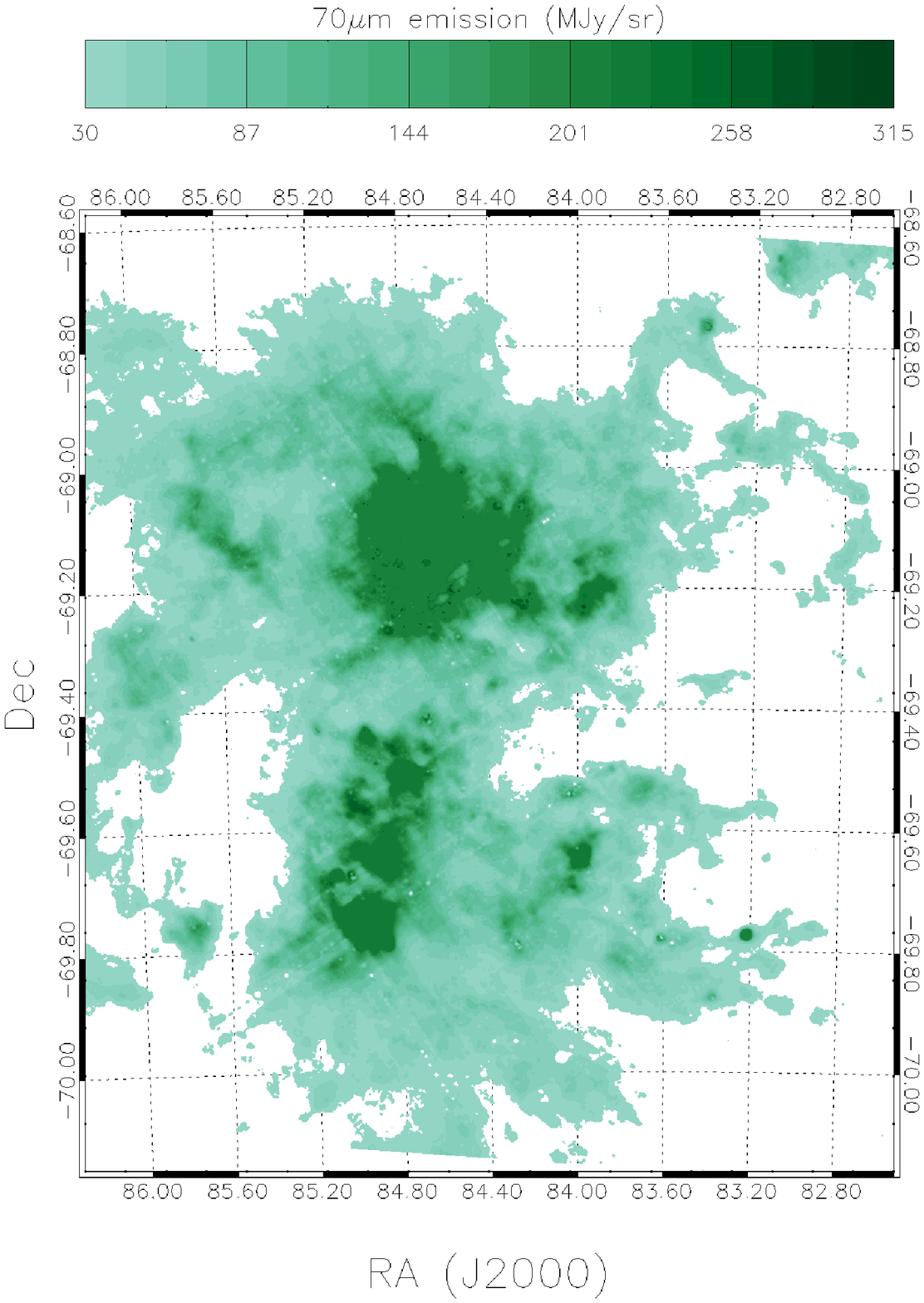}}
  {\label{fig:h1map1}\includegraphics[width=0.32\textwidth,clip=true]{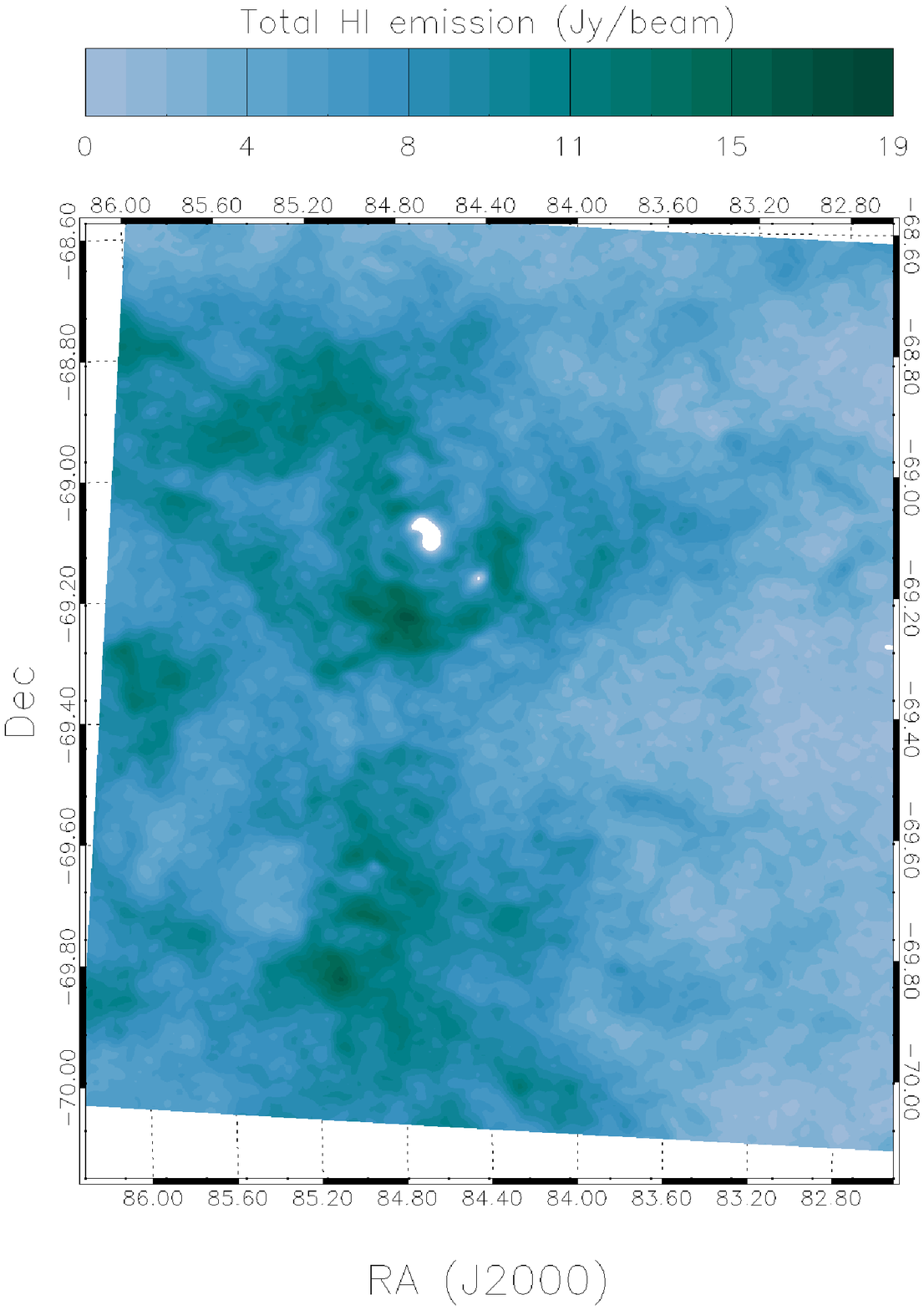}}\\
 {\label{fig:mipsrcmap1}\includegraphics[width=0.32\textwidth,clip=true]{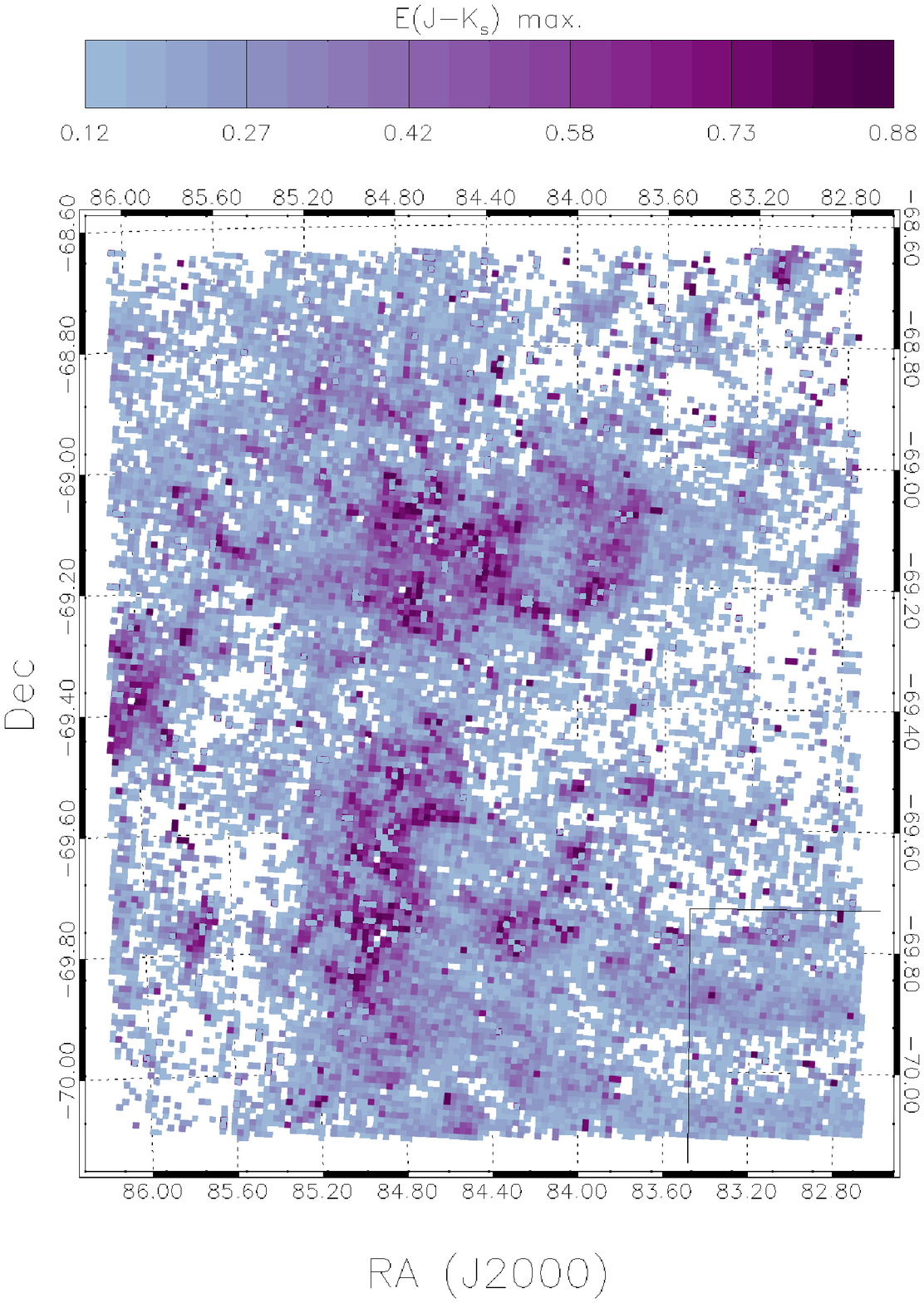}}
 {\label{fig:mipsrcmap2}\includegraphics[width=0.32\textwidth,clip=true]{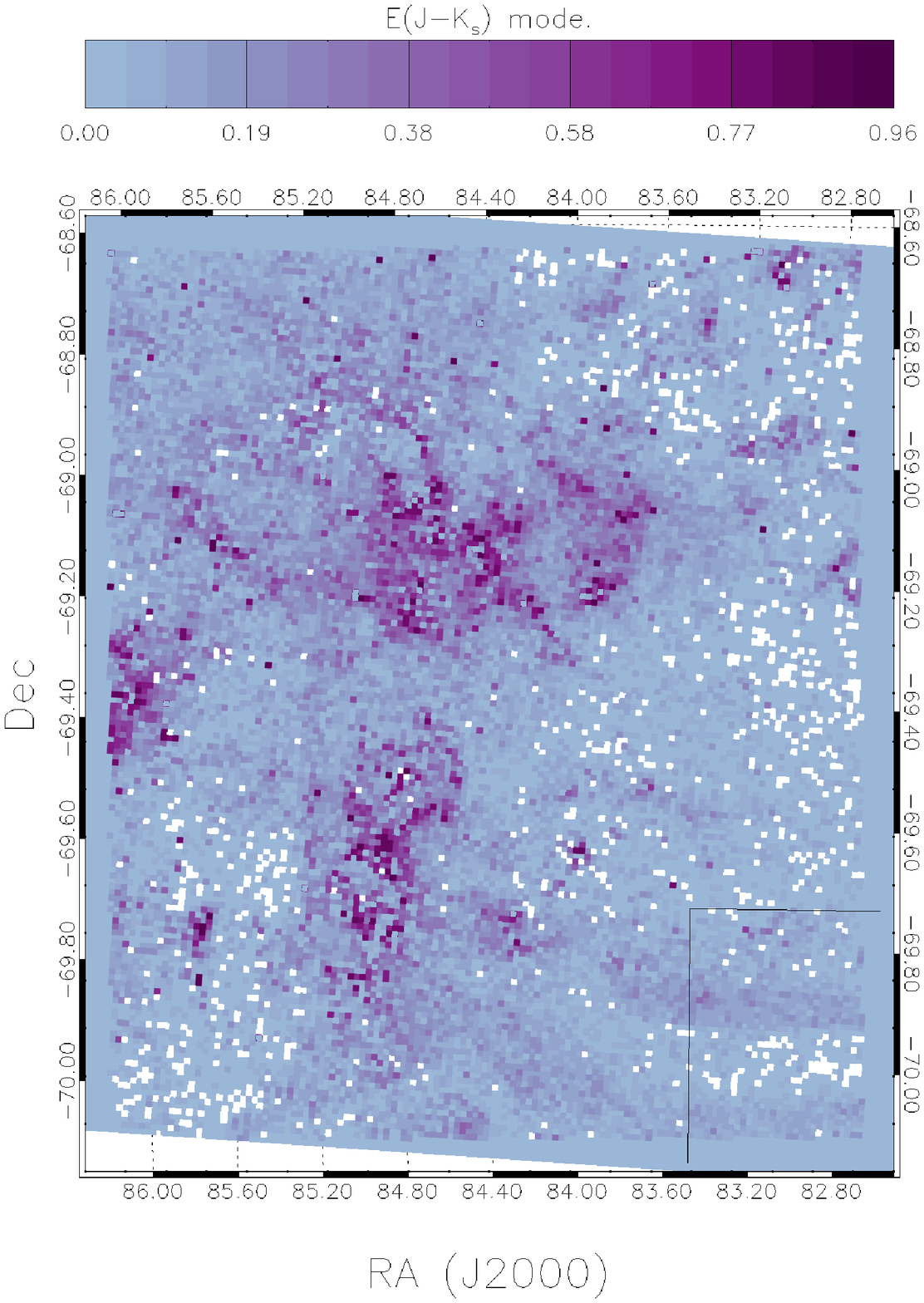}}
  {\label{fig:mipsrcmap3}\includegraphics[width=0.32\textwidth,clip=true]{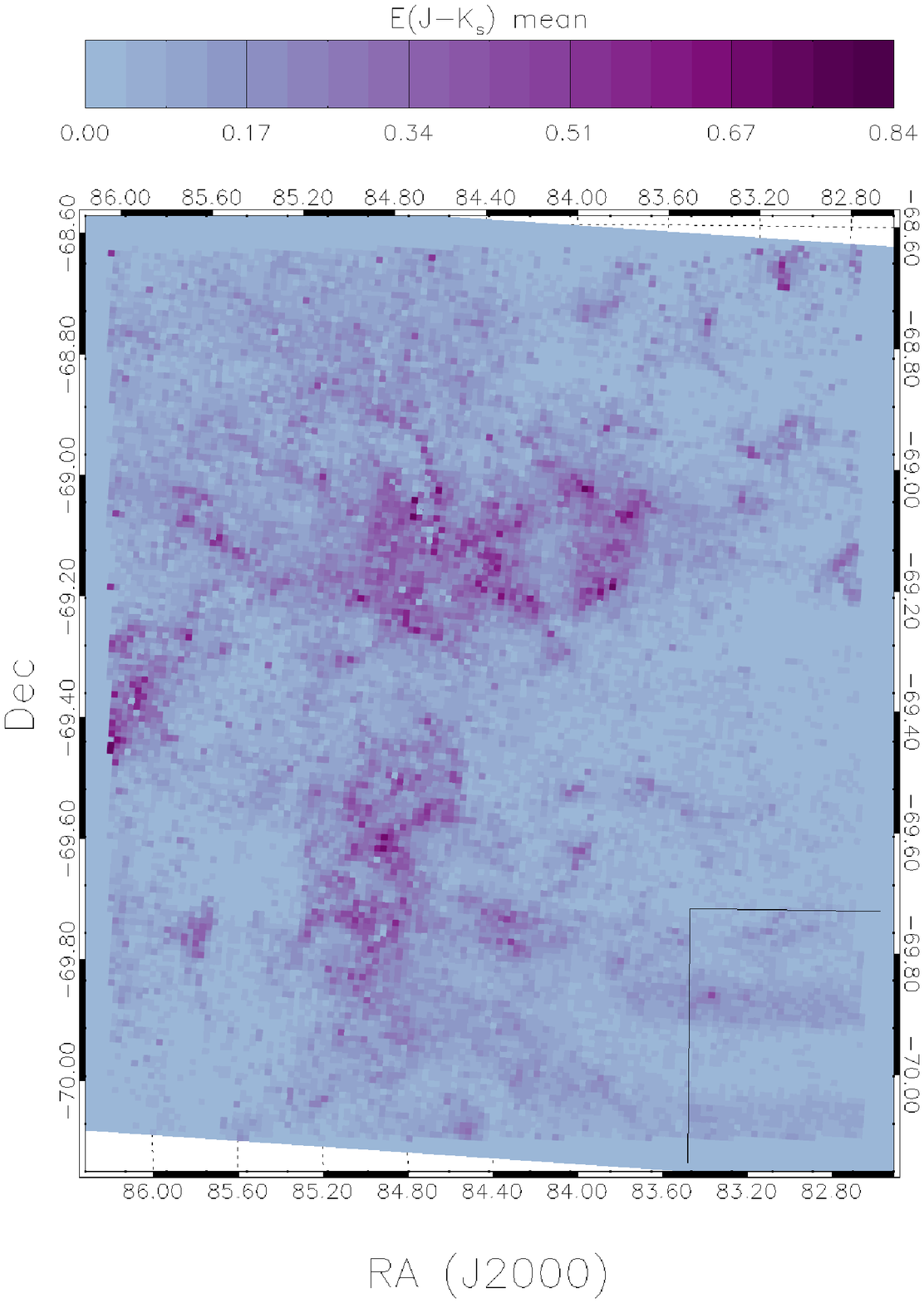}} \\
 \caption{Top--left: The $24\,\mu$m FIR maps with emission scale of $1$--$2471$ MJy\,sr$^{-1}$. Top--middle: The FIR $70\,\mu$m map with emission scale of $30$--$315$  MJy\,sr$^{-1}$. Top--right: Total H\,{\sc i} emission (Jy\,beam$^{-1}$). Sources outside stated limits are omitted from these maps.
Bottom: The peak (left), mode (middle) and median (right) extinction map from our data for the tile using the group method. Sources outside of scale are omitted from these maps.}
\label{fig:mipsmaps}
\end{figure*}

\textnormal{The first comparison we perform is by directly looking at the reddening and the emission maps. These are shown in Figure \ref{fig:mipsmaps}. There is good correlation between emission peaks and high extinction. For example, the bright FIR emission around 30 Doradus\footnote{RA=$84\rlap{.}^{\circ}8$, Dec=$-69\rlap{.}^{\circ}1$} and the southern molecular ridge\footnote{RA=$85\rlap{.}^{\circ}0$, Dec=$-69\rlap{.}^{\circ}75$} is matched by high reddening. With regard to correlation between lower extinction and emission, the non-mapped emission regions (those with FIR emissions below 1 MJy\,sr$^{-1}$ at $24\,\mu$m and below 15 MJy\,sr$^{-1}$ at $70\,\mu$m) correspond to reddening slices $1$--$3$\footnote{colour excess of $E(J-K_\mathrm{s}){}\leq0.105$ mag and $A_V{}\leq{}0.605$ mag}.}

However there \textnormal{are two types of regional differences we observe; regions where high extinction is found but there is less emission and the opposite where strong emission is found without corresponding extinction.}

\textnormal{Two prominent regions with high extinction and less emission are found in the east\footnote{RA=$86^{\circ}$, Dec=$-68\rlap{.}^{\circ}4$} and a crescent west of R136\footnote{RA$=84\rlap{.}^{\circ}0$, $-69\rlap{.}^{\circ}0\leq$Dec$\leq{}-69\rlap{.}^{\circ}2$}. If we assume the FIR emission source is the sole cause of extinction then this would mean that this emission has a much stronger effect in this region. A more likely explanation is that these are regions where the dust clouds have small scale structure that is diluted by the FIR emission, but revealed by the RC stars (we see this in all our reddening maps). When examining these regions in prior comparisons, we find optical extinction (in Figure \ref{fig:hascomp}) is around $60\%$--$80\%$ of our NIR extinction and the measured H\,{\sc i} map (top--right panel of Figure \ref{fig:mipsmaps}) agrees with the eastern region but only the bottom half of the crescent. Interference patterns caused by the nearby 30 Doradus (evident from the ring like structure seen in the measured H\,{\sc i} map) may be a possibly cause. The eastern region contains known molecular clouds (Figure \ref{fig:molcpot}) while the crescent region has molecular clouds in the top and bottom regions but not the middle. These molecular clouds regions were also based on the H\,{\sc i} data.}

\textnormal{Two examples of the opposite effect (bright FIR emission but lower extinction), are found around RA=$83\rlap{.}^{\circ}2$, Dec=$-69\rlap{.}^{\circ}8$ and RA=$84\rlap{.}^{\circ}0$, Dec=$-69\rlap{.}^{\circ}6$. These are fairly small structures (too small to be detected in the regions sampled in Figure \ref{fig:colsli}). Also the narrow sightlines of the RC mean that compact clouds will not be reflected in extinction if there is not a RC star detected behind them and the likelihood of missing such clouds increases for smaller structures. Given the second region is only partially missed suggests this is the case. Finally, the first region also lies just within the detector 16 region (which as a whole does not correlate well with the features seen in the FIR maps).}

\textnormal{Another comparison we perform is by comparing emission with extinction on a region-by-region and star-by-star basis (like was performed for the H\,{\sc i} data in Section \ref{30d:dis:r}). To account for decreasing source density at higher emissions we plot emission on a logarithmic scale while the contours are plotted the same as in Figure \ref{fig:30d:comphydro} except the bin sizes now cover $1\%$ of the x-axis range and $2\%$ of the y-axis range. For the group methods we plot linear fits (which appear curved due to log scale) for each comparison where the gradient is twice as steep for the mean comparison as it is for the maximum, accounting for the fact that on average half of RC stars are in front of the dust. These comparisons exclude sources from within the detector 16 region and regions where there is zero or negative emission (caused by bright source subtraction). We supplement each comparison with a boxcar smoothed version.}

\begin{figure}
\centering
\resizebox{\hsize}{!}{\includegraphics[angle=0,width=0.48\textwidth]{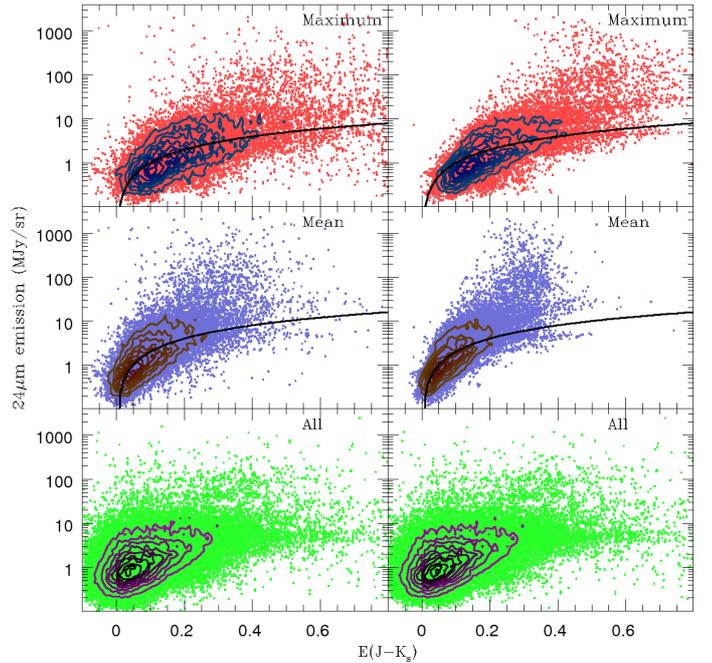}}
 \caption{$24\mu$m emission vs. $E(J-K_\mathrm{s})$ extinction for individual method (bottom) and group method means (middle) and maximum (top). Contours represent increasing source density. Right panel applies boxcar smoothing to the data. \textnormal{The linear fits applied are: Emission$=20\times\,E(J-K_\mathrm{s})$ MJy/sr to the mean and Emission$=10\times\,E(J-K_\mathrm{s})$ MJy/sr to the maximum.}}
\label{fig:comp24s}
\end{figure}

\begin{figure}
\centering
\resizebox{\hsize}{!}{\includegraphics[angle=0,width=0.48\textwidth]{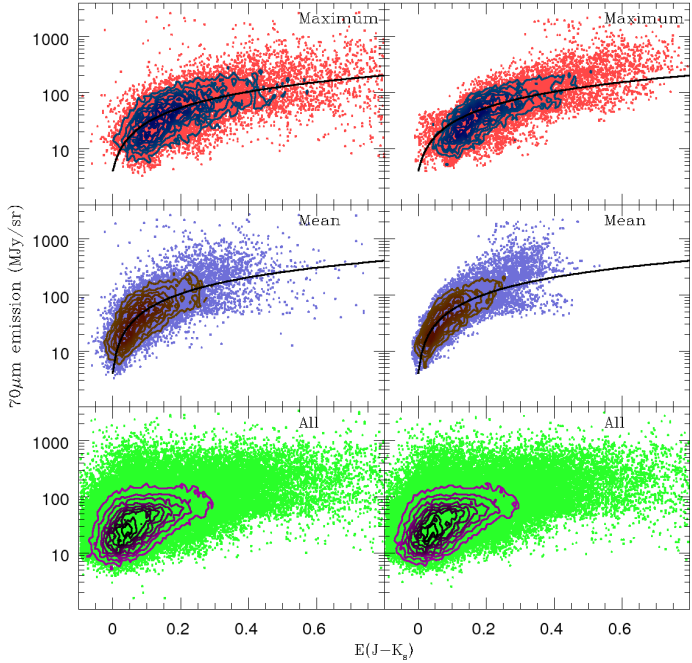}}
 \caption{$70\mu$m emission vs. $E(J-K_\mathrm{s})$ extinction for individual method (bottom) and group method means (middle) and maximum (top). Contours represent increasing source density. Right panel applies boxcar smoothing to the data. \textnormal{The linear fits applied are: Emission$=(500\times\,E(J-K_\mathrm{s})+4)$ MJy/sr to the mean and Emission$=(250\times\,E(J-K_\mathrm{s})+4)$ MJy/sr to the maximum.}}
\label{fig:comp70s}
\end{figure}

\textnormal{Figures \ref{fig:comp24s} and \ref{fig:comp70s} compare the $24\,\mu$m and $70\,\mu$m emission to the $E(J-K_\mathrm{s})$ extinction. The individual methods in the bottom panels consistently show low source density ($<10\%$) above $E(J-K_\mathrm{s})=0.3$ mag. A tail of sources around 10 MJy\,sr$^{-1}$ is seen above $E(J-K_\mathrm{s})=0.5$ mag in the $24\,\mu$m emission. The $70\,\mu$m emission distribution is similar to the H\,{\sc i} (Figure \ref{fig:30d:comphydro}) with high emission found for much of the extinction range.}

\textnormal{We see for the group methods that the chosen linear fits fit the contour ranges well. However we should focus on where this is not the case because this can help us identify regions of interest. In general these regions are more common in the $24\,\mu$m emission than in the $70\,\mu$m emission.} 

\begin{figure*}
\centering
 {\label{fig:compmap24}\includegraphics[width=0.32\textwidth,clip=true]{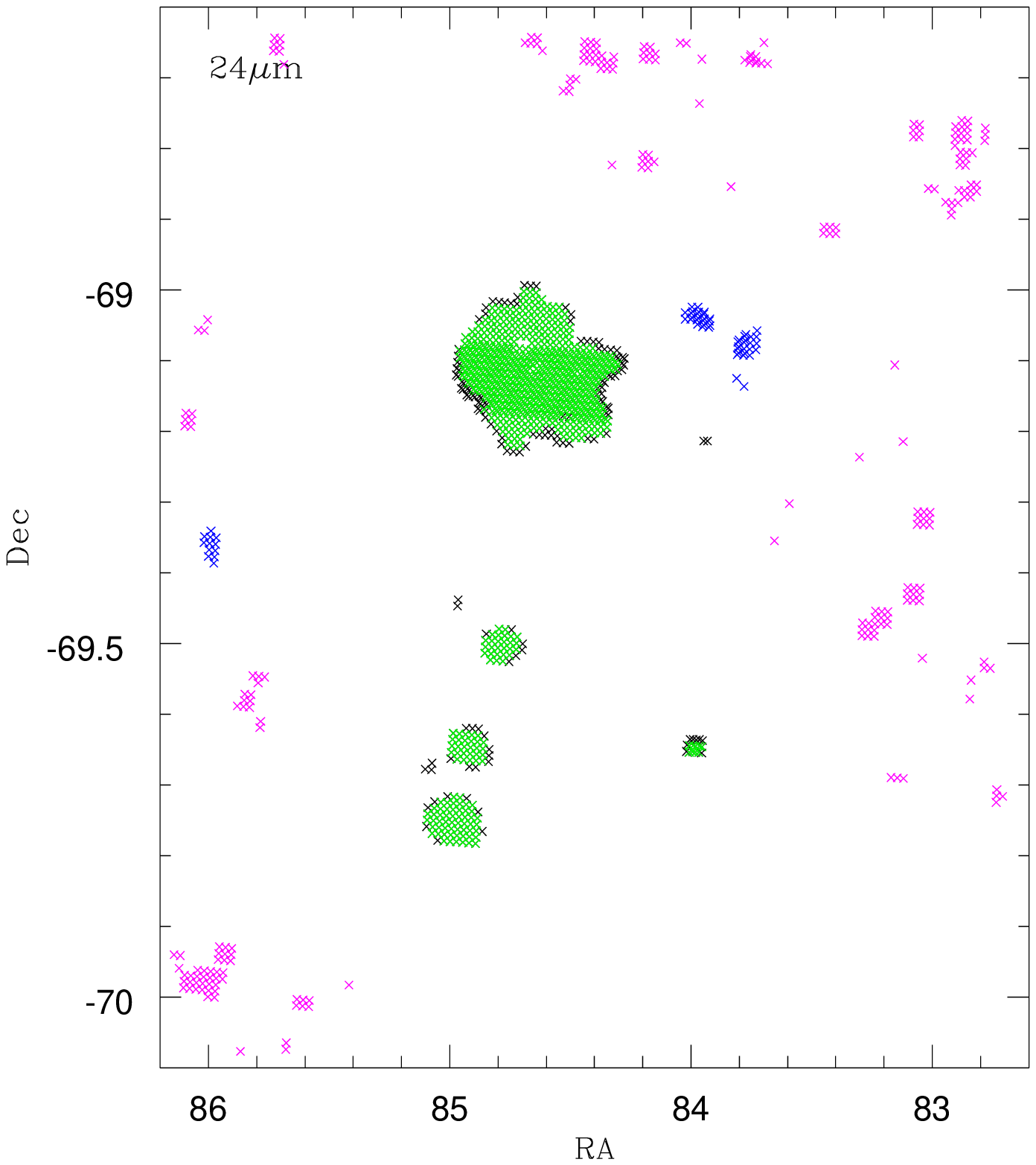}}
 {\label{fig:compmapregs}\includegraphics[width=0.32\textwidth,clip=true]{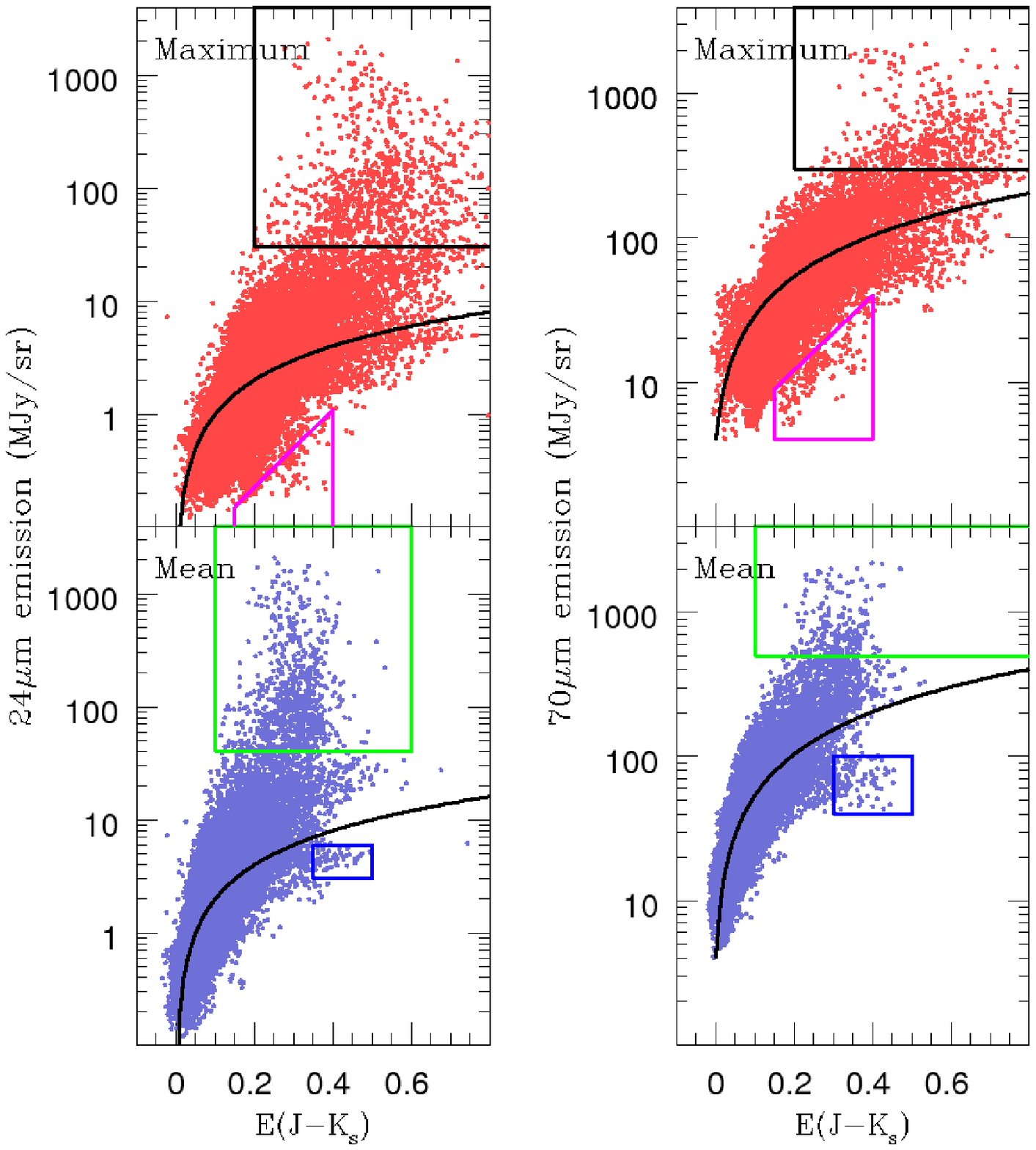}}
  {\label{fig:compmap70}\includegraphics[width=0.32\textwidth,clip=true]{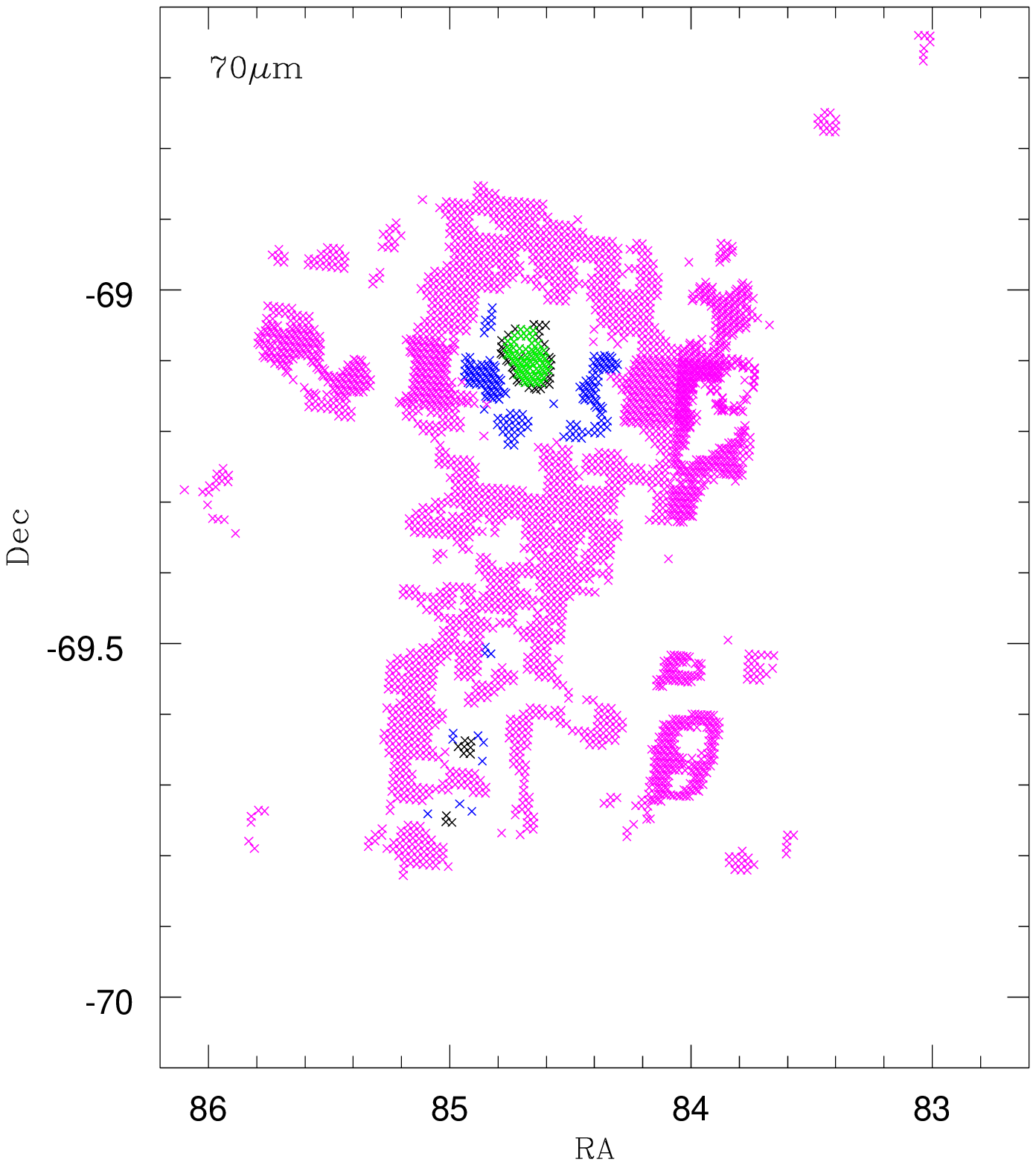}}\\
 \caption{Regions of higher and lower than linear increase in emission for $24\,\mu$m (left) and $70\,\mu$m (right). Smoothed Emission vs. $E(J-K\mathrm{s})$ with selection regions outlined shown in adjacent panels. Green selection largely overlaps black selection.}
\label{fig:compmap}
\end{figure*}

\textnormal{We have mapped some of these regions in Figure \ref{fig:compmap}. The extinction vs. emission ranges depicted in each colour are shown in the panels adjacent to the maps in the figure. Regions below the linear fit line represent where the emission does not detect the effects seen in reddening while above the linear fit is where RC stars do not probe this. We notice the H\,{\sc ii} regions where massive hot stars lie are responsible for the strongest emission (much stronger than a linear increase would suggest). The heat produced by these stars have the greatest effect on the small dust grains (probed by the $24\,\mu$m emission) and this effect does not lead to reddening increases on the same scale which is why these regions are brighter in the $24\,\mu$m emission. Regions with very low emission compared to maximum extinction (shown in magenta) are scattered around the tile edges at $24\,\mu$m but at $70\,\mu$m they generally outline the warm H\,{\sc ii} regions suggesting the two are linked. The fact the maximum extinction is greater also, suggests the dust content is structured on smaller scales in these regions, this is backed by this not being observed for the mean. Alternatively, the radiation field from the H\,{\sc ii} emission affects the $24\,\mu$m more greatly (which is seen by the larger green regions). The regions where there are linear tails in the emission (blue selection) are found in the $24\,\mu$m emission around the previously mentioned crescent and eastern regions of high emission and lower extinction. In the $70\,\mu$m these tails surround the H\,{\sc ii} regions instead. These tails are evident in the mean so effects in these regions are affecting some or most of the RC stars suggesting this dust content is a larger structure. These observations suggest the dense dust is found around the H\,{\sc ii} regions (and it may be that we are unable to probe the highest extinction of these regions, which explains why the reddening appears smaller) with the density decreasing around them but having small-scale structure.}

However, it should be considered that dust grains can have their radii and emissivity modified from being coated by other elements causing their extinction law to change. When this occurs this can lead to an overestimate of the column density and thus, an apparent non-linear increase with respect to reddening such as seen with the H$_2$O ice \citep{oliveira09}. We did observe a non-linear increase to the $24\mu$m emission, though we attribute this to heating which makes ice grains an unlikely candidate. Even so, the reddening, as expected, is not exclusively caused by a single form (composition, size \textnormal{and temperature}) of dust grain.

\subsection{Reddening around R136} \label{30d:r136}
\textnormal{In Section \ref{ssec:dens} and in Figure \ref{fig:map}, we noticed that RC stars were lacking within the R136 region. In this subsection we explore making use of other stellar tracers to compliment the extinction map for this region.}

\begin{figure}
\resizebox{\hsize}{!}{\includegraphics[angle=0]{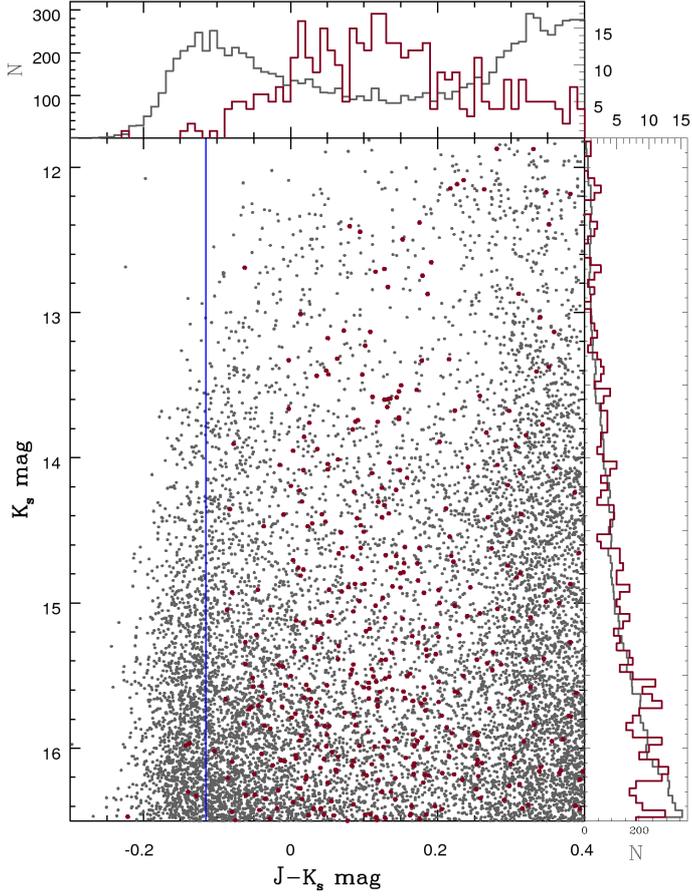}}
\caption{CMD of all stars (grey), stars around R136 (burgundy). Also plotted are histograms of colour and magnitude (\textnormal{bin sizes are the same as used in Figure \ref{fig:cmd}}). Magnitude distribution is similar for both sets. Colour distribution shows an offset of $\sim0.2$ mag in the R136 region indicating reddening. Vertical blue line at $(J-K_\mathrm{s})_0=-0.115$ mag is the \textnormal{baseline} colour.}
\label{fig:30d:r136cmd}
\end{figure}

In Section \ref{ssec:dens} we noticed a high density of stars around the R136 region in the CMD selection brighter than $K_\mathrm{s}=16.5$ mag (a CMD range which we defined as being of intermediate age). In the CMD shown in Figure \ref{fig:30d:r136cmd} we focus on this part of the CMD, ignoring the RGB. The sources around R136 (in \textnormal{burgundy}) are mainly found at $(J-K_\mathrm{s})\simeq0.2$ mag \textnormal{(much bluer than the RGB)} and resemble the reddened population of \textnormal{young} sources (in grey) at $(J-K_\mathrm{s})\simeq0$. \textnormal{The histograms in Figure \ref{fig:30d:r136cmd} show that the magnitude of stars around R136 follows the same pattern as the rest of the tile. The colour on the other hand, is offset confirming the earlier observation.}

\textnormal{This offset can be used to estimate the intrinsic colour of the population} by assuming that the bulk of the population in the tile is not affected by reddening and so adopting a baseline colour of $(J-K_\mathrm{s})=-0.115$ mag. \textnormal{However, as this is estimated from the CMD it is affected by foreground Galactic extinction and so to account for this we must also subtract this, $(J-K_\mathrm{s})=0.04$ mag, finding an intrinsic colour\footnote{\textnormal{Note that isochrone tracks (including version 2.4 of the Padova isochones, \citealt{bressan12}) do not cover this region of the CMD.}} of $(J-K_\mathrm{s})_0=-0.155$ mag.}

The reddening we can probe \textnormal{using} these stars is limited \textnormal{to $A_V\simeq2.9$ mag} due to contamination from the redder supergiant and RGB populations. However, this should be sufficient for R136 itself as it is not highly reddened (although its surroundings are). We use the reddening to firstly, contrast the effect reddening has on young and intermediate aged populations, and secondly, better examine this region as the young population gives use more coverage \textnormal{in the centre} compared to the RC stars.

\begin{figure}
\resizebox{\hsize}{!}{\includegraphics[angle=0,clip=true]{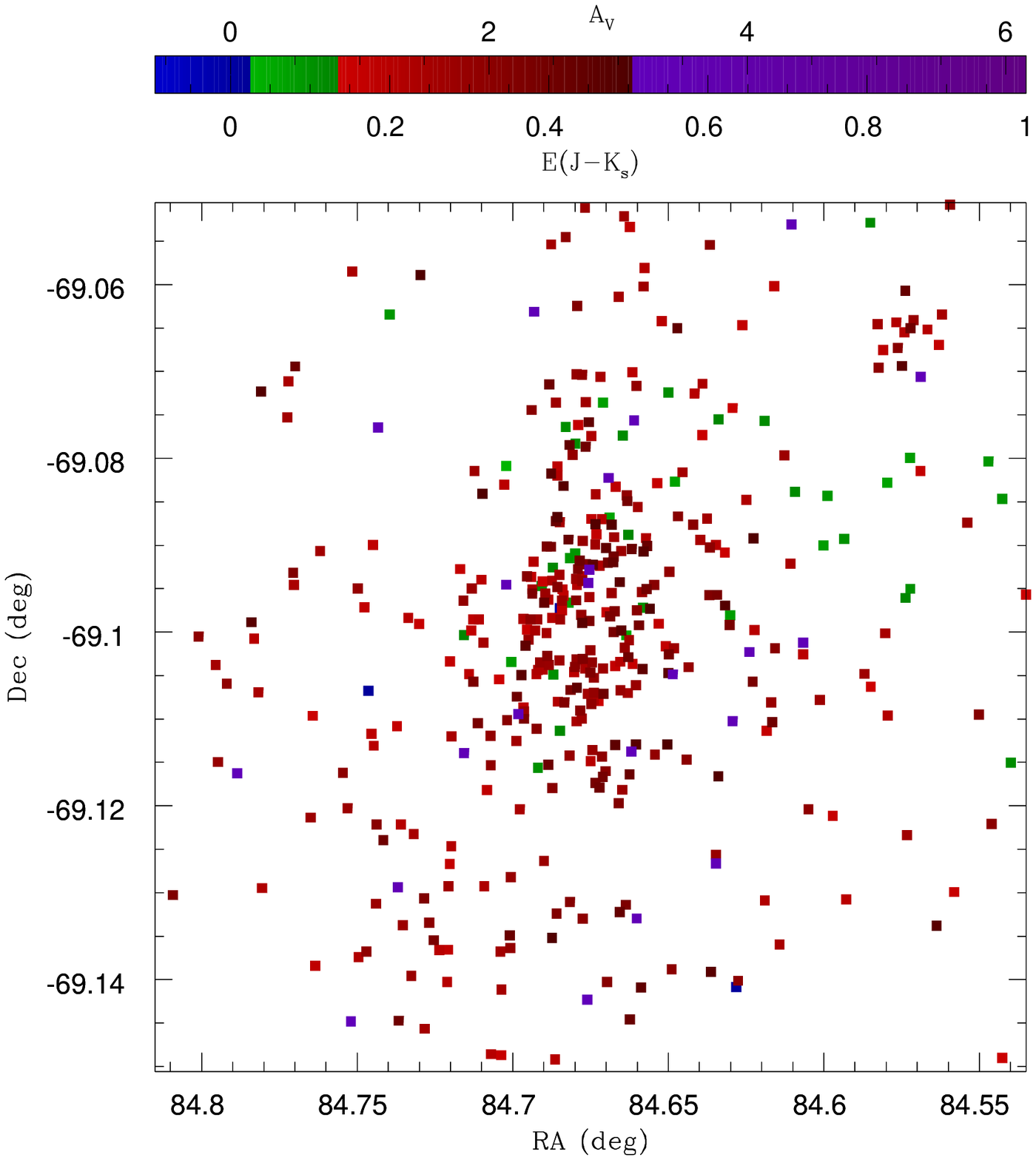}}
\resizebox{\hsize}{!}{\includegraphics[angle=0,clip=true]{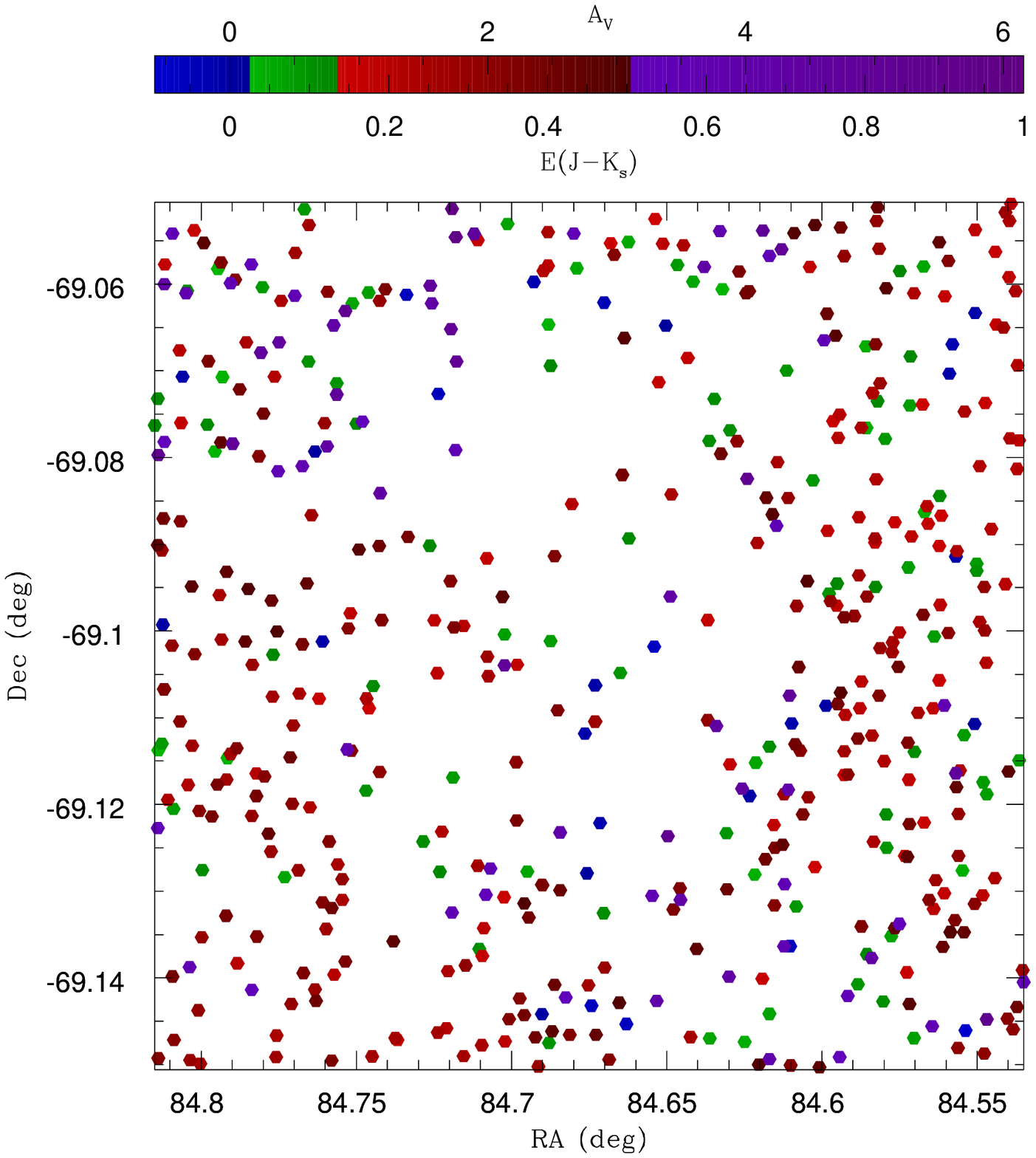}}\\
\caption{Reddening map of R136 region using CMD selection from Fig. \ref{fig:30d:r136cmd} (top, squares) and RC stars (bottom, circles). The reddening scale is consistent with Figure \ref{fig:map}.}
\label{fig:30d:r136map}
\end{figure}

The reddening map we produce is shown in Figure \ref{fig:30d:r136map} with the R136 selection plotted as \textnormal{squares (top panel, total 407 sources)} and the RC stars plotted as \textnormal{circles (bottom panel, total 500 sources)}. We use the same scale as Figure \ref{fig:map}. \textnormal{Distribution-wise the younger population is found more towards the centre than the RC stars and the reddening probed is slightly lower but covers a narrower range. The mean and median reddening (given in Table \ref{tab:r136rc}) are very similar for the two populations (with a difference of around $0.01$ mag). The larger mean in the RC stars and hence the larger difference between the mean and median is due to the RC stars probing a larger extinction range.}

\begin{table}
\centering
\caption{Mean and Median reddening for R136 selection and RC stars.}
\label{tab:r136rc}
\begin{tabular}{ccc}
\hline
\hline
Selection & \multicolumn{2}{c}{$E(J-K_\mathrm{s})$ (mag)} \\
 & Mean & Median \\
\hline
R136 & 0.288 & 0.280 \\
RC & 0.297 & 0.271 \\
\hline
\end{tabular}
\end{table}

\begin{figure}
\resizebox{\hsize}{!}{\includegraphics[angle=0]{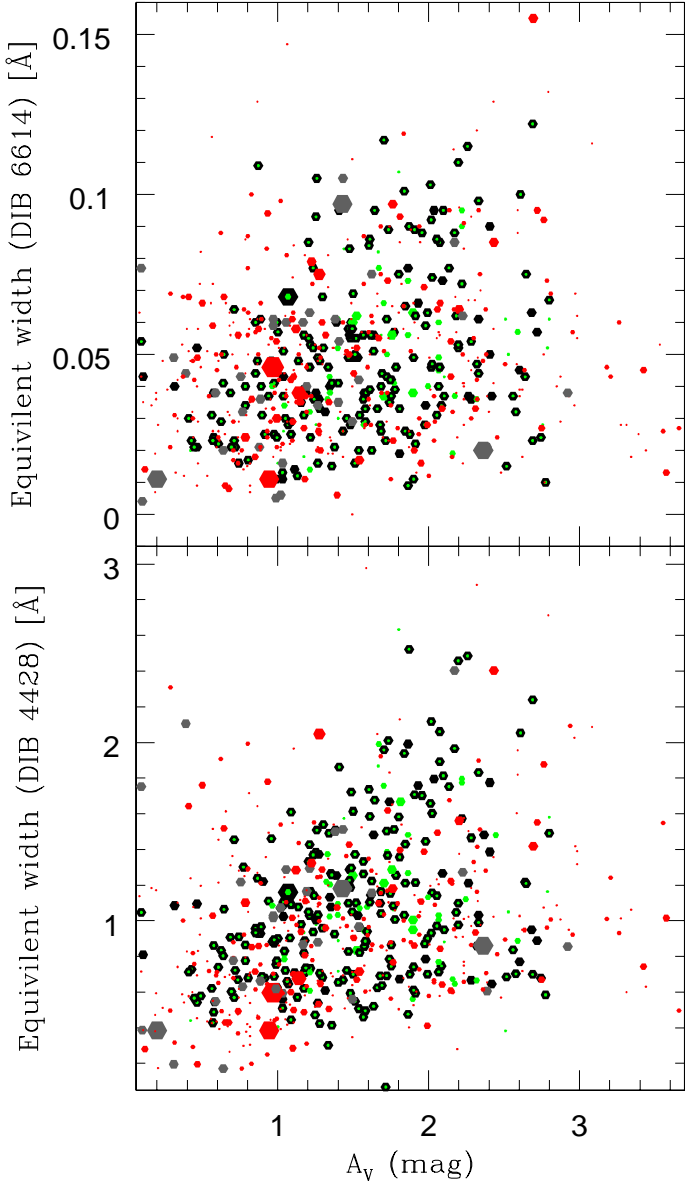}}
\caption{\textnormal{O- and B-type star DIB $6614\AA$ (top), DIB $4428\AA$ (bottom) equivalent widths ($\AA$) vs. average extinction ($A_V$, mag) of nearby young stars and nearby RC stars. Average extinction of young stars within 1\arcsec and 5\arcsec are shown in black and green. Average extinction of RC stars within 5\arcsec and 15\arcsec are shown in grey and red.
Greater number of stars in selection is depicted by larger points. Green and red points use smaller scale than black points and grey points.}}
\label{fig:30d:r136db}
\end{figure}

Another comparison that can be made is with the Diffuse Interstellar Bands (DIBs) analysed in \citet{vanloon13}. They have compared DIB equivalent widths (combining the Galactic and LMC components) to $A_V$ measured from the Very Large Telescope Flames Tarantula Survey (VFTS) sample (which mostly contains O- and B-type stars). This is shown in Figure 14 of \citet{vanloon13} (hereafter, Figure vL). We compare th\textnormal{e DIB equivalent widths} with the \textnormal{reddening} of our younger \textnormal{(R136)} and intermediate \textnormal{(RC)} aged star populations. For our younger selection, we select stars within a 1\arcsec\ radius and 5\arcsec\ radius, averaging extinction where there is more than 1 star. We also do the same for the RC, but due to fewer stars we have to use radii of 5\arcsec\ and 15\arcsec\ instead. Figure \ref{fig:30d:r136db} shows the results of DIBs $4428\AA$ and $6614\AA$ equivalent widths (with Galactic and LMC components combined) with wider radii points plotted in green \textnormal{and red (for young and RC stars respectively)} and the smaller plotted in black \textnormal{and grey (for young and RC stars respectively)}. The point sizes are based on number of sources within radii (larger radii contain more sources) where green points use a smaller scale to black points.

Comparing Figure \ref{fig:30d:r136db} to Figure vL we notice both plots have a lot of scatter and also the $A_V$ range probed by our young stars is slightly smaller. The RC stars on the other hand are even more scattered. Looking at the data extremes and best fit of Figure vL and our young stars, there is great similarity in the data range \textnormal{and the best fit up to} $A_V=2$ mag but after that our data becomes a bit narrower and steeper in comparison. This may be due to drop-off from \textnormal{a} lack of sources. However, this steeper gradient suggests the young stars \textnormal{lie} further behind the dust than the O- and B-type stars. The RC stars cover a wider extinction range and similar emission range. This suggests their population is more mixed with regards to position in dust.
\textnormal{However what occurs at the high and low ends of extinction is interesting. At high extinction we see the DIBs are weak and there are few RC stars. These RC stars may be seen behind dense structure while the OB stars are seen though less dense structure. At low extinction we see more RC stars where a few regions have strong DIB absorption which suggests that these RC stars are in front or in between the bulk of the dust.}

This subsection shows that it is possible to supplement the RC sample with other stellar populations within some regions. However, different populations probe the ISM differently and have different spatial positions in relation to the ISM.


\section{Conclusions}
\label{30d:conc}
Using the VMC observations we selected RC stars for the 30 Doradus region in the LMC. Isochrones were combined with SFH to determine the intrinsic colour for a RC star for calculation of extinction. We converted this extinction into $A_V$ values and produced a map of this made available in several formats. We are able to use the RC stars to probe extinction up to $A_V=6$ mag. The map has been used to:
\begin{itemize}
\item Compare with the optical reddening map (also based on RC stars) of \citet{haschke11} (which we converted to $A_V$). We found the most disagreements at the extreme ends of the data as infrared is less affected by low extinction and optical being too heavily affected by high extinction. In general though we found \citeauthor{haschke11} for a given value had $75\%$ of the $A_V$ we found.
\item Compare H\,{\sc i} column densities \citep{kim98} by converting the $A_V$ into total H column density. We encountered limitations in the H\,{\sc i} ISM (at very high densities hydrogen becomes molecular) but were able to use this to determine the atomic--molecular transition starts at column densities of $N_H\,\simeq\,4\times10^{21}$ cm$^{-2}$ and is largely complete at $N_H\,\simeq\,6\times10^{21}$ cm$^{-2}$. Also, we were able to map out molecular cloud regions based on the total column density and excluding regions of atomic hydrogen.
\item Compare with the Spitzer SAGE dust emission \citep{meixner06} finding overall agreement with some differences in the detail, due presumably to temperature variations of the emitting grains. In particular an eastern region is prominent in extinction maps and the H\,{\sc i} (as high extinction and dense regions, respectively) but is not a feature in the \textnormal{M}IR emission maps. This may highlight some small scale structure. In general we found the $70\mu$m emission a better tracer than the $24\mu$m emission. This may be due to RC stars not heating dust to the extent warm stars would and/or RC stars not being found in the proximity of warm stars (which would warm the dust).
\end{itemize}

Finally, we take another look at reddening in the R136 region using an abundance of young stars present in the region. We see the young stars have similar reddening to the RC stars and use them to examine the relationship between $A_V$ and DIB equivalent width using data from the FLAMES Tarantula survey \citep{vanloon13} finding similar results to what they found when using O-and B-type stars. This shows it is possible to use other populations to supplement the reddening map.

The future work lies in extending the method onto other tiles in the VMC survey, and for this tile, dereddening the RC stars, to determine three dimensional structure.

\begin{acknowledgements}
BLT acknowledges an STFC studentship awarded to Keele University.
RdG acknowledges research support through grant 11073001 from the National Natural Science Foundation of China (NSFC).
The VISTA Data Flow System pipeline processing and science archive are described in \citet{irwin04} and \citet{cross12}. We have used data from the v20120126 VSA data release, which is described in detail in Paper I. 
\textnormal{We thank the anonymous referee for his/her comments that helped improve the presentation of the paper.}
\end{acknowledgements}

\bibliographystyle{aa}
\bibliography{30d}

\appendix
\section{Red Clump peak variances} \label{ssec:rcvar}
\begin{figure}
\resizebox{\hsize}{!}{\includegraphics[width=\textwidth,clip=true]{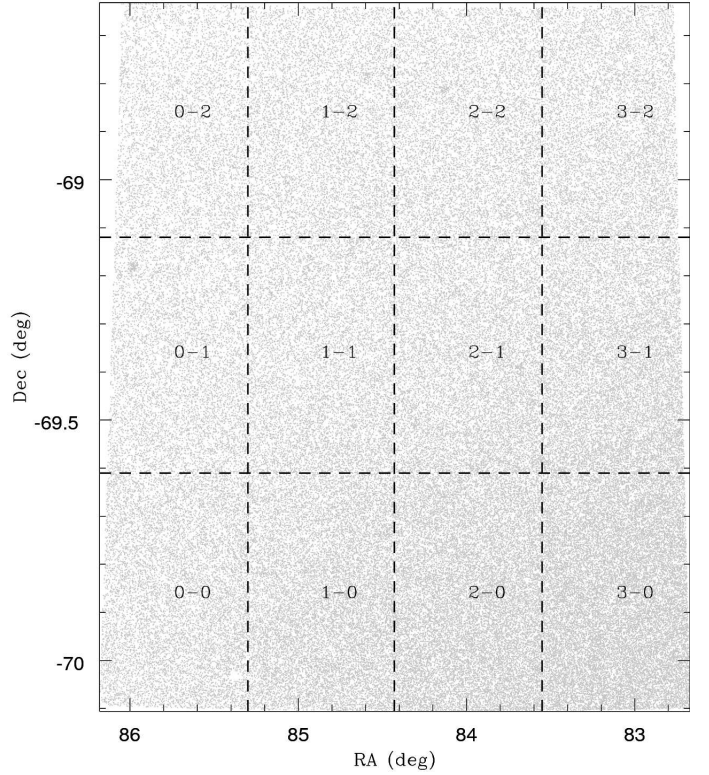}}
\caption{Field $6\_6$ split into a $4\times3$ array for analysis of RC variance (shown in Figure \ref{fig:gramcomp}). Region names labelled.}
\label{fig:varmap}
\end{figure}

\begin{figure*}
\begin{centering}
\includegraphics[height=\textwidth,angle=-90,clip=true]{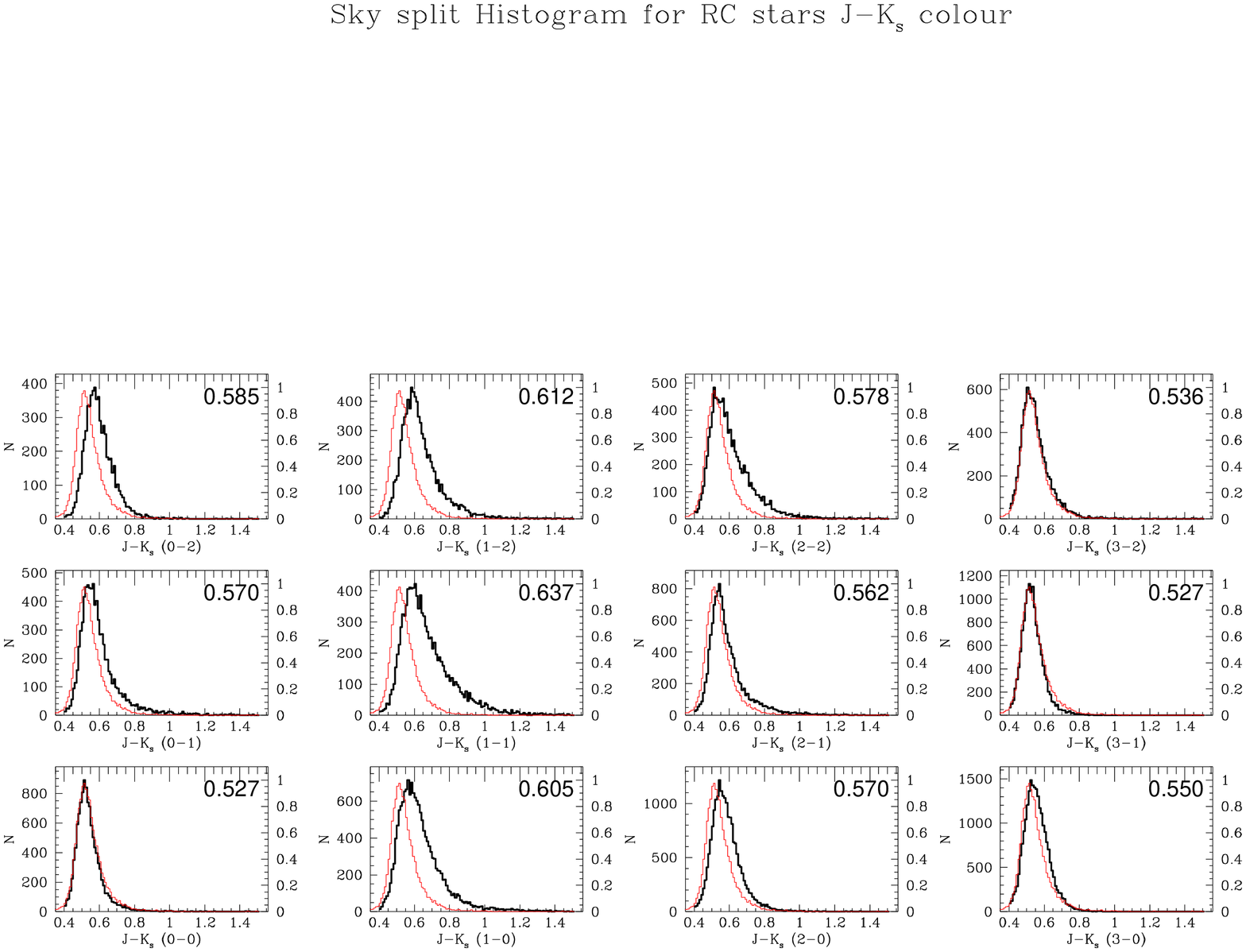}\\
\includegraphics[height=\textwidth,angle=-90,clip=true]{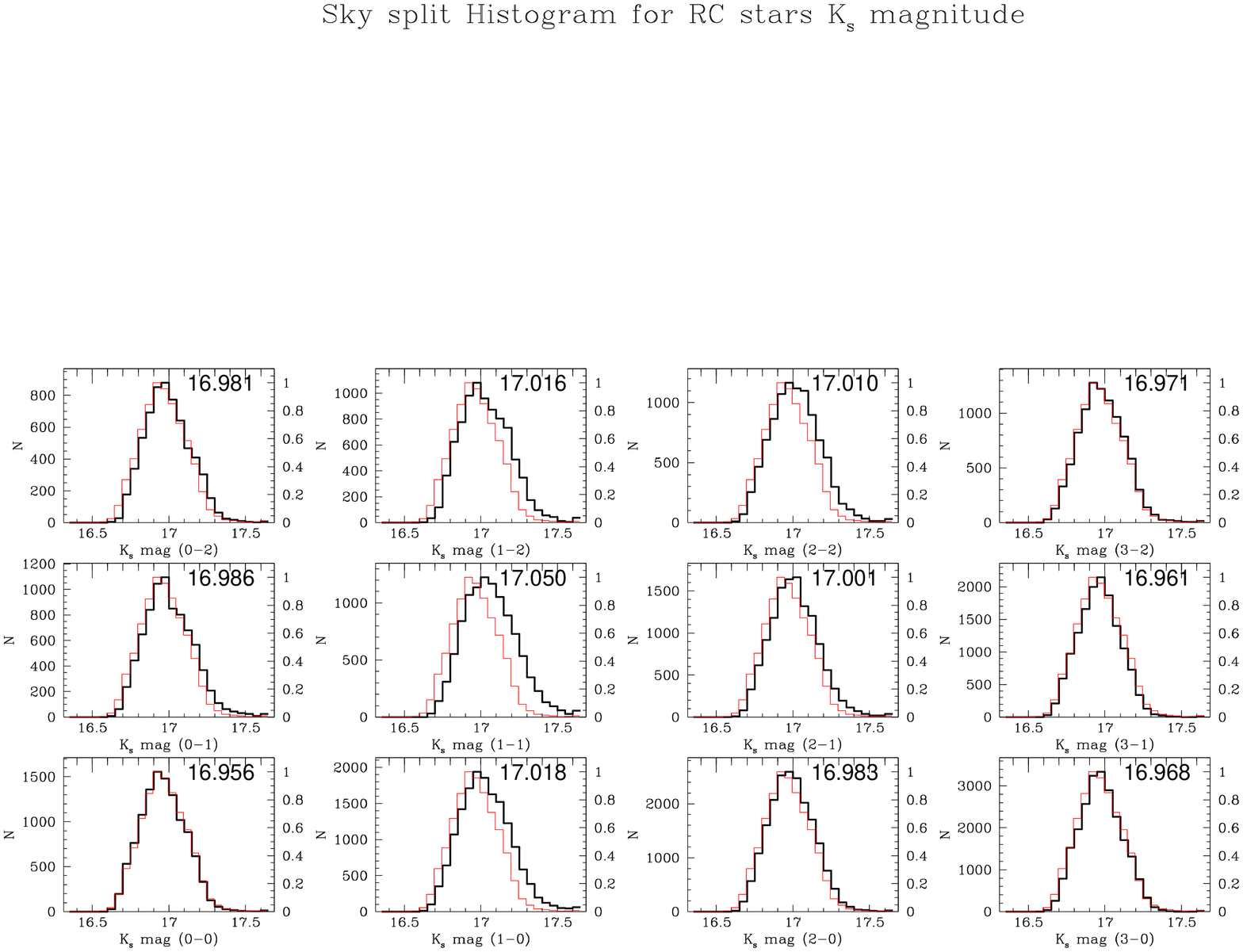}
\caption{Histograms of the $(J-K_\mathrm{s})$ colour (top 3 rows, bin size of $0.01$ mag) and $K_\mathrm{s}$ magnitude (bottom 3 rows, bin size $0.05$ mag) across the field (regions in parentheses defined in Table \ref{tab:clumppos}). Value in top--right is the median. Overplotted in red is the normalised mean of regions $0$-$0$ and $3$-$2$.}
\label{fig:gramcomp}
\end{centering}
\end{figure*}

We analyse how the RC varies across the tile by splitting the tile into a $4\times3$ array ($12$ regions; shown in Figure \ref{fig:varmap}). We locate the CMD area of highest density, corresponding to the most populated bin (using the same bin sizes as before: $0.01$ mag in $J-K_\mathrm{s}$ and $0.05$ mag in $K_\mathrm{s}$; this limits their accuracy) and compare this with the median for colour and magnitude for each region; the values are in Table \ref{tab:clumppos}. Figure \ref{fig:gramcomp} shows the histogram distribution of the RC stars in colour (top) and magnitude (bottom). For each region the normalised (with respect to the peak) average of regions $0$-$0$ and $3$-$2$ are overplotted in dashed red as a low extinction comparison. From this, it is seen that the regions with higher median colours have their source distribution extended in the redder component rather in addition to a complete offset (i.e. they are positively skewed). This suggests for a given region not all of the stars are affected by reddening but there is a multi-layer structure with some stars lying in front of the cause of the extinction. When comparing the contour peaks, which mirror the histogram peaks, to the median values, the difference is generally larger when the extinction is higher. The clearest example of this effect is in region $1$-$1$ where 30 Doradus lies with a difference of $0.067$ mag. Regions found to the north--west and south ($1$-$0$, $0$-$1$, $0$-$2$, $1$-$2$) have similar difference. All these regions have strong H\,{\sc i} and H$\alpha$ emissions which may be related to the cause of this effect.

\begin{table}
\caption{Red Clump peak position across field $6\_6$ from median in CMD region and contour map peak position.}
\label{tab:clumppos}
\[
\begin{tabular}{l@{ }c@{ }c@{ }l@{ }c@{ }c@{ }l@{ }}
\hline
\hline
\noalign{\smallskip}
Region & \multicolumn{2}{c}{Centre (deg, J2000)} & \multicolumn{2}{c}{Median} & Contour\tablefootmark{a} & N$_\mathrm{RC}$ \\
 & RA & Dec & $(J-K_\mathrm{s})$ & $K_\mathrm{s}$ & $(J-K_\mathrm{s})$ & \# \\
\hline
0-0 & 85.74 & $-$69.87 & 0.527 & 16.956 & 0.52 & 11396\\
0-1 & 85.74 & $-$69.37 & 0.540 & 16.986 & 0.56 & 8168 \\
0-2 & 85.74 & $-$68.87 & 0.585 & 16.981 & 0.57 & 6389 \\
1-0 & 84.87 & $-$69.87 & 0.605 & 17.018 & 0.56 & 15894\\
1-1 & 84.87 & $-$69.37 & 0.637 & 17.050 & 0.57 & 10619\\
1-2 & 84.87 & $-$68.87 & 0.612 & 17.016 & 0.60 & 8613 \\
2-0 & 83.99 & $-$69.87 & 0.570 & 16.983 & 0.55 & 19677\\
2-1 & 83.99 & $-$69.37 & 0.562 & 17.001 & 0.54 & 12763\\
2-2 & 83.99 & $-$68.87 & 0.578 & 17.010 & 0.51 & 9658 \\
3-0 & 83.11 & $-$69.87 & 0.550 & 16.968 & 0.52 & 23180\\
3-1 & 83.11 & $-$69.37 & 0.527 & 16.961 & 0.53 & 14534\\
3-2 & 83.11 & $-$68.87 & 0.536 & 16.971 & 0.53 & 9432 \\
\noalign{\smallskip}
\hline
\end{tabular}
\]
\tablefoot{\tablefoottext{a}{For contour method the peak in magnitude is always $K_\mathrm{s}=16.95$ mag.}}
\end{table} 

\section{Extinction laws} \label{30d:law}
In section \ref{30d:av} we mentioned that there are alternative extinction laws that could have been used instead (or rather, would have to have been used had the $E(J-K_\mathrm{s})$ coefficients not already been calculated for our data in a model independent way). Here, we discuss these in more detail.

Table 6 of \cite{schlegel98} gives $A_J/A_V = 0.276$ and $A_K/A_V = 0.112$. Therefore, $E(J-K) = 0.164\times A_V$. These \citeauthor{schlegel98} values are for $JHK$ for UKIRT's IRCAM3, while our observations are in VISTA $YJK_\mathrm{s}$. Both the UKIRT and VISTA photometry are in a Vega magnitude system.

\begin{figure}
\resizebox{\hsize}{!}{\includegraphics[angle=0,width=\textwidth]{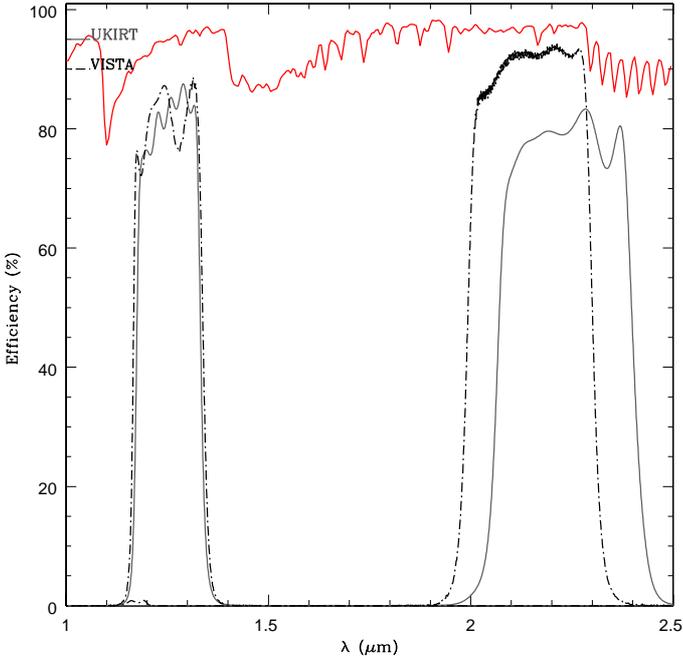}}
\caption{Comparison of wavelength vs. efficiency for the transmission of UKIRT (IRCAM3) $J$ and $K$ filters (in grey) and VISTA $J$ and $K_\mathrm{s}$ filters (in dashed black). Curves from JAC and ESO, respectively. Synthetic spectra of a typical RC star in red.}
\label{fig:kcomp}
\end{figure}

However, UKIRT's IRCAM3 and VISTA have different filter curves and system transmissions in detail, particularly in $K/K_\mathrm{s}$. This causes the post-calibration magnitudes to vary slightly. In Figure \ref{fig:kcomp}, the filter curves and synthetic spectra of a RC star are shown (produced using model atmospheres from \citealt{castelli04}). The $J$ band widths are very similar, the main difference being at about $1.27\,\mu$m where the VISTA transmission dips while the UKIRT one rises. The $K_\mathrm{s}$ passband curve begins before the $K$ passband, finishes before it and the overall width is nearly identical. The extinction coefficients do vary slightly with spectral type \citep{vanloon03}. This variance is small in the NIR (0.004 mag from A0 to M10 star in the $J$ band). 

The $\lambda_\mathrm{eff}$ and central wavelength values are given in Table \ref{tab:avrec}. The ratios to $A_V$ are derived from the $R_V=3.1$ curve meaning the $A_{K_\mathrm{s}}/A_V$ value is derived using equations 1, 2a and 2b in \citet{cardelli89}; the results are shown in the rightmost column of Table \ref{tab:avrec}.

\begin{table}
\centering
\caption{Extinction ratio $A_{\lambda}/{A_V}$ for VISTA passbands.}
\label{tab:avrec}
\[
\begin{tabular}{cccccc}
\hline
\hline
Passband & $\lambda_\mathrm{central}$ ($\mu$m) & $\lambda_\mathrm{eff}\,(\mu$m) & a(x) & b(x) & $A_{\lambda}/A_V$\\ 
\hline
VISTA $J$ & 1.252\tablefootmark{a} & 1.254 & 0.3987 & $-0.3661$ & 0.2806 \\
VISTA $K_\mathrm{s}$ & 2.147\tablefootmark{a} & 2.149 & 0.1675 & $-0.1538$ & 0.1179 \\
UKIRT\tablefootmark{b} $J$ & 1.250\tablefootmark{c} & 1.266 & 0.3926 & $-0.3605$ & 0.2764 \\
UKIRT\tablefootmark{b} $K$ & 2.200\tablefootmark{c} & 2.215 & 0.1595 & $-0.1465$ & 0.1123 \\
\hline
\end{tabular}
\]
\tablefoot{References:\\
\tablefoottext{a}{\url{http://casu.ast.cam.ac.uk/surveys-projects/vista/technical/filter-set/vista-filter-set}}
\tablefoottext{b}{UKIRT IRCAM3}
\tablefoottext{c}{\citet{schlegel98}}
} 
\end{table}

Using the VISTA values the conversion becomes $E(J-K_\mathrm{s}) = 0.1627\times A_V$, a difference of $\sim0.01$ mag with band-by-band differences of $\sim0.05$ mag when comparing the two systems. A drawback of this method is that it assumes the filter is well represented by the $\lambda_\mathrm{eff}$ point, rather than taking into account the full filter transmission curve. When comparing with the result from Section \ref{30d:av}; $E(J-K_\mathrm{s}) = 0.16237\times A_V$, we see this simplification has had a rather small effect on the conversion, finding $102\%$ of the used value and a maximum $A_V$ difference of $\sim-0.013$ mag.

Some of the difference between UKIRT and VISTA values might be accounted for by how the $\lambda_\mathrm{eff}$ is calculated because the VISTA $\lambda_\mathrm{eff}$ values were calculated using equation $3$ of \citet{fukugita96} (which was defined in \citealt{schneider83}) and accounting for the Quantum Efficiency curve while \citet{schlegel98} did not state how the UKIRT $\lambda_\mathrm{eff}$ was calculated (only that it represents the point on the extinction curve with the same extinction as the full passband).

\section{Using the $Y$ band} \label{30d:app}
\begin{figure}
\resizebox{\hsize}{!}{\includegraphics[angle=0]{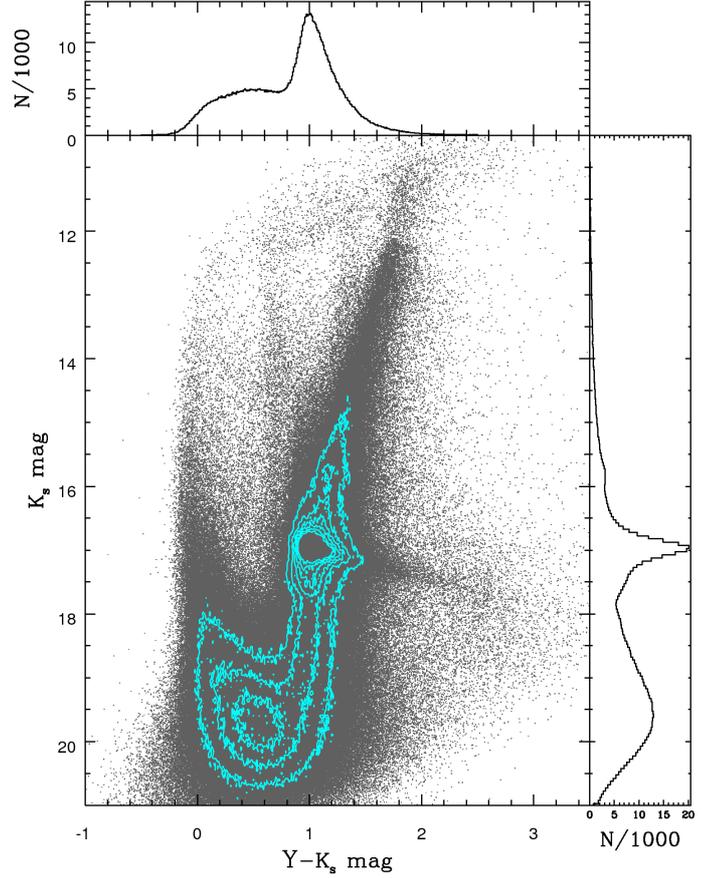}}
\caption{$K_\mathrm{s}$ vs. $(Y-K_\mathrm{s})$ CMD of LMC field $6\_6$ accompanied by \textnormal{cyan} contours representing density and histograms of $(Y-K_\mathrm{s})$ (bin size $0.01$ mag) and $K_\mathrm{s}$ (bin size $0.05$ mag). It can be seen that this field is largely made up of main sequence and RC stars.}
\label{fig:ykcmd}
\end{figure}

\begin{figure}
\resizebox{\hsize}{!}{\includegraphics[angle=0,width=\textwidth]{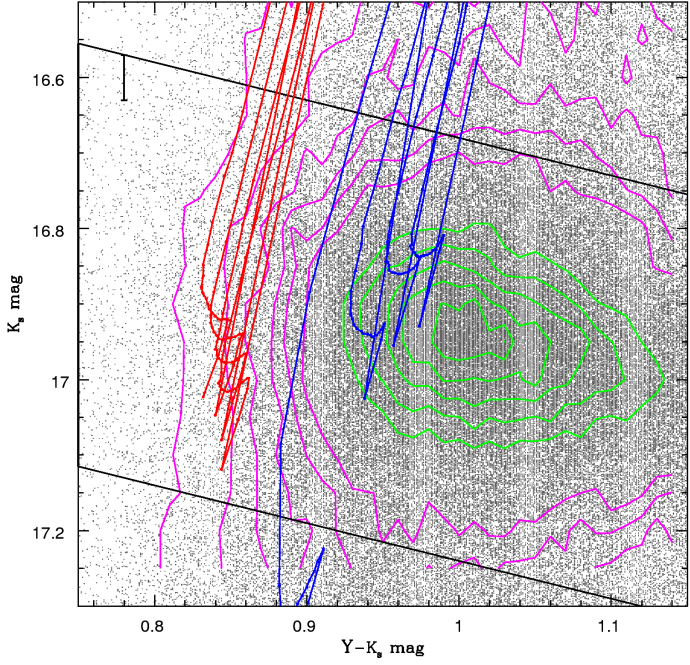}}
\caption{$(Y-K_\mathrm{s})$ vs. $K_\mathrm{s}$ CMD with isochrone lines for the helium burning sequence for the same two metallicities and age ranges as Section \ref{30d:red}. Younger, more metal rich (log$\,(t/{\mathrm{yr}})=9.0-9.4$, $Z=0.0125$) are in blue and the older, more metal poor (log$\,(t/{\mathrm{yr}})=9.4-9.7$, $Z=0.0033$) are in red. Error bars represent error in LMC distance. Contours are plotted in \textnormal{magenta} and green and have the same levels as in Figure \ref{fig:rcloc}. The older population can be seen to be at the $5\%$ contour level.}
\label{fig:isoliy}
\end{figure}

Figure \ref{fig:ykcmd} shows a CMD of $K_\mathrm{s}$ mag vs. $(Y-K_\mathrm{s})$. It is very similar to the  $K_\mathrm{s}$ vs. $(J-K_\mathrm{s})$ CMD seen in Figure \ref{fig:cmd} with the difference of an extended colour axis due to a larger wavelength difference between the two bands. Focusing on the region of the RC we overplot the isochrones used in Section \ref{30d:red}. This is shown in Figure \ref{fig:isoliy}. We see that the older isochrones lie bluer than the RC. This confirms, as expected, that the $Y$ band is more affected by extinction towards the MCs than $J$ or $K_\mathrm{s}$ are. When applying $A_V=0.249$ mag \citep{schlegel98} as foreground Galactic extinction the $(Y-K_\mathrm{s})$ and $(J-K_\mathrm{s})$ are in a consistent position in the context of the contour levels (around the $35\%$ level). This shows that the $Y$ band, being closer to optical is more affected by extinction which would be a boon for low extinction regions because the contrast between extinctions would be greater. However, for a high extinction region it offers no advantage and has the disadvantage of stars being fainter in $Y$ than in $K_\mathrm{s}$. Colour--colour diagrams do not aid in selection of RC stars.

\begin{figure}
\resizebox{\hsize}{!}{\includegraphics[angle=0]{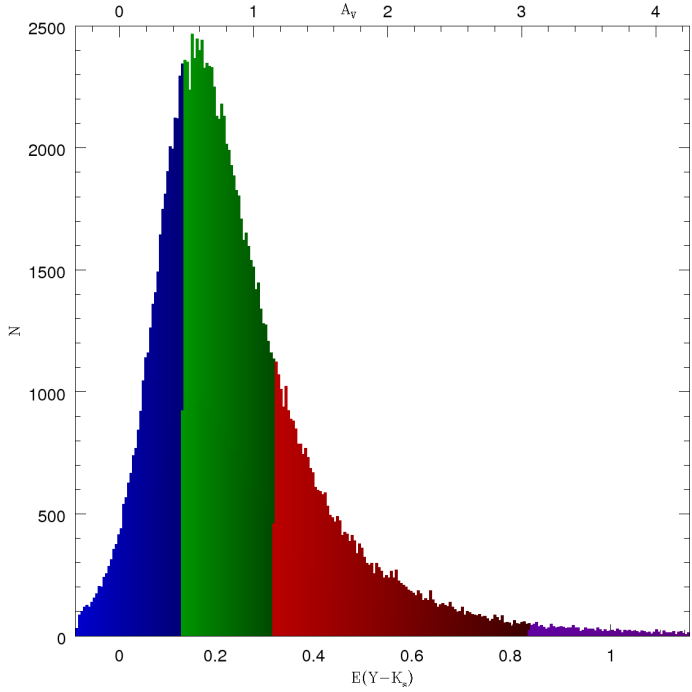}}
\caption{Histogram of $E(Y-K_\mathrm{s})$ (bottom) and $A_V$ (top) distribution with bin size of $0.005$ mag. Green is the region covering $50\%$ of the sources centred on the median where darker blue is lower than this and brighter red higher and \textnormal{purple} extremely high.} 
\label{fig:ycolgram}
\end{figure}

The RC selection box is defined like it is in Section \ref{30d:rcloc}. For $(Y-K_\mathrm{s})$ \textnormal{the} reddening vector is shallower (with a gradient of $0.5$) and the colour limits are $0.75<(Y-K_\mathrm{s})<2$ mag. Combined with an intrinsic colour of $E(Y-K_\mathrm{s})_0=0.84$ mag and $E(Y-K_\mathrm{s})=0.2711\,\times\,A_V$ the furthest we can probe is $A_V=4.28$ mag. The histogram for the RC selection is shown in Figure \ref{fig:ycolgram}.  Comparing with Figure \ref{fig:colgramexcite} we see the peak is at a slightly higher $A_V$ in the selection (due to greater foreground Galactic extinction probed). The percentage of stars relative to the peak at $A_V=1$ mag and $A_V=2$ mag is very similar in both histograms.

\end{document}